\documentclass[a4paper,11pt]{article}
\pdfoutput=1 

\usepackage{jcappub} 
\usepackage[T1]{fontenc} 

\newcommand\myeq{\stackrel{\mathclap{\normalfont\mbox{$\star$}}}{=}}
\newcommand\myeqs{\stackrel{\mathclap{\normalfont\mbox{$\star \star$}}}{=}}

\newcommand{\ta}{\tilde{a}}
\newcommand{\vk}{\vec{k}}
\newcommand{\vq}{\vec{q}}

\newcommand{\vx}{\vec{x}}

\newcommand{\tT}{\Theta}
\newcommand{\td}{\delta}
\def\r{\rho}

\def\d{{\partial}}
\def\nn{{\nonumber}}

\newcommand{\hinvMpc}{\,h\, {\rm Mpc}^{-1}\,}

\title{\LARGE Tree-Level Bispectrum in the Effective Field Theory of Large-Scale Structure extended to Massive Neutrinos}

\author[a,b,c]{Roger de Belsunce}
\author[b,c]{and Leonardo Senatore}

\affiliation[a]{Institute for Particle Physics and Astrophysics, Department of Physics, \\
ETH Zurich, Wolfgang-Pauli-Strasse 27, 8093 Zurich, Switzerland}
\affiliation[b]{Stanford Institute for Theoretical Physics, \\
Stanford University, Stanford, CA 94305, USA}
\affiliation[c]{Kavli Institute for Particle Astrophysics and Cosmology, \\
Stanford University and SLAC, Menlo Park, CA 94025, USA}

\emailAdd{rmvd2@cam.ac.uk}
\emailAdd{senatore@stanford.edu}

\abstract{We compute the tree-level bispectrum of dark matter in the presence of massive neutrinos in the mildly non-linear regime in the context of the effective field theory of large-scale structure (EFTofLSS). For neutrinos, whose typical free streaming wavenumber ($k_{\rm fs}$) is longer than the non-linear scale ($k_{\mathrm{NL}}$), we solve a Boltzmann equation coupled to the effective fluid equation for dark matter. We solve perturbatively the coupled system by expanding in powers of the neutrino density fraction ($f_{\nu}$) and the ratio of the wavenumber of interest over the non-linear scale ($k/k_{\mathrm{NL}}$) and add suitable counterterms to remove the dependence from short distance physics. For equilateral configurations, we find that the total-matter tree-level bispectrum is approximately $16f_{\nu}$ times the dark matter one on short scales ($k > k_{\rm fs}$). The largest contribution stems from the back-reaction of massive neutrinos on the dark matter growth factor. On large scales ($k < k_{\rm fs}$) the contribution of neutrinos to the bispectrum is smaller by up to two orders of magnitude.}

\arxivnumber{1804.06849}

\begin{document}
\maketitle
\flushbottom

\section{Introduction}
The standard model of cosmology predicts the formation and statistical distribution of large-scale structure (LSS) through gravitational amplification of density perturbations; it describes the universe as a spatially flat $\Lambda$CDM cosmology. The cosmic microwave background (\textsc{cmb}) has been measured with great precision and significantly increased our understanding of the primordial universe. In the next few years we are going to explore all the remaining information within reach. The next leading source of cosmological information will probably be the upcoming LSS surveys such as \textsc{Euclid}\footnote{\url{http://sci.esa.int/euclid/}} and the ground-based Large Scale Synoptic Telescope (\textsc{lsst}\footnote{\url{http://www.lsst.org}}). The three-dimensional structure of LSS will provide us with a lot of information about the cosmological standard model, in particular, in comparison to the two-dimensional nature of the \textsc{cmb}. The amount of data one can extract from the cosmological surveys scales as the cube of the number of modes, the largest reliably observable wavenumber $k_{\mathrm{max}}$. Therefore, it is essential to have an accurate analytic prediction of LSS formation at the smallest scales possible. 

The \textit{effective field theory of large-scale structure} (EFTofLSS), initially proposed in \cite{Baumann:2010tm, Carrasco:2012cv,Porto:2013qua}, has been introduced to augment the predictive power for studying LSS in the mildly non-linear regime. It provides an accurate perturbative framework to describe the statistical distribution of the cosmological LSS of the universe at long distances. The basic idea of the EFTofLSS approach is to include additional terms in the equation of motions for long wavelengths in order to account for the effect of short distance physics at those wavelengths. The equations of motion for dark matter are solved perturbatively in an expansion in powers of the wavenumber of interest over the non-linear scale, $k/k_{\mathrm{NL}}$. In theory, the EFTofLSS is arbitrarily accurate when going to very high perturbative order. Including counterterms, that systematically account for the uncontrollable non-linear short distance physics, is crucial for describing the dynamics of the system. The perturbative treatment of neutrinos is particularly complex if done in a way such that theoretical errors are under control, it is therefore important to be aware of the former development of the EFTofLSS where some of the ideas, that are needed in the context of neutrinos, first appear. So far, in the context of the EFTofLSS the following calculations and comparisons with simulations have been performed: 
the dark-matter power spectrum computations at one- and two-loop orders as well as the comparison with simulations have been presented in \cite{Carrasco:2012cv, Carrasco:2013sva, Carrasco:2013mua, Carroll:2013oxa, Senatore:2014via, Baldauf:2015zga, Foreman:2015lca, Baldauf:2015aha, Cataneo:2016suz, Lewandowski:2017kes}. Further theoretical work, especially the thorough understanding of renormalization \cite{Carrasco:2012cv,Pajer:2013jj,Abolhasani:2015mra} (including more subtle components such as lattice-running \cite{Carrasco:2012cv}), a procedure to obtain the numerical value of the counterterms through simulations \cite{Carrasco:2012cv,McQuinn:2015tva}, the discussion and understanding of non-locality in time \cite{Carrasco:2013sva, Carroll:2013oxa,Senatore:2014eva} such as dealing with subtleties of the velocity field \cite{Carrasco:2013sva,Mercolli:2013bsa} have been presented. The IR-resummed EFTofLSS has been introduced in \cite{Senatore:2014vja, Baldauf:2015xfa, Senatore:2017pbn,Lewandowski:2018ywf} to correctly describe the baryon acoustic oscillation (BAO) peak which requires an IR-resummation of the long displacement fields. Baryons and baryonic effects including comparisons with simulations have been presented in \cite{Lewandowski:2014rca}. For higher order $n$-point functions the dark-matter bispectrum has been computed at one-loop in \cite{Angulo:2014tfa, Baldauf:2014qfa} (including a comparison with simulations), the trispectrum at one-loop in \cite{Bertolini:2016bmt} and the dark matter displacement field in \cite{Baldauf:2015tla} which again included a comparison with simulations. These developments permitted a deeper understanding of the dark matter dynamics leading to the computation of the first observable (lensing) in \cite{Foreman:2015uva}. To further increase the predictive power of the EFTofLSS, biased tracers have been studied in \cite{ Senatore:2014eva, Mirbabayi:2014zca, Angulo:2015eqa, Fujita:2016dne, Perko:2016puo, Nadler:2017qto} (see also \cite{McDonald:2009dh}). \cite{Senatore:2014vja, Angulo:2015eqa} extend these studies to the halo power spectrum and bispectrum including all cross correlations with the dark matter field. Initially, redshift space distortions for dark matter were developed in \cite{Senatore:2014vja, Lewandowski:2015ziq} and subsequently applied to biased tracers in \cite{Perko:2016puo}. Furthermore, the EFTofLSS framework has been extended to dark energy \cite{Lewandowski:2016yce,Lewandowski:2017kes,Cusin:2017wjg,Bose:2018orj} and non-gaussianities \cite{Angulo:2015eqa, Assassi:2015jqa, Assassi:2015fma, Bertolini:2015fya, Lewandowski:2015ziq, Bertolini:2016hxg}. Comparisons between the EFTofLSS approach and simulations have been very promising since it was possible to go to greater $k$ than with the formerly available analytic techniques. Hence, the EFTofLSS should be applied to survey data in future work for precise data analysis.

In the last decades, neutrino oscillation experiments have provided conclusive evidence that neutrinos have non-zero masses (for a review, see \cite{Lesgourgues:2006nd}). As the universe expands, neutrinos slow down and become non-relativistic at late times due to their non-vanishing mass, and start clustering. This affects the dynamics of dark matter clustering. In order to increase the accuracy of the predictions, the effect of massive neutrinos has to be included into the analytic description of dark matter clustering. Although their effect is highly suppressed on large scales and relatively small on short scales due to their small contribution to the total energy density (about 1\%), it is expected that the upcoming LSS surveys will be sensitive enough to measure their absolute masses \cite{Audren:2012vy, Cerbolini:2013uya}. Such a measurement would be a remarkable achievement for the field of cosmology because it would represent the second example where a coefficient of the Lagrangian of the standard model is measured directly from cosmology (the first example being the cosmological constant). 

Previously, the EFTofLSS has been extended to massive neutrinos by computing the one-loop power spectrum in \cite{Senatore:massive_neutrinos}. We further develop the analytic treatment of dark matter clustering in the presence of massive neutrinos by computing the leading order three-point correlation function, the tree-level bispectrum in Fourier space. In this paper we use a different approach than in the literature by involving only \textit{controlled approximations}. These approximations are proportional to a small parameter and the subleading corrections can, at least in principle, be computed in order to reach the required accuracy. The dynamics of dark matter clustering are affected by massive neutrinos through the gravitational force. Therefore, we set up a coupled system of equations and solve it perturbatively. First, we construct the collisionless neutrino Boltzmann equation from the phase space evolution of the number density of neutrinos. Second, we assume collisionless cold dark matter (CDM) particles in the Newtonian limit and smooth out the short scale physics ($k_{\mathrm{NL}}^{-1} \sim 10 \mathrm{Mpc}$). This yields a fluid-like coupled system of equations for dark matter at long distances. The evolution equations of the effective dark matter fluid include an effective stress tensor $\tau^{\mu \nu}$ which is sourced by short modes and accounts for the short distance non-linearities. 

We solve this coupled set of equations perturbatively by expanding in the smallness of their relative energy density $f_{\nu}$ and in the ratio of the wavenumber of interest over the wavenumber associated to the non-linear scale $k/k_{\rm NL}$. The non-linear scale is defined as the breakdown point for perturbation theory. The neutrino Boltzmann equation is solved at linear level in $f_{\nu}$. Solving it at non-linear order leads to products of gravitational fields evaluated at the same location that are sensitive to unknown short-distance non-linearities. When perturbatively solving the Boltzmann equation, we split the neutrino population into \textit{fast} and \textit{slow} neutrinos based on their different expansion parameters. Fast neutrinos have a free streaming wavenumber smaller than the non-linear scale, $k_{\rm fs} \lesssim k_{\rm NL}$, for which each interaction corresponds to a small correction. However, as explained in \cite{Senatore:massive_neutrinos}, one should use the renormalized expressions for the product of the gravitational fields that appear in the perturbative solutions, as they are sensitive to non-linear physics. This ensures the correct perturbative expansion for fast neutrinos at long wavelengths. On the contrary, slow neutrinos have a free streaming wavenumber larger than the non-linear scale, $k_{\rm fs} \gtrsim k_{\rm NL}$. When considering interactions with gravitational fields with wavenumbers shorter than the non-linear scale, the expansion parameter is larger than order one and the perturbative solution does not hold anymore.  However, as argued in \cite{Senatore:massive_neutrinos}, for slow-neutrino perturbations with wavenumbers longer than the non-linear scale, one can consider slow neutrinos as an effective fluid, and, therefore, the effect of short distance physics can be corrected by the inclusion of a speed-of-sound like counterterm ($c_s$). In our tree-level calculations no counterterms are required as no UV sensitive terms arise when performing the integrals, contrary to the case of the one-loop power spectrum computed in \cite{Senatore:massive_neutrinos}. For the one-loop bispectrum, counterterms will be necessary beyond those ones that are present in the stress tensor and that renormalize the dark matter only bispectrum \cite{Angulo:2014tfa, Baldauf:Bispectrum}. The new counterterms at one-loop level will be similar to the ones appearing in the one-loop power spectrum for neutrinos calculated in  \cite{Senatore:massive_neutrinos}.

Briefly summarized, our computation of the tree-level total-matter bispectrum yields that the contribution of neutrinos, for $k \gtrsim k_{\rm fs}$, scales as $16f_{\nu}$ times the dark matter tree-level bispectrum. As we explain, the enhancement with respect to $f_{\nu}$ stems mainly from modifying the Poisson equation at linear level due to the presence of massive neutrinos. In fact, dark matter grows in the presence of massive neutrinos, at distances shorter than $k_{\rm fs}$,  with a fraction of non-relativistic matter that does not cluster. Hence, the exponent of the growth factor is corrected by a factor proportional to $f_{\nu}$. This correction builds up with time, and leads to an enhancement from matter-radiation equality until present time of: $\sim \frac{1}{2}\log{\frac{a_{\rm eq}}{a_0}} \sim 4$. It accounts for approximately 90\% of the final result\footnote{We warn the reader that this enhancement due to the logarithms is quantitatively not very large: $\sim 4$. This means that this effect can be overcast by the diagrams that do not receive this contribution, which are in general numerous.}. For wavenumbers lower than the free streaming length, $k\lesssim k_{\rm fs}$, the contribution is suppressed by up to two orders of magnitude because neutrino perturbations ($\td_{\nu}$) behave as CDM perturbations ($\td_{\rm dm}$) on very large scales. Notice that the suppression is only of order one for the one-loop power spectrum on large scales \cite{Senatore:massive_neutrinos}, because the diagrams are affected by the difference of dark matter and neutrino overdensities even at wavenumbers higher than the free streaming scale, where it is non-zero.  \\

\textit{Comparison with former literature:} Until now, two main approaches have been used to quantify the effect of massive neutrinos on LSS. With increasing accuracy in describing the specificities of neutrinos, they have been included in numerical simulations in \cite{Villaescusa-Navarro:2013pva, Baldi:2013iza, Castorina:2013wga, Castorina:2015bma, Zennaro:2016nqo, Banerjee:2016zaa, Banerjee:2018bxy}. In theoretical models, partly using perturbative models, such as \cite{Hu:1998kj, Saito:2008bp, Wong:2008ws, Lesgourgues:2009am, Shoji:2009gg, Shoji:2010hm, Lesgourgues:2011rh, Upadhye:2013ndm, Blas:2014hya, Castorina:2015bma, Fuhrer:2014zka, Levi:2016tlf} neutrinos have been included with varying levels of accuracy. In \cite{AliHaimoud:2012vj} a combination of both methods has been used by numerically solving for dark matter and analytically modeling neutrinos. In the theoretical context, a relevant paper is \cite{Saito:2008bp}. They solve the linearized collisionless Boltzmann equation yielding linear-order neutrino perturbations when computing the non-linear power spectrum in the presence of massive neutrinos. Their linear predictions for the neutrino perturbations are obtained using a Boltzmann code. These perturbations act as a source for CDM and the authors limit themselves to include only the non-linear corrections to the CDM component. Although \cite{Saito:2008bp} estimates the systematic error stemming from the linear neutrino approximation, in the context of the power spectrum, to be small, they state that solving the full collisionless Boltzmann equation would be the accurate solution. In comparison to all former approaches the novelty of the EFTofLSS approach used in the present paper (discussed in Sec. \ref{sec:EFTofLSSmN}) is, that the full collisionless Boltzmann equation is solved and at the same time suitable counterterms are added that accurately describe the effect of short distance non-linearities on neutrinos both in the Boltzmann equation and in the rest of the fluid equations.

An analytical treatment of the tree-level bispectrum in a cosmology with massive neutrinos has been presented in \cite{Fuhrer:2014zka} and \cite{Levi:2016tlf}.
\cite{Fuhrer:2014zka} presents a test of the two-fluid and linear neutrino approximation for the leading order three-point correlation function. They show that the two-fluid model fails to recover the leading order bispectrum for the whole neutrino mass range. Moreover, the linear neutrino perturbation approximation, obtained from solving the linearized Boltzmann equation, fails to recover the large scale bispectrum  when going to larger neutrino masses, $m_{\nu}$. As the neutrino perturbations behave as CDM perturbations on very large scales, a lack of power for small $k$ in the bispectrum, as a result of the non-linear evolution of all matter components, is observed. Moreover, they test, in their own words, an ad-hoc hybrid approach, whose derivation is not justified, which uses the Green's function of a Boltzmann equation and the non-linear vertices of a two-fluid system which was able to approximate their leading order bispectrum for the entire neutrino mass range. They use the tree-level total-matter bispectrum (CDM + massive neutrino cosmology) obtained from solving in the SPT framework the non-linear Gilbert's equation as a benchmark where they find that the neutrino contribution scales, for $k \gtrsim k_{\rm fs}$, as $\sim 13 f_{\nu}$ times the dark matter tree-level bispectrum. Since our calculation is at tree-level and no high wavenumbers arise in the present calculation, none of the counterterms introduced by the EFTofLSS formalism are actually necessary, and our approach reduces to the one of \cite{Fuhrer:2014zka}. They only analyze the equilateral configuration and our results for the neutrino contribution to the total-matter tree-level bispectrum differ quantitatively by 18\% in this limit. Although our numerical results differ, we are confident with our calculation since we performed several powerful checks (discussed in Sec. \ref{sec:cross_checks}), gave a deep physical explanation and carefully discuss the shape dependence of the bispectrum.

\cite{Levi:2016tlf} uses the approach that CDM behaves as an effective fluid and take their perturbation's linear $k$-dependence into account. As in \cite{Saito:2008bp}, they approximate the neutrinos as linear, and include only the dark matter non-linearities. They also make a rough approximation to estimate the size of the neglected corrections from including the non-linearities in the neutrinos (in our paper, we compute this correction, and find that it is roughly 10\% instead of the estimated 1\%). They obtain for their perturbative approach that the contribution of neutrinos, for $k \gtrsim k_{\rm fs}$, scales as $\sim 13.5 f_{\nu}$ times the dark matter tree-level bispectrum for equilateral configurations. This result for the neutrino contribution to the total-matter tree-level bispectrum deviates by 15\% from our computation for the equilateral configuration. Their analysis of the shape dependence of the tree-level bispectrum yields, in agreement with our result, that the neutrino contribution to the total matter bispectrum is the same in the equilateral and flat configurations. However, the difference between our results is much larger for the squeezed configuration as they observe a steep enhanced suppression, which is not observed by \cite{Ruggeri:2017dda} and us, which will be discussed in more depth in Sec. \ref{sec:Conclusion}. More in detail, what in the present paper, in the context of the bispectrum, we will determine and identify as log-enhanced diagrams, and show to give the largest contribution ($\sim 90\%$ of the total neutrino contribution), are the ones that are computed in \cite{Levi:2016tlf}\footnote{\label{footnote:Levi}For equilateral configurations, when restricted to this subset of diagrams, our results disagree quantitatively, with the ones of \cite{Levi:2016tlf} by approximately 10\%. The log-enhanced diagrams account for approximately 90\% of our total neutrino contribution. Hence, their final result, i.e. the computation of the neutrino contribution $\Delta B_{f_{\nu}} / B_{f_{\nu} = 0}$, deviates by 15\% in the equilateral and much more in the squeezed limit (which will be discussed in Sec. \ref{sec:Conclusion}) from our final result. Notice that the suppression of the non-log-enhanced diagrams is not a simple parametric suppression, as its quantitative value depends on the number of legs of the correlation function and the order at which it is computed. In fact, they become more important during calculations with a higher number of legs or loops, as explained in \cite{Senatore:massive_neutrinos}.}. 

More recently, \cite{Ruggeri:2017dda} presented measurements of the bispectrum from N-body simulations that include massive neutrinos as particles. However, as in \cite{Saito:2008bp} and \cite{Castorina:2015bma}, \cite{Ruggeri:2017dda} does not solve the full Boltzmann equation for neutrinos but assumes neutrinos to contribute at linear level in the perturbative calculation and CDM density perturbations to contribute up to the one-loop level but do not include the EFTofLSS counterterms. This is not parametrically accurate, but, at least for the leading order bispectrum, offers a numerically good approximation. They analyze the shape dependence of the tree-level bispectrum: for equilateral configurations our results for the neutrino contribution differ by 11\%. For the squeezed configurations the difference is slightly larger and is approximately 15\% for the neutrino contribution to the total matter bispectrum. We will further comment on the squeezed limit in Sec. \ref{sec:Conclusion}.  

The two-fluid approaches and solutions to the linearized Boltzmann equation are not arbitrarily accurate descriptions of the physics of massive neutrinos at short distances but capture some of their main aspects up to a certain level of accuracy. The same discussion as in footnote \ref{footnote:Levi} applies for the quantitative difference between \cite{Ruggeri:2017dda} and our final result for the neutrino's contribution and, hence, it is important to have an arbitrarily accurate formalism (rather than a quite accurate one where the errors are not under exact control), that allows us to safely compute correlation functions of neutrinos. This is what we will do in this paper by solving the full collisionless Boltzmann equation and analyzing the various kinematic configurations. It would be interesting to perform a more detailed quantitative comparison with the aforementioned approximate techniques previously introduced in the literature, but we leave this to future work. \\

The structure of the paper is as follows: in Sec. \ref{sec:EFTofLSSmN} we review the EFTofLSS extended to massive neutrinos by constructing the neutrino Boltzmann equation and the fluid-like dark matter equations. We compute the tree-level dark matter bispectrum in Sec. \ref{sec:BispectrumCalc}. We discuss the results and the shape dependence of the tree-level bispectrum in one and two dimensions in Sec. \ref{sec:BispectrumResults}. We conclude and summarize the paper in Sec. \ref{sec:Conclusion}. We derive the dark matter growth factor in the presence of massive neutrinos in App. \ref{sec:growth_factor_DM_appendix}.

\section{The EFTofLSS extended to Massive Neutrinos \label{sec:EFTofLSSmN}}
We will review the system of coupled equations to describe the effect of massive neutrinos through the gravitational force on dark matter derived in \cite{Senatore:massive_neutrinos}. These equations can be solved to arbitrary accuracy in the EFTofLSS framework by going to sufficiently high perturbative order in the expansion parameters\footnote{We mean this in the sense of an asymptotic series.}.

\subsection{Collisionless Neutrino Boltzmann Equation}
The collisionless Boltzmann equation could in principle be used to describe the evolution of neutrinos if we knew the physics down to the smallest astrophysical scales. In the LSS context, we are interested in length scales much smaller than the horizon where massive neutrinos are already non-relativistic and start to cluster. We use comoving coordinates and define the comoving momentum $\vq = a m\ d \vx / d\tau$ with $\tau$ the conformal time, $a(\tau)$ the scale factor and $m$ the mass. The comoving velocity is $\vec{v} = \vq /m =a\ d \vx / d\tau$. The neutrino Boltzmann equation is given by (see for example \cite{AliHaimoud:2012vj, Bertschinger:1988mu}):
\begin{equation} \label{eq:NeutrinoBoltzmannEq}
\frac{\partial f(\vec{x}, \vec{v}, \tau)}{d\tau} + \frac{v^i}{a} \frac{\partial f(\vec{x}, \vec{v}, \tau)}{\partial x^i} -a\frac{\partial \Phi}{\partial
x^i} \frac{\partial f(\vec{x}, \vec{v}, \tau)}{\partial v^i} = 0 \ ,
\end{equation}
where $\Phi$ is the gravitational potential and $f(\vx, \vec{v}, \tau)$ the number density of neutrinos in phase space with $\vx$ being the coordinate position. As we solve our equations at late enough times, the relativistic corrections to equation \eqref{eq:NeutrinoBoltzmannEq} are negligible. The mass density of massive neutrinos relative to the total matter density is defined as
\begin{equation}
f_{\nu} \equiv \dfrac{\r_{\rm \nu}^{(0)}}{\r_{\rm dm}^{(0)}+\r_{\rm \nu}^{(0)}} = \dfrac{\Omega_{\nu}}{\Omega_{\rm m}} 
\end{equation}
and stays constant once the neutrinos have become non-relativistic.

The effective fluid-like approach for dark matter holds for distances longer than the non-linear scale. It is easy to include the effect of neutrino overdensities by modifying the Poisson equation. We therefore obtain the following set of equations \cite{Baumann:2010tm, Carrasco:2012cv, Carrasco:2013mua, Senatore:massive_neutrinos}:
\begin{subequations}
\begin{align}
&\dfrac{\d^2}{a^2}\Phi = \dfrac{3}{2}H_0^2 \dfrac{a_0^3}{a^3} \left(\Omega_{\rm dm,0} \delta_{\rm dm}+\Omega_{\nu, 0} \delta_\nu\right)\simeq \dfrac{3}{2}H_0^2 \dfrac{a_0^3}{a^3} \Omega_{\rm NR,0} \left(\delta_{\rm dm}+f_\nu \left(\delta_\nu-\delta_{\rm dm}\right)\right) \ ,\\ 
&\frac{1}{a}\frac{\d \delta_{\rm dm}}{\d \tau} + \frac{1}{a\rho^{(0)}_{\rm dm}}\d_i\pi^i_{\rm dm} = 0\ , \\ 
&\frac{1}{a}\frac{\d\pi^i_{\rm dm}}{\d\tau} + 4 H\pi^i_{\rm dm}+\frac{1}{a}\d_j\left(\frac{\pi^i_{\rm dm}\pi^j_{\rm dm}}{\rho^{(0)}_{\rm dm}(1+\delta_{\rm dm})}\right)+\frac{1}{a}\rho^{(0)}_{\rm dm} (1+\delta_{\rm dm}) \d^i\Phi= -\frac{1}{a}\d_j\tau^{ij}\ ,
\end{align}
\end{subequations}
where $\Omega_{\rm NR} = (\r^{(0)}_{\rm dm}(t_0) + \r^{(0)}_{\nu}(t_0))/\r_{\rm tot}(t_0)$ is the energy fraction in non-relativistic matter at present time $a_0$. $\pi^i_{\rm dm}$ is the dark matter momentum and $\tau^{ij}$ is the effective stress tensor for dark matter which is sourced by the short modes \cite{Baumann:2010tm, Carrasco:2012cv}. The stress tensor depends on some time dependent kernel function $K_1(\tau, \tau')$, the tidal tensor of gravity $\d_i\d_j\Phi$, the gradient of velocites, and the location of the fluid element $x_{\rm fl}(\vx, \tau, \tau')$. It incorporates the effect of uncontrolled short distance non-linearities in the perturbative solution for dark matter at long distances. Schematically, the effective stress tensor is given at linear order by \cite{Carrasco:2013mua}:
\begin{equation}
\tau^{ij} = \int^{\tau} d{\rm \tau'} \ \lbrace K_1(\tau, \tau') \td_{\rm dm}(x_{\rm fl}(\vx, \tau, \tau'), \tau') + \dots\rbrace\ ,
\end{equation}
where $\dots$ stands for higher order terms. The location of the fluid element $\vx_{\rm fl}$ is iteratively defined as a function of the dark matter velocity 
\begin{equation}
\vx_{\rm fl}(\vx, t, t') = \vx - \int_{\tau(t')}^{\tau(t)} d{\rm \tau''} \vec{v}_{\rm dm} (\tau'', x_{\rm fl}(\vx, \tau, \tau''))\ ,
\end{equation}
where the dark matter velocity is given by $\vec{v}_{\rm dm} = d\vx/d\tau$ which is not a comoving velocity as previously defined for neutrinos. The effective stress tensor is sufficient to correctly include the effect of uncontrolled short distance non-linearities on the perturbative solution for dark matter in a universe only containing dark matter. The extension to a universe with baryons was done in \cite{Lewandowski2016:Greensfunction}, showing that stellar and galactic dynamics can be correctly parametrized in the EFTofLSS framework.

In order to compute the leading order three-point correlation function in a universe with dark matter and neutrinos, we have to define the dark matter and neutrino observables of which we will take the expectation values\footnote{The following procedure can be applied to any N-point correlation function computation. Here, we focus on the tree-level bispectrum case.}. The energy density of the system is given by the sum of the energy densities of dark matter and neutrinos $\r(\vx, t) = \r_{\mathrm{dm}}(\vx, t) + \r_{\mathrm{\nu}}(\vx, t)$. Thus, the overdensities induced by dark matter and neutrinos are given by
\begin{equation} \label{eq:density_eq}
\td = (1-f_{\nu})\td_{\mathrm{dm}} + f_{\nu}\td_{\nu} = \td_{\mathrm{dm}} + f_{\nu}
\td_{\mathrm{diff}}\ ,
\end{equation}
with $\td_{\rm diff} = \td_{\nu}-\td_{\rm dm}$. The energy density of neutrinos is given by $\r_{\nu}(\vx, t)=\r_{\nu}^{(0)}(t)\int d^3v f(\vec{v}, \vx, t)$ with the neutrino distribution, $f(\vec{v}, \vx, t)$, normalized to unity. We define the overdensity field as $\td = \r/\r^{(0)}-1$.

For the computation of the tree-level bispectrum, we compute the expectation value of three density fields to leading order\footnote{With the current accuracy and sensitivity of cosmological LSS surveys, going to higher order in $f_{\nu}$ is a purely academic challenge. The current leading order approach is off by $f_{\nu}^2\td^2 \lesssim 10^{-5}$ with $\td$ being the dark matter overdensity at the length scale of interest.} in $f_{\nu}$ which is given by:
\begin{align} \label{eq:threePCF}
\langle \td(\vk_1, a) & \td(\vk_2, a) \td(\vk_3, a) \rangle = \\
&= (2\pi)^3\td_D^{(3)}(\vk_1+\vk_2+\vk_3)B(k_1,k_2,k_3,a) \nonumber \\
&= \langle \td_{\rm dm}(\vk_1, a)\td_{\rm dm}(\vk_2, a) \td_{\rm dm}(\vk_3, a)\rangle + 3f_{\nu}\langle \td_{\rm diff}(\vk_1, a)\td_{\rm dm}(\vk_2, a)\td_{\rm dm}(\vk_3, a)\rangle + \mathcal{O}(f_{\nu}^2) \nonumber\\
&= (2\pi)^3 \td_D^{(3)}(\vk_1+\vk_2+\vk_3) \left( B_{\rm dm, dm, dm}(k_1,k_2,k_3, a) + f_{\nu} \left\lbrace B_{\rm diff, dm, dm}(k_1,k_2,k_3, a) \right.\right. \nn\\
&\left.\left.\qquad  +B_{\rm dm, diff, dm}(k_1,k_2,k_3,a) + B_{\rm dm, dm, diff}(k_1,k_2,k_3,a) \right\rbrace \right) + \mathcal{O}(f_{\nu}^2) \nn
\end{align} 
As usually we go to $k$-space and solve the Boltzmann equation perturbatively in powers of $f_{\nu}$ and $k/k_{\rm NL}$. To perturbatively solve the Boltzmann equation, we need the Fourier transforms of the unperturbed neutrino and dark matter distributions, $\tilde{f}_{\nu}^{[0]}$ and $\tilde{f}_{\rm dm}^{[0]}$, respectively. The Fourier transformed dark matter distribution equals to unity because we assume exactly cold dark matter particles. The Fourier transform of the distribution $\td_{\rm diff}$ is given by the difference of the distribution for neutrinos and dark matter:
\begin{equation}
\tilde f^{[0]}(q)=\tilde f^{[0]}_\nu(q)-\underbrace{\tilde f^{[0]}_{\rm dm}(q)}_{\rm CDM}=\tilde f^{[0]}_\nu(q)-1.
\end{equation}
The calculations in the present paper will be performed using $\tilde f^{[0]}(q)$. For the Fermi distribution we can find a series expansion of the Fourier transform of the neutrino distribution \cite{Bertschinger:1988mu}. Here we use the fitting function derived in \cite{AliHaimoud:2012vj} that is accurate to 3\%. It is advantageous to use, as the numerical evaluation is significantly faster:
\begin{equation}
\tilde f^{[0]}_{\nu_i}(q)=\cfrac{0.0407 \left(q \cfrac{k_B T_{\nu,0}}{m_{\nu_i} c}\right)^4+0.0168 \left(q \cfrac{k_B T_{\nu,0}}{m_{\nu_i} c}\right)^2+1}{1.6787 \left(q \cfrac{k_B T_{\nu,0}}{m_{\nu_i} c}\right)^{4.1811}+0.1467 \left(q \cfrac{k_B T_{\nu,0}}{m_{\nu_i} c}\right)^8+2.1734 \left(q \cfrac{k_B T_{\nu,0}}{m_{\nu_i} c}\right)^2+1}\ ,
\end{equation}
with $T_{\nu,0}$ the temperature of the unperturbed neutrino distribution, $m_{\nu_i}$ the mass of the neutrino species, $c$ the speed of light and $k_B$ the Boltzmann constant. As there are three degenerate neutrino species, the sum over all three species has to be considered to obtain the total neutrino distribution function:
\begin{equation}
\tilde f^{[0]}_{\nu}(q)=\frac{\sum_i  m_{\nu_i}\tilde f^{[0]}_{\nu_i}(q)}{\sum_i  m_{\nu_i}}.
\end{equation}

\subsection{Perturbative Solution of the Neutrino Boltzmann Equation}
Before solving the previously derived neutrino Boltzmann equation, we want to develop some intuition on what controls the perturbative expansion. The perturbative solution of the neutrino Boltzmann equation amounts to solving the trajectory of a set of particles with known initial conditions. We want to estimate the deviation from a straight trajectory in a region of size $\sim 1/k$. This would be the same as determining `backwards' the aforementioned problem; to determine the initial conditions of a particle from a certain current position and velocity. The trajectory of a particle is given by:
\begin{align}
x(t)-x(t_{\rm in}) &\sim \int^t dt_1 \left( v-\int^{t_1}dt_2 \ \d \Phi \left(x-\int^{t_2}dt_3 \left( v-\int^{t_3}dt_4 \ \d \Phi \left(x-\dots \right)\right)\right)\right)\\
&\sim \underbrace{v\Delta t}_{\td x^{(1)}} + \underbrace{\d \Phi (\Delta t)^2}_{\td x^{(2)}} + \d ^2 \Phi \ v (\Delta t)^3 + \dots \ , \nn
\end{align}
where we considered only one Hubble time and neglected the scale factor. The ratio of the first two terms, $\frac{\td x^{(2)}}{\td x^{(1)}}\sim \frac{\d \Phi \Delta t}{v}$ , lets us estimate the impact the gravitational force has on the trajectory of the particles. The crossing time $\Delta t$ of a region with size $1/k$ is given by the minimum of the Hubble time and the velocity of the particle times the wavenumber: $\Delta t \sim {\rm Min}\lbrack 1/H, 1/(kv)\rbrack$. Clearly, we have to differentiate between slow and fast neutrinos. Let us consider `relatively-slow' neutrinos\footnote{`relatively' because it depends on the wavenumber $k$.}, which have a velocity $v \lesssim H/k$ and cross a region $v/H \sim 1/k_{\rm fs}$. In particular, this means that $k \lesssim k_{\rm fs}$. Since the whole region is not crossed in a Hubble time, we can go to a local inertial frame and Taylor expand $\Phi$ around its minimum and evaluate it at a distance $1/k_{\rm fs}: \d \Phi \sim \d ^2 \Phi \Delta x \sim \d ^2 \Phi/k_{\rm fs}$. With $\Delta t \sim 1/H$ and $k \lesssim k_{\rm fs}$ we obtain:
\begin{equation} \label{eq:fast_nu_estimate}
\dfrac{\td x^{(2)}}{\td x^{(1)}}\sim \dfrac{1}{vH}\dfrac{\d ^2 \Phi}{k_{\rm fs}} \sim \dfrac{\d^2 \Phi}{H^2} \sim \td(k) \ ,
\end{equation}
where we used $\d \Phi \sim H^2\td$. For `relatively-fast' neutrinos, $k \gtrsim k_{\rm fs}$, the region crossing takes $\Delta t \sim 1/(kv)$ and the influence on the trajectory through the gravitational field is given by:
\begin{equation}
\dfrac{\td x^{(2)}}{\td x^{(1)}}\sim \dfrac{1}{kv}\dfrac{\d \Phi}{v} \sim \dfrac{ \Phi}{v^2} \sim \td(k) \left( \dfrac{k_{\rm fs}}{k} \right)^2.
\end{equation}
As the gravitational potential is approximately scale invariant, we can expand around this parameter in our universe, which is always smaller than one. We define the non-linear velocity as $v_{\rm NL} \equiv H/k_{\rm NL}$ and with $\Phi \sim v_{\rm NL}^2$, we can write the expansion parameter in equation \eqref{eq:fast_nu_estimate} as $v^2_{\rm NL}/v^2$.

This estimate shows that we can split the fast-neutrino population, which has a free streaming wavenumber smaller than the non-linear scale ($k_{\rm fs}\lesssim k_{\rm NL}$), into two subpopulations when perturbatively solving the Boltzmann equation: first, `relatively-slow' fast neutrinos with $k \lesssim k_{\rm fs}$ and, second, `relatively-fast' ones with $k \gtrsim k_{\rm fs}$. In fact, as the expansion parameter is always smaller than unity for fast neutrinos, $k_{\rm fs}\lesssim k_{\rm NL}$, we can expand in $\td(k)$, for $k \lesssim k_{\rm fs}$, and in $\Phi / v^2$, for $k \gtrsim k_{\rm fs}$. This is even possible for wavenumbers inside the non-linear scale, even though, only after using the correct gravitational potential.

On the contrary, for slow neutrinos, which have a free streaming wavenumber larger than the non-linear scale ($k_{\rm fs}\gtrsim k_{\rm NL}$), there are wavenumbers $k \gtrsim k_{\rm NL}$ where the expansion parameter is larger than order one. In this context, an effective stress tensor term, as explained in \cite{Senatore:massive_neutrinos}, is added to account for the effect of short distance non-linearities.

Now we can perturbatively solve the neutrino Boltzmann equation given in eq. \eqref{eq:NeutrinoBoltzmannEq} which is given in Fourier space by:
\begin{equation}
\frac{\partial f(\vec{k}, \vec{v}, \tau)}{\d\tau} + i \frac{\vec{v}\cdot\vec{k}}{a} f(\vec{k}, \vec{v}, \tau) -a \left[ \frac{\partial \Phi}{\partial x^i} \frac{\partial f(\vec{x}, \vec{v}, \tau)}{\partial v^i} \right]_{\vec{k}} = 0\ ,
\end{equation}
where $[\mathcal{O}(\vx)]_{\vk}$ indicates that we take the $\vk$-component of the Fourier transform of $\mathcal{O}(\vx)$. The gravitational potential $\Phi$ is sourced by neutrinos only proportional to $f_{\nu}$. The gravitational field in the neutrino Boltzmann equation can be regarded as an external source, when solving it to linear order\footnote{Going to higher order in $f_{\nu}$ works analogously. We can then write $\Phi$ as an expansion in $f_{\nu}$: $\Phi = \sum_{i=0}^n f_{\nu}^i \Phi _i$. Then the equations are solved up to the desired order in $f_{\nu}$. As mentioned earlier, with the current accuracy of cosmological LSS surveys the leading order expansion in $f_{\nu}$ is sufficient. Hence, going to higher N-point correlation functions will be a purely academic challenge.} in $f_{\nu}$. This assumption allows us to neglect the back-reaction of the gravitational potential in a universe with massive neutrinos onto the neutrino dynamics. We can write $f = \sum_{i=0}^{n} f^{ \left[ i \right] }$ with $i$ indicating the order at which the gravitational force $\d \Phi$ has been included. With our choice of the definition of the velocity, being a comoving one, the zeroth order solution for $f(\vk, \vec{v}, \tau)$ is time independent. At $n$-th order the solution is given by the recursive structure \cite{Senatore:massive_neutrinos}:
\begin{align} \label{eq:FNeutrinoPert}
f^{\lbrack n\rbrack}(\vec{k}, \vec{v}, \tau) 
&= \int_0^{\tau} \mathrm{d \tau '} G_R(\tau, \tau '; \vec{v}, k) \left[ a(\tau ') \left( \frac{\partial}{\partial x^i} \Phi(\vec{x}, \tau ')\right) \frac{\partial}{\partial v^i} f^{\lbrack n-1 \rbrack} (\vec{x}, \vec{v}, \tau ') \right]_{\vec{k}}\ , 
\end{align}
where $G_R(\tau, \tau '; \vec{v}, \vec{k})$ is the analytically known retarded Green's function from $\tau '$ to $\tau$. It's analytic solution in \textit{super-conformal} time $s(\tau)$ is given by:
\begin{equation}
G_R(s, s'; \vec{v}, \vec{k}) = e^{-i\vec{k}\cdot \vec{v}(s-s_i)} \ , \qquad s(\tau) = \int^{\tau} \frac{\mathrm{d\tau '}}{a(\tau ')} = \int^{a_0} \frac{1}{a'^2\mathcal{H}(a')} \mathrm{da'} \ .
\end{equation}
We performed a substitution for the super-conformal time using the conformal time relation $\frac{\mathrm{d\tau}}{\mathrm{dt}} = a(t)^{-1}$. Thus, the $n$-th order solution, expressed in terms of $s(\tau)$, is given by:
\begin{equation} \label{eq:recursive_sol_nB_superconformaltime}
f^{\lbrack n\rbrack}(\vec{k}, \vec{v}, s) 
=\int_0^{\tau} \mathrm{d s'} a(s')^2 e^{-i \vec{k}\cdot \vec{v} (s-s')} \left[ \left( \frac{\partial}{\partial x^i} \Phi(\vec{x}, s')\right) \frac{\partial}{\partial v^i} f^{\lbrack n-1 \rbrack} (\vec{x}, \vec{v}, s') \right]_{\vec{k}}\ ,
\end{equation}
which involves a convolution of fields in Fourier space. In the present paper the validity of the equation is only ensured for $k \ll k_{\rm NL}$ but the fields could include very high wavenumbers. In \cite{Senatore:massive_neutrinos}, counterterms have been included to correct for these effects as well. At tree-level, we do not have to include any additional terms to account for this effect. Hence, we can proceed with the perturbative solution of \eqref{eq:NeutrinoBoltzmannEq}. At first order the solution to the neutrino Boltzmann equation is given by:
\begin{align}
f^{\lbrack 1\rbrack}(\vec{k}, \vec{v}, s) = e^{-i\vec{k}\cdot \vec{v}(s-s_i)}\delta f(\vec{k}, \vec{v}, s_i)+ \int_{s_i}^s \mathrm{d s '} a(s')^2 e^{-i \vec{k}\cdot \vec{v} (s-s')} i \frac{\vec{v}\cdot \vec{k}}{v} \Phi(\vec{k}, s') \frac{\partial f^{\lbrack 0 \rbrack} (v)}{\partial v}\ ,
\end{align}
where $s_i$ is the initial time of evaluation and $\delta f(\vk, \vec{v}, s_i)$ the perturbation of the neutrino distribution at time $s_i$. The first term coming from the initial conditions gives a negligible contribution\footnote{This approximation is valid because even using the assumption that $\frac{\td \r_{\nu}}{\r_{\nu}} (\vk, s_i) \sim \frac{\td \r_{\rm dm}}{\r_{\rm dm}} (\vk, s_i)$ at initial time $s_i$, the overdensities coming from the initial conditions at present time are of order $\leq 10^{-3}\frac{\td \r_{\rm dm}}{\r_{\rm dm}} (a_0)$ when starting at a redshift of $z \sim 10$. This is because, when summing over the velocities, each different velocity contributes with a different phase, leading to a cancellation. The error shrinks by going to even higher redshift. In the numerical evaluations $z_i=100$ is used which lets us safely use this approximation.}, as estimated in \cite{Senatore:massive_neutrinos}. Moreover, we used $\d_{v^i}f^{\lbrack 0 \rbrack}(v) = \frac{v^i}{v} \ \d_{v}f^{\lbrack 0 \rbrack}(v)$. 

At second order the solution is given by:
\begin{align}
f^{\lbrack 2 \rbrack}(\vec{k}, \vec{v},s) &= \int_0^s \mathrm{ds_1} a(s_1)^2 e^{-i\vec{k}\cdot \vec{v}(s-s_1)}\bigg \lbrack \left( \frac{\partial}{\partial x^i}\Phi(\vec{x},s_1)\right)\frac{\partial}{\partial v^i}f^{\lbrack 1\rbrack}(\vec{x}, \vec{v}, s_1))\bigg \rbrack_{\vec{k}}\nn\\
&= \int \frac{\mathrm{d^3q_1}}{(2\pi)^3} \int \frac{\mathrm{d^3q_2}}{(2\pi)^3} (2\pi)^3 \delta_D^{(3)}(\vec{k}-\vec{q_1}-\vec{q_2}) \\
&\quad \quad \int_0^s \mathrm{ds_1} a(s_1)^2 e^{-i(\vec{q_1}+\vec{q_2})\cdot \vec{v}(s-s_1)} i q_1^{i_1}\Phi(\vec{q_1}, s_1)\nn\\
&\qquad \quad \frac{\partial}{\partial v^{i_1}} \int_0^{s_1} \mathrm{ds_2} a(s_2)^2 e^{-i\vec{q_2}\cdot \vec{v}(s_1-s_2)} i q_2^{i_2}\Phi(\vec{q_2}, s_2)\frac{\partial f^{\lbrack 0 \rbrack}(v)}{\partial v^{i_2}}\ .\nn
\end{align}

\subsection{Renormalization of Neutrino Correlation Functions}
One might wonder if the expansion parameter is within the convergence radius of the perturbative series. As mentioned earlier, we can naturally split our neutrino population into fast, $k_{\rm fs} \lesssim k_{\rm NL}$, and slow, $k_{\rm fs} \gtrsim k_{\rm NL}$, neutrinos. For fast neutrinos we can certainly trust our perturbative expansion as the expansion parameter is less than one for all wavenumbers. For slow neutrinos this is not necessarily true since the expansion parameter is of order one for $k\gtrsim k_{\rm NL}$. In both cases, the solution is determined by the gravitational field, which is unknown to us for wavenumbers larger than the non-linear scale, $k\gtrsim k_{\rm NL}$. Nevertheless, these wavenumbers affect the solution at $k \ll k_{\rm NL}$. In fact, long wavelength perturbations can be sourced by a product of fields with some having high wavenumbers $q_i \gtrsim k_{\rm NL}$. These modes are not under perturbative control and, hence, their momentum and time dependence is unknown. In general, the perturbative solution will have mistakes originating from this part of phase space because the momentum and time integrals cannot be performed correctly. This sensitivity to high wavenumbers is mapped in real space to a product of fields evaluated at the same location, demonstrating their sensitivity to uncontrolled short-distance physics. Since we are interested in the $k\ll k_{\rm NL}$ regime, we can renormalize the products of fields by adding suitable counterterms to correctly account for the contribution from uncontrolled short distance non-linearities, as explained in \cite{Senatore:massive_neutrinos}. These counterterms renormalize the perturbative expressions for the case of fast neutrinos in the one-loop power spectrum computation in \cite{Senatore:massive_neutrinos}. However, for slow neutrinos additional counterterms have to be included as the perturbative series does not hold when inserting interactions at $k\gtrsim k_{\rm NL}$ \footnote{The splitting between slow and fast neutrinos is done only at the level of the counterterms and not in the way we solve the equations. This means that for the slow and fast neutrinos the full Boltzmann equation is solved. While this is necessary for fast neutrinos, it is sufficient but non-necessary for slow neutrinos. These could have, in principle, been solved using a fluid equation. However, since the splitting between fast and slow neutrinos is not sharply defined, it is ultimately impossible to use a different equation for the slow neutrinos, because this would require an imprecise splitting of the population. Using the perturbative equations for slow neutrinos creates additional mistakes that are not corrected by the counterterms we use when solving the Boltzmann equations. For this reason, for slow neutrinos, we add a fluid-like counterterm that corrects for these mistakes. Now, the order-one ambiguity between the splitting for fast and slow neutrinos is therefore relegated to an order one difference in the numerical value of this counterterm, which is unknown anyway, apart of an order of magnitude estimate.}. In the present paper no counterterms have to be included in the tree-level calculations. The renormalization formalism is described in more detailed in App.  \ref{sec:Renormalization_Correlation}.

\section{Bispectrum Calculations at Tree-Level \label{sec:BispectrumCalc}}
In order to extract information from the LSS of the universe, we compute $N$-point correlation functions between density fields. In the present paper the tree-level bispectrum, the equivalent of the three-point correlation function in Fourier space, is computed. From equation \eqref{eq:threePCF} one can see that the calculation is split into computing $B_{\mathrm{diff, dm, dm}}$ to zeroth order in $f_{\nu}$, which we do in Sec. \ref{sec:OrdinaryB_diff_dm_dm}, and $B_{\mathrm{dm, dm, dm}}$ to first order in $f_{\nu}$, which we do in Sec. \ref{sec:B_dm_dm_dm}. The goal of this paper is to compare the tree-level correction to the total-matter bispectrum in the presence of massive neutrinos, $\Delta B_{\rm tree-level, total}$, which is the sum of $B_{\rm diff, dm, dm}$ and $\Delta B_{\rm dm, dm, dm}$ to the dark matter bispectrum $B_{\rm tree, f_{\nu}=0}$ for different configurations. With the above derived tools, we can now start to compute correlation functions between density fields. 

In Fourier space, the bispectrum is defined as:
\begin{equation}
\langle\td(\vk_1, a)\td(\vk_2, a)\td(\vk_3, a)\rangle \equiv (2\pi)^3\td_D^{(3)}(\vk_1+\vk_2+\vk_3)B(k_1,k_2,k_3,a) \ .
\end{equation}
Due to momentum conservation, the bispectrum $B(\vk_1, \vk_2, \vk_3,a)$ is not a function of three independent vectors but of the moduli of the momenta, $B(k_1, k_2, k_3,a)$. At zeroth order in $f_{\nu}$, which is the pure dark-matter case, the first non-trivial contribution to the bispectrum stems from the first non-linear contribution to $\td^{(1)}_{\rm dm}$, which is the second order overdensity field of dark matter, $\td^{(2)}_{\rm dm}$. As only Gaussian initial conditions are considered, the expectation value of three linear dark matter fields is zero. Hence, the leading order three-point correlation function in Fourier space,  the tree-level bispectrum is given by:
\begin{equation} \label{eq:bispectrum_tree_level}
B_{\rm tree, \ f_{\nu}  =  0}(k_1, k_2, k_3, a) = 2F_2(\vec{k_1}, \vec{k_2})P_{11}(k_1,a)P_{11}(k_2,a) + 2 \mathrm{\ cycl. \ perm.} \ ,
\end{equation}
where $F_2(\vk_1, \vk_2)$ is a kernel function from perturbation theory and the linear power spectrum, $P_{11}(k,a)$, is given by the two-point correlator of two $\td^{(1)}(\vk)$ (for a review, see \cite{Bernardeau:2001CPT}).

\subsection{Contribution to $B_{\mathrm{diff, dm, dm}}$ \label{sec:OrdinaryB_diff_dm_dm}}

\begin{figure}[t!]
\centering
\subfigure[Contracting $f^{[1,1]}$ with $\td_{\rm dm}^{(1)}$ and $\td_{\rm dm}^{(2)}$ yields $B^{ \lbrack 1 , 1 \rbrack  \lvert (1) \lvert (2)}_{\mathrm{diff, dm, dm}}$ \label{fig:B_diff_dm_dm_1}]{\includegraphics[width = 0.31\textwidth]{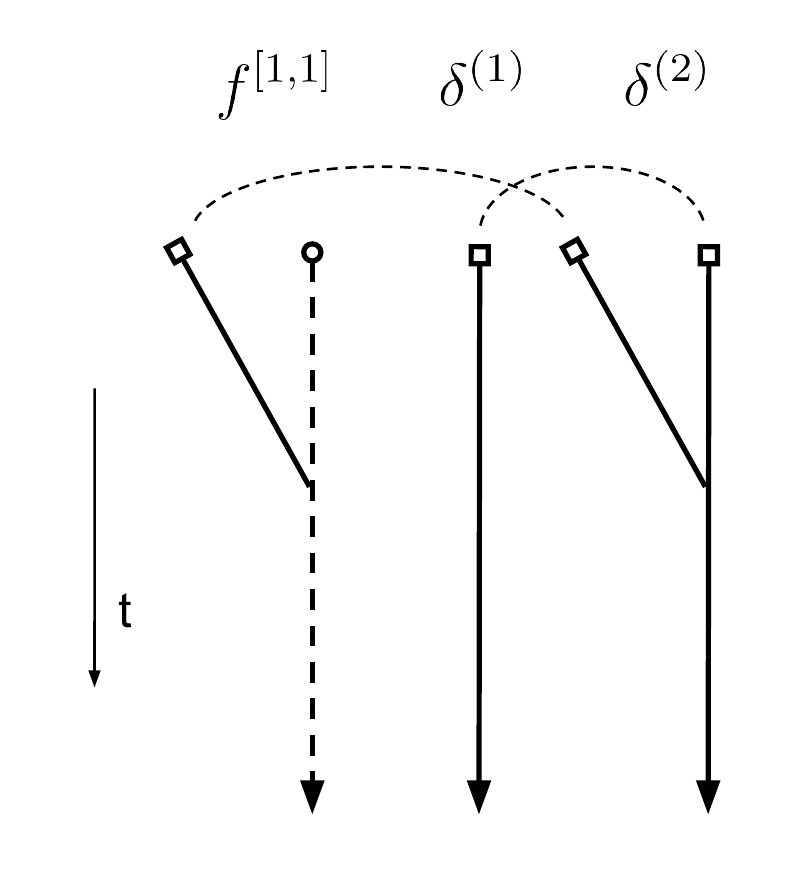}} \quad
\subfigure[Contracting $f^{[1,2]}$ with two $\td_{\rm dm}^{(1)}$ yields $B^{ \lbrack 1 , 2 \rbrack  \lvert (1) \lvert (1)}_{\mathrm{diff, dm, dm}}$ \label{fig:B_diff_dm_dm_2}]{\includegraphics[width = 0.30\textwidth]{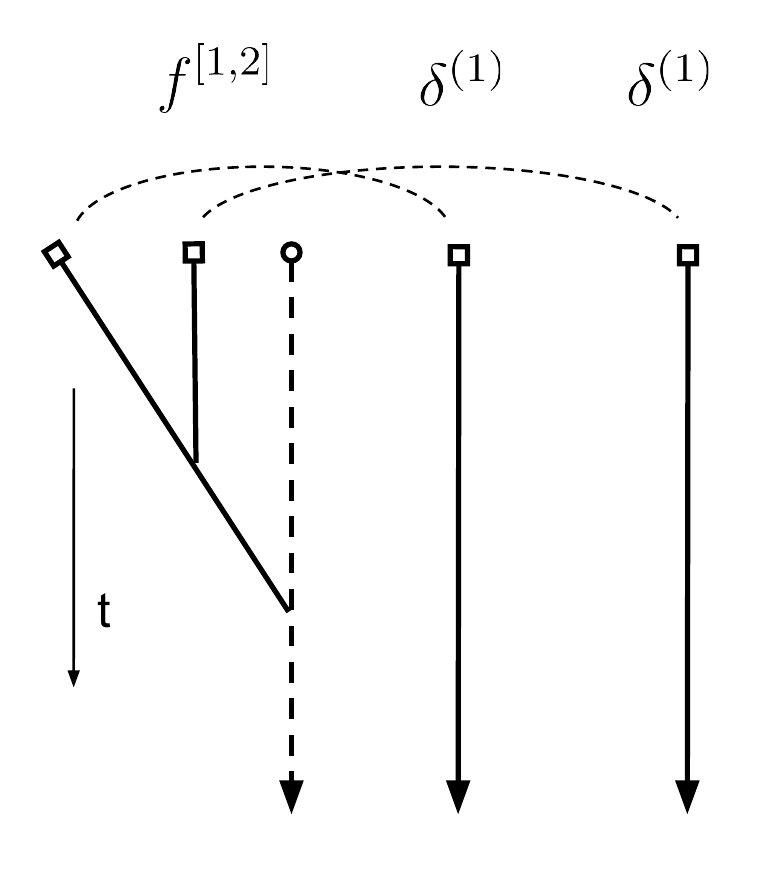}} \quad
\subfigure[Contracting $f^{[2,11]}$ with two $\td_{\rm dm}^{(1)}$ yields $B^{ \lbrack 2 , 11 \rbrack  \lvert (1) \lvert (1)}_{\mathrm{diff, dm, dm}}$ \label{fig:B_diff_dm_dm_3}]{\includegraphics[width = 0.305\textwidth]{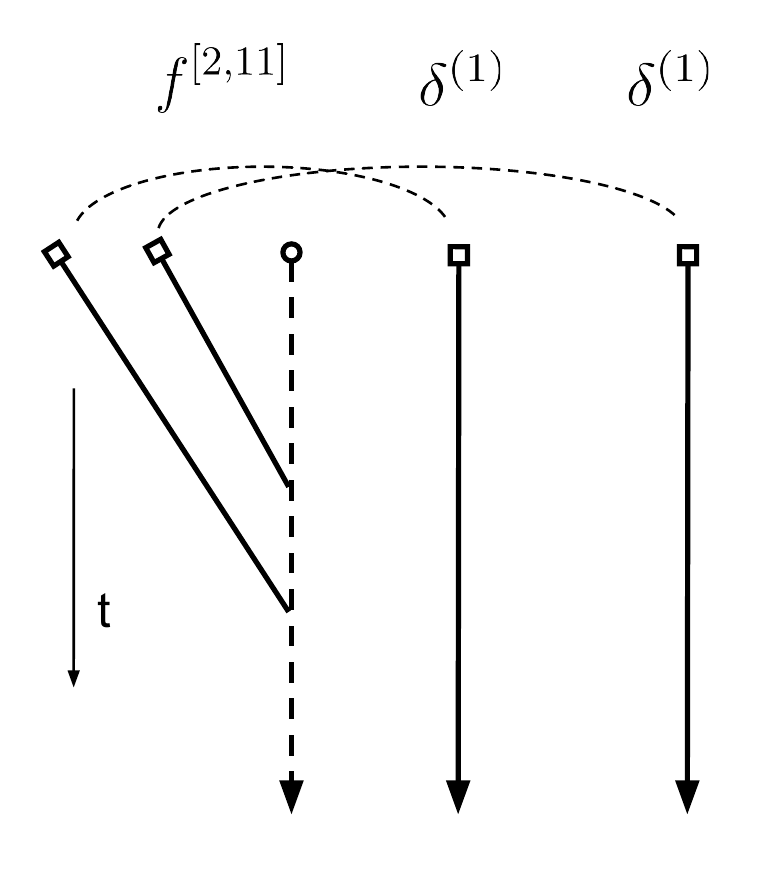}}
\caption[]{Diagrammatic representation of the perturbative solution of neutrino distributions. Dashed arrows represent the neutrino, while continuous lines represent the dark matter fields. The circles represent the background initial neutrino distribution $f^{\left[ 0 \right]}$, while the squares represent the initial dark matter fields, of which we take correlations to create the bispectra diagrams, $B_{\mathrm{diff, dm, dm}}$, later on. The dashed lines between the dark matter fields indicate the performed contractions using Wick's theorem. For the tree-level bispectrum computation, we use dark matter fields at first, $\td^{(1)}_{\rm dm}$, and second order, $\td^{(2)}_{\rm dm}$. The perturbative solution for the neutrino distributions that use linear dark matter fields to first and second order are $f^{[1,1]}$ and $f^{[2,11]}$, respectively. $f^{[1,2]}$ is a second order solution, obtained from using a non-linear dark matter field at second order on the $f^{[1]}$ solution. 
\label{fig:perturbative_sol_NB}}
\end{figure}

The tree-level bispectrum can be computed using the perturbative solutions $f^{\left[ 1 \right]}$ and $f^{\left[ 2 \right]}$ of the neutrino distribution function given in equation  \eqref{eq:FNeutrinoPert}. For the tree-level calculation, at zeroth order in $f_{\nu}$, we are interested in the first, $\Phi^{(1)}$, and second, $\Phi^{(2)}$, order gravitational potential. The neutrino side ($\td_{\rm diff}$) is evaluated to order $n = j_1 + \dots + j_i$ using the $f^{\lbrack i , j_1 \dots j_i \rbrack}$ perturbative solution. Our notation is as follows: $f^{\lbrack i , j_1 \dots j_i \rbrack}$ represents the solution obtained by using the $i$-th order perturbative solution of equation \eqref{eq:FNeutrinoPert}, denoted $f^{\lbrack i\rbrack}$, and replacing each $\Phi$ in it with $\Phi^{(j_1)}$ to $\Phi^{(j_i)}$. We denote with $\Phi^{(j)}$ the gravitational potential at order $j$ in perturbation theory. Moreover, the time ordering of the different insertions is encoded in the notation: $\Phi^{(j_i)}$ being the first and $\Phi^{(j_1)}$ the last inserted interaction. Hence, we use for the perturbative solution of the neutrino distribution the form $\lbrack i, j_1 +\dots+j_i\rbrack$. Since we are interested in computing the three-point correlation function, we contract the perturbative solutions of the neutrino distribution $f^{\lbrack i , j_1 \dots j_i \rbrack}$ with two dark matter fields, each evaluated to order $m_1$ and $m_2$, denoted by $\td^{(m_1)}_{\rm dm}$ and $\td^{(m_2)}_{\rm dm}$, respectively. Therefore, we will use the notation $B^{ \lbrack i , j_1 \dots j_i \rbrack  \lvert (m_1) \lvert (m_2)}_{\mathrm{diff, dm, dm}}$. At tree-level, there are three diagrams for $B_{\rm diff, dm, dm}$ that have to be computed.

Figure \ref{fig:perturbative_sol_NB} shows the perturbative solutions for the neutrino distributions. The first diagram is shown in figure \ref{fig:B_diff_dm_dm_1}; it consists of computing the expectation value of $\langle f^{\lbrack 1,1 \rbrack}\delta^{(1)}_{\mathrm{dm}} \delta^{(2)}_{\mathrm{dm}}\rangle$.  We take the gravitational potential to first order, $\Phi^{(1)}$, in $f^{\lbrack 1\rbrack}$ and contract it with a linear, $\td^{(1)}_{\rm dm}$, and a second order, $\td^{(2)}_{\rm dm}$, dark matter field. The first diagram is therefore given by:
\begin{align} \label{eq:B1112}
&B^{ \lbrack 1 , 1 \rbrack  \lvert (1) \lvert (2)}_{\mathrm{diff, dm, dm}} (k_1, k_2, k_3, a_0)= \nn \\ 
&= \bigg  \langle D(a_1) \int \mathrm{d^3v} \int_0^s ds' a(s')^2 e^{-i \vec{k} \cdot \vec{v} (s-s')} \bigg\lbrack \frac{\partial}{\partial x^i} \Phi^{( 1)}(\vec{x}, s') \frac{\partial}{\partial v^i} f^{\lbrack 0 \rbrack}(v) \bigg\rbrack_{\vec{k}} D(a_0)^3\nn  \\
&\qquad \left.\vphantom{\int}\int \frac{\mathrm{d^3q_2}}{(2 \pi )^3} \dots \int \frac{\mathrm{d^3q_4}}{(2 \pi )^3} F_2(\vec{q}_3, \vec{q}_4) \delta_D(\vec{q_2}-\vec{k_2}) \delta_D(\vec{q_3}+\vec{q_4}-\vec{k_3}) \delta^{(1)}(\vec{q_2}, a_0)\delta^{(1)}(\vec{q_3}, a_0) \delta^{(1)}(\vec{q_4}, a_0) \bigg \rangle \right.\nn \\
&\myeq \bigg  \langle \int \mathrm{d^3v} \int_{q_1} (2 \pi )^3 \delta_D^{(3)}(\vec{q_1}-\vec{k_1})\int^{a_0} \frac{da_1}{a_1^2 \mathcal{H}(a_1)} a_1^2 e^{-i \vec{q_1} \cdot \vec{v} (s-s(a_1))} \left(\frac{3}{2}\right) \frac{iq_1^{i_1}}{-q_1^2} H^2 a_1^2 \Omega_{\mathrm{dm}}(a_1)  \frac{\partial}{\partial v^i} f^{\lbrack 0 \rbrack}(v)\nn  \\
&\qquad \left.\vphantom{\int} D(a_0)^3 D(a_1) \int_{q_2} \int_{q_3} \int_{q_4}  F_2(\vec{q}_3, \vec{q}_4) \delta_D(\vec{q_2}-\vec{k_2}) \delta_D(\vec{q_3}+\vec{q_4}-\vec{k_3})\delta^{(1)}(\vec{q_1}) \delta^{(1)}(\vec{q_2})\delta^{(1)}(\vec{q_3}) \delta^{(1)}(\vec{q_4}) \bigg \rangle \right.\nn \\
&\myeqs  \left(\frac{3}{2}\right) D(a_0)^3\int^{a_0} \mathrm{da_1}  D(a_1) \mathcal{H}(a_1) \Omega_{\mathrm{dm}}(a_1) \ (s-s(a_1))  \\
&\qquad \qquad \vphantom{\int} \times \tilde{f}^{\lbrack 0 \rbrack}(\vec{k_1}(s-s(a_1)) \ 2 \cdot F_2(\vec{k_1}, \vec{k_2})P_{11}(k_1,a_0)P_{11}(k_2,a_0) + 2 \mathrm{\ cycl. \ perm.} \ ,\nn 
\end{align}
where in equation \eqref{eq:B1112} the Poisson equation in Fourier space has been used in $\star$, given by:
\begin{align}
\partial_i \Phi_k = \frac{\partial_i }{\partial^2} \partial^2 \Phi_k = \frac{i k^{i_1}}{-k^2} \left( \frac{3}{2} \right) \mathcal{H}^2(a) \Omega_{\mathrm{dm}}(a) \delta_{\mathrm{dm}}(\vec{k},a_0)\ ,
\end{align}
and an integration by parts has been performed in $\star \star$. The first term vanishes at the boundaries since there are no particles with infinite velocity and $f^{\lbrack 0\rbrack}(v)$ is zero at the origin. Moreover, for notational convenience, we wrote the density fields at present time as $\delta^{(n)}(\vec{q_1},a_0)=\delta^{(n)}(\vec{q_1})$ and used the shorthand notation for the momentum integrals: $\int \frac{d^3q}{(2 \pi)^3} = \int_q$. The lowest order kernel is defined as $F_1(\vq_1) = 1$ and, for simplicity of the equation, has been neglected in the equations \cite{Bernardeau:2001CPT}. We remind the reader, that $\tilde{f}^{\lbrack 0 \rbrack}$ is the Fourier transform of the relative distribution of the difference of neutrinos and dark matter.

The next diagram is shown in figure \ref{fig:B_diff_dm_dm_2}; it consists of computing to first order in the neutrino density distribution the expectation value of $\langle f^{\lbrack 1,2 \rbrack}\delta^{(1)}_{\mathrm{dm}} \delta^{(1)}_{\mathrm{dm}}\rangle$. We take the gravitational potential to second order, $\Phi^{(2)}$, in $f^{\lbrack 1\rbrack}$ and contract it with two linear order dark matter fields, $\td^{(1)}_{\rm dm}$. The second diagram is therefore given by:
\begin{align}
B^{ \lbrack 1 , 2 \rbrack \lvert (1) \lvert (1)}_{\mathrm{diff, dm, dm}} &(k_1, k_2, k_3, a_0)=\nn \\ 
&=  \left(\frac{3}{2}\right) D(a_0)^2\int^{a_0} \mathrm{da_1}  D(a_1)^2 \mathcal{H}(a_1) \Omega_{\mathrm{dm}}(a_1) \ (s-s(a_1))  \\
&\qquad \qquad \vphantom{\int} \times \tilde{f}^{\lbrack 0 \rbrack}(\vec{k_1}(s-s(a_1)) \ 2 \cdot F_2(\vec{k_2}, \vec{k_3})P_{11}(k_2,a_0)P_{11}(k_3,a_0) + 2 \mathrm{\ cycl. \ perm.} \nn 
\end{align}

The next diagram is shown in figure \ref{fig:B_diff_dm_dm_3}; it consists of computing to second order in the neutrino density distribution the expectation value of $\langle f^{\lbrack 2,11 \rbrack}\delta^{(1)}_{\mathrm{dm}} \delta^{(1)}_{\mathrm{dm}}\rangle$. We take the gravitational potential twice at linear order, $\Phi^{(1)}$, in $f^{\lbrack 2\rbrack}$ and contract it with two linear order dark matter fields, $\td^{(1)}_{\rm dm}$. The third diagram is therefore given by:
\begin{align}
B^{ \lbrack 2,11 \rbrack \lvert (1) \lvert (1)}_{\mathrm{diff, dm, dm}} &(k_1, k_2, k_3, a_0, a_1)=\nn \\ 
&= D(a_0)^2 \int^{a_0} \mathrm{da_1}\int^{a_1} \mathrm{da_2} \left(\frac{3}{2}\right)^2 \mathcal{H}(a_1) \Omega_{\mathrm{dm}}(a_1) \mathcal{H}(a_2) \Omega_{\mathrm{dm}}(a_2)  D(a_1) D(a_2) \\
&\qquad \quad \Bigg\{ \left[ \frac{\vec{-k_2}(-\vec{k_2}-\vec{k_3})}{k_2^2}(s-s(a_1))\right] \left[\frac{-\vec{k_3}(-\vec{k_2}-\vec{k_3})}{k_3^2} (s-s(a_1))+(s(a_1)-s(a_2))\right] \nn \\
&\qquad \qquad \vphantom{\int} \times \tilde{f}^{\lbrack 0 \rbrack}\left( (-\vec{k_2}-\vec{k_3})(s-s(a_1))- \vec{k_3}(s(a_1)-s(a_2)) \right) \times 2P_{11}(k_2,a_0)P_{11}(k_3,a_0) \Bigg\}_{\vec{k_2} \leftrightarrow \vec{k_3}} \nn \\
&\qquad \qquad \quad \vphantom{\int} + 2 \mathrm{\ cycl. \ perm.} \nn 
\end{align}
The sum of the three diagrams shown in figure \ref{fig:perturbative_sol_NB} makes up $B_{\rm diff, dm, dm}$. 

\subsection{Contribution to $B_{\mathrm{dm, dm, dm}}$ \label{sec:B_dm_dm_dm}}

The overdensities in neutrinos exert an effect on the gravitational potential  which in turn affects the dark matter overdensity. We call the effect on $B_{\rm dm, dm, dm}$, linear in $f_{\nu}$, $\Delta B_{\rm dm, dm, dm}$. In other words, in our notation, the total bispectrum takes the form:
\begin{align}
B(k_1,k_2,k_3,a) &= B_{\rm tree, f_{\nu}=0}(k_1,k_2,k_3,a) + f_{\nu} \Delta B_{\rm dm,dm,dm}(k_1,k_2,k_3,a)+\\
& \quad \quad + f_{\nu} \left( B_{\rm diff, dm, dm}(k_1,k_2,k_3,a) + B_{\rm dm, diff, dm}(k_1,k_2,k_3,a) + \right. \nn \\
& \left. \quad \quad \quad + B_{\rm dm, dm, diff}(k_1,k_2,k_3,a) \right) + {\cal{O}}(f_{\nu}^2) \nn \ ,
\end{align}
where $B_{\rm tree, f_{\nu}=0}$ was defined in equation \eqref{eq:bispectrum_tree_level}. In order to compute these diagrams contributing to the tree-level bispectrum, the equations of motions of dark matter in the presence of neutrinos are solved perturbatively. This set of equations is given by the following form (using the notation and derivations given in  \cite{Senatore:massive_neutrinos,Lewandowski2016:Greensfunction}):
\begin{subequations} \label{eq:EoM_delta_B_dmdmdm}
\begin{align}
& \frac{\d^2}{a^2} \Phi = \frac{3}{2} H_0^2 \frac{a_0^3}{a^3} (\Omega_{\mathrm{dm,0}}\delta_{\mathrm{dm}} + \Omega_{\nu,0}\delta_{\nu}) \approx \frac{3}{2} H_0^2 \frac{a_0^3}{a^3} \Omega_{\mathrm{NR,0}}(\delta_{\mathrm{dm}} + f_{\nu}\delta_{\rm diff})\ , \\
&\frac{1}{a} \frac{\d \delta_{\mathrm{dm}}}{\d a} + \frac{1}{a}\theta_{\mathrm{dm}} + \frac{1}{a} \d_i (\delta_{\mathrm{dm}}v^i_{\mathrm{dm}}) = 0\ , \\
&\frac{1}{a} \frac{\d \theta_{\mathrm{dm}}}{\d a} +H\theta_{\mathrm{dm}}+ \frac{1}{a}\d^2 \Phi + \frac{1}{a} \d_i (v^j_{\mathrm{dm}} \d_j\delta^i_{\mathrm{dm}}) = 0\ ,
\end{align}
\end{subequations}
where we used the following definitions\footnote{We are still doing the computations without counterterms as no sensitivities to uncontrolled short-distance non-linearities arise during the calculations at tree-level. Moreover, we neglected the velocity vorticity that is generated by the counterterms, and, hence, negligible at leading order \cite{Carrasco:2013sva, Mercolli:2013bsa, Senatore:massive_neutrinos}.}: $\theta_{\mathrm{dm}}=\d_iv^i_{\mathrm{dm}}(\vec{x},t)$, $\pi^i_{\mathrm{dm}}(\vec{x},t) =\rho(\vec{x},t)v^i_{\mathrm{dm}}(\vec{x},t)$ and $\Omega_{\mathrm{NR,0}}$ is the present energy fraction in non-relativistic matter. With the standard growth factor $f_g = \frac{d \log D_0}{d \log a}$ where $D_0$ is the growth factor in absence of neutrinos, we define
\begin{equation}
\Theta_{\mathrm{dm}}\equiv-\frac{\theta_{\mathrm{dm}}}{f_g \mathcal{H}}\ .
\end{equation}
This gives us a compact way of rewriting the equations of motions\footnote{For simplicity of the equations, we dropped the dark matter subscript $_{\rm dm}$ whenever it does not create confusion.} in terms of $\Theta$:
\begin{subequations}\label{EoMdm}
\begin{align}
&a \frac{\d\delta_{\vec{k}}}{\d a} - f_g\Theta_{\vec{k}} = f_g \int_{q_1} \int_{q_2} (2 \pi)^3 \delta_D^{(3)}(\vec{k}-\vec{q_1}-\vec{q_2}) \alpha(\vec{q_1},\vec{q_2})\Theta_{\vec{q_1}}\delta_{\vec{q_2}}\ , \\
&a \frac{\d\Theta_{\vec{k}}}{\d a} - f_g\Theta_{\vec{k}} + \frac{3}{2} \frac{\Omega_{\mathrm{dm}}(a)}{f_g}(\Theta_{\vec{k}}-\delta_{\vec{k}}-f_{\nu}\delta_{\nu,\vec{k}}) =\\
&\qquad \qquad \quad \quad = f_g \int_{q_1} \int_{q_2} (2 \pi)^3 \delta_D^{(3)}(\vec{k}-\vec{q_1}-\vec{q_2}) \alpha(\vec{q_1},\vec{q_2}) \Theta_{\vec{q_1}} \Theta_{\vec{q_2}}\ ,\nonumber
\end{align}
\end{subequations}
where $\alpha$ and $\beta$ are kernels depending on the momentum vectors:
\begin{equation}
\alpha(\vec{q_1},\vec{q_2}) = 1+\frac{\vec{q_1} \cdot \vec{q_2}}{q_1^2}\ ,  \qquad \qquad \beta(\vec{q_1}, \vec{q_2}) = \frac{|\vec{q_1}+\vec{q_2}|^2 \vec{q_1}\cdot\vec{q_2}}{2q_1^2q_2^2} \ .
\end{equation}
At zeroth order in $f_{\nu}$, we can solve the above equations iteratively; $\delta^{(n)}_{\vk}$ and $\Theta^{(n)}_{\vk}$, at any perturbative order, are given by \cite{Senatore:massive_neutrinos, Lewandowski2016:Greensfunction}:
\begin{subequations} \label{eq:dtGreen}
\begin{align}
 \delta^{(n)}_{\vk}&=\int^a_0 d\ta \bigg(G^{\delta}_{1}(a,\ta)S^{(n)}_1(\ta,\vk)+G^{\delta}_{2}(a,\ta)S^{(n)}_2(\ta,\vk)\bigg)\ ,  \\
  \Theta^{(n)}_{\vk}&=\int^a_0 d\ta \bigg(G^{\Theta}_{1}(a,\ta)S^{(n)}_1(\ta,\vk)+G^{\Theta}_{2}(a,\ta)S^{(n)}_2(\ta,\vk)\bigg)\ .
\end{align}
\end{subequations}
The solution is given in terms of two density Green's functions, $G^{\delta}_{1}(a,\ta)$, $G^{\delta}_{2}(a,\ta)$, and two velocity Green's functions, $G^{\Theta}_{1}(a,\ta)$ and $G^{\Theta}_{2}(a,\ta)$ as well as the source terms of the non-linear continuity and Euler equations at $n$-th order, $S_1^{(n)}(a, \vec{k})$ and $S_2^{(n)}(a, \vec{k})$, respectively. They are given by:
\begin{subequations} \label{eq:source}
\begin{align}
&S_1^{(n)}(a,\vk)=f_{g}(a)\sum\limits_{m=1}^{n-1}\int \frac{d^3q}{(2\pi)^3}  \alpha(\vq,\vk-\vq)  \tT^{(m)}_{\vq}(a)\td^{(n-m)}_{\vk-\vq}(a)\ ,      \\
&S_2^{(n)}(a,\vk)=f_{g}(a)\sum\limits_{m=1}^{n-1}\int \frac{d^3q}{(2\pi)^3} \beta(\vq,\vk-\vq) \tT^{(m)}_{\vq}(a)\tT^{(n-m)}_{\vk-\vq}(a)\ .
\end{align}
\end{subequations}
The explicit expressions for the Green's functions are given in App. C of \cite{Lewandowski2016:Greensfunction}.

At first order in $f_{\nu}$, the contribution of $\td_{\nu}$ to the evolution of dark matter is related to the influence it has through the gravitational potential. This is denoted by $\delta_{\mathrm{dm \leftarrow \nu}}^{(n)}$ and $\Theta_{\mathrm{dm \leftarrow \nu}}^{(n)}$, respectively. The notation $\mathrm{dm \leftarrow \nu}$ stands for induced dark matter perturbations originating from the neutrino perturbations. The only difference between $\delta_{\mathrm{dm \leftarrow \nu}}^{(n)}$ and $\Theta_{\mathrm{dm \leftarrow \nu}}^{(n)}$ is the used Green's function after the transition from a neutrino to a dark matter field, denoted by the mixing vertex $\frac{3}{2} \frac{\Omega_{\mathrm{dm}}}{f_g}$. The induced perturbations of $\td_{\nu}$ in $\td_{\rm dm}$ and $\Theta_{\rm dm}$ are given by: 
\begin{subequations}\label{eq:nu_induced_pert}
\begin{align} 
&\delta^{(n)}_{\mathrm{dm \leftarrow \nu}}(\vec{k}, a)= f_{\nu} \int^a \mathrm{d\tilde{a}} \ G^{\delta}_{2}(a,\ta)\ \frac{3}{2} \frac{\Omega_{\mathrm{dm}}}{f_g(\tilde{a})} \delta_{\nu}^{(n)}(\vk,\ta),  \\
&\Theta^{(n)}_{\mathrm{dm \leftarrow \nu}}(\vec{k}, a)= f_{\nu} \int^a \mathrm{d\tilde{a}} \ G^{\Theta}_{2}(a,\ta)\ \frac{3}{2} \frac{\Omega_{\mathrm{dm}}}{f_g(\tilde{a})} \delta_{\nu}^{(n)}(\vk,\ta).
\end{align} 
\end{subequations}
Now, we can compute a part of the desired tree-level bispectra by directly contracting $\delta^{(n)}_{\mathrm{dm \leftarrow \nu}}$ and $\Theta^{(n)}_{\mathrm{dm \leftarrow \nu}}$ with the adequate linear and second order dark matter fields. We can derive the remaining diagrams, by using the solutions in the perturbative expression resulting from equation \eqref{EoMdm}. In particular, this means that we can insert the expression in equation \eqref{eq:nu_induced_pert} in all possible ways into equation \eqref{eq:dtGreen} to obtain a perturbation at the relevant order and then perform the necessary contractions. Thus, two different kinds of diagrams have to be computed: first, the ones where we evolve non-linearly the neutrinos, and then the neutrino field is converted into a dark matter field, without any additional interactions, called \textit{integral diagrams}. Second, the dark matter field perturbation induced by the conversion of the neutrino perturbation is evolved non-linearly. We call these diagrams \textit{non-integral diagrams}.

\subsubsection*{Integral diagrams}
\begin{figure}[t!]
\centering
\subfigure[Contracting $f^{[1,1]}$ with $\td_{\rm dm}^{(1)}$ and $\td_{\rm dm}^{(2)}$ yields $\Delta B^{ \lbrack 1 , 1 \rbrack  \lvert (1) \lvert (2)}_{\mathrm{dm, dm, dm}}$. \label{fig:B_dm_dm_dm_4}]{\includegraphics[width = 0.3\textwidth]{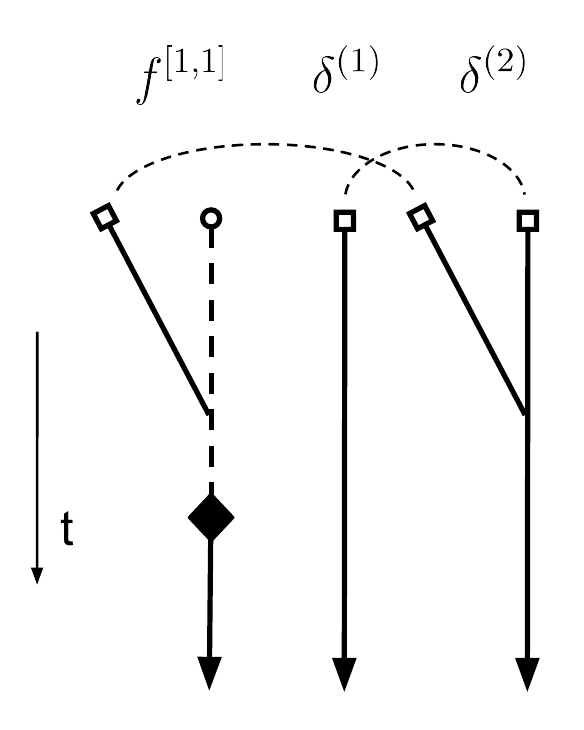}} \quad 
\subfigure[Contracting $f^{[1,2]}$ with two linear dark matter fields, $\td_{\rm dm}^{(1)}$, yields $\Delta B^{ \lbrack 1 , 2 \rbrack \lvert (1) \lvert (1)}_{\mathrm{dm, dm, dm}}$. \label{fig:B_dm_dm_dm_5}]{\includegraphics[width = 0.3\textwidth]{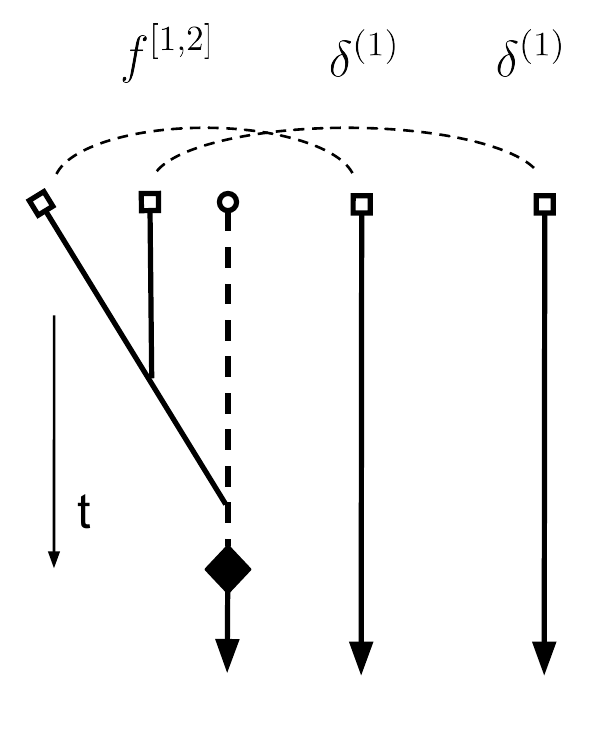}}\quad
\subfigure[Contracting $f^{[2,11]}$ with two linear dark matter fields, $\td_{\rm dm}^{(1)}$, yields $\Delta B^{ \lbrack 2 , 11 \rbrack \lvert (1) \lvert (1)}_{\mathrm{dm, dm, dm}}$. \label{fig:B_dm_dm_dm_6}]{\includegraphics[width = 0.32\textwidth]{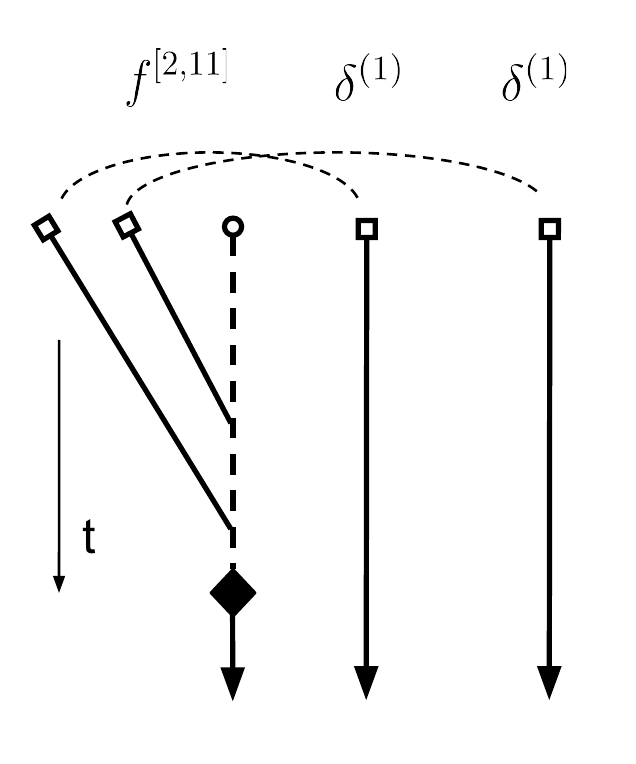}}
\caption[]{Diagrammatic representation of the integral diagrams contributing to $\Delta B_{\mathrm{dm, dm, dm}}$. Perturbative solution for linearly induced, shown in figure \ref{fig:B_dm_dm_dm_4} and \ref{fig:B_dm_dm_dm_6}, and the non-linearly induced, shown in figure \ref{fig:B_dm_dm_dm_5}, dark matter perturbations from the neutrino perturbations, $\td_{\rm dm \leftarrow \nu}^{(n)}$. The filled diamond represents the insertion of the mixing vertex, $\frac{3}{2} \frac{\Omega_{\rm dm}}{f_g}$, which converts a neutrino field (dashed line) into a dark matter field (continuous line). The diagrams with $\theta_{\rm dm \leftarrow \nu}^{(n)}$ are analogous except that after the mixing vertex a different Green's function has to be used. \label{fig:B_dm_dm_dm_integral_diag}}
\end{figure}

These diagrams are formed by directly contracting the fluctuations in $\delta^{(n)}_{\mathrm{dm \leftarrow \nu}}$ and $\theta^{(n)}_{\mathrm{dm \leftarrow \nu}}$ with the dark matter fields at the required order, as shown in figure \ref{fig:B_dm_dm_dm_integral_diag}. These diagrams allow us to recycle the results from the previous section. We only need to add a neutrino mixing vertex and the adequate Green's function, $G_2^{\delta}(a, \ta)$ or $G_2^{\theta}(a, \ta)$, and integrate over time. This yields the contribution, called the integral diagrams:
\begin{align} \label{eq:integral_diag_formula}
\Delta B^{\left[n, i_1 \dots i_n\right] \lvert (m_1)\lvert (m_2)}_{\mathrm{dm,\ dm,\ dm}} (k_1, k_2, k_3, a_0) = \int^{a_0}\mathrm{d\tilde{a}} \ G_2^{\delta}(a_0, \tilde{a}) \ \frac{3}{2} \frac{\Omega_{\mathrm{dm}}(\tilde{a})}{f_g(\tilde{a})} B^{\left[n, i_1 \dots i_n\right] \lvert (m_1) \lvert (m_2)}_{\rm diff,\ dm,\ dm}(k_1, k_2, k_3,  a_0, \tilde{a})\ ,
\end{align}
where $B^{\left[n, i_1 \dots i_n\right] \lvert (m_1) \lvert (m_2)}_{\rm diff,\ dm,\ dm}$ corresponds to the diagrams computed in Sec. \ref{sec:OrdinaryB_diff_dm_dm}. However, due to a physical (and technical) subtlety we compute the diagram shown in figure \ref{fig:B_dm_dm_dm_4} not by using equation \eqref{eq:integral_diag_formula} due to an arising logarithmic divergence. We will discuss this in more depth in the following paragraphs. 

\subsubsection*{Non-integral diagrams}
\begin{figure}[t!]
\centering
\includegraphics[width = 0.4\textwidth]{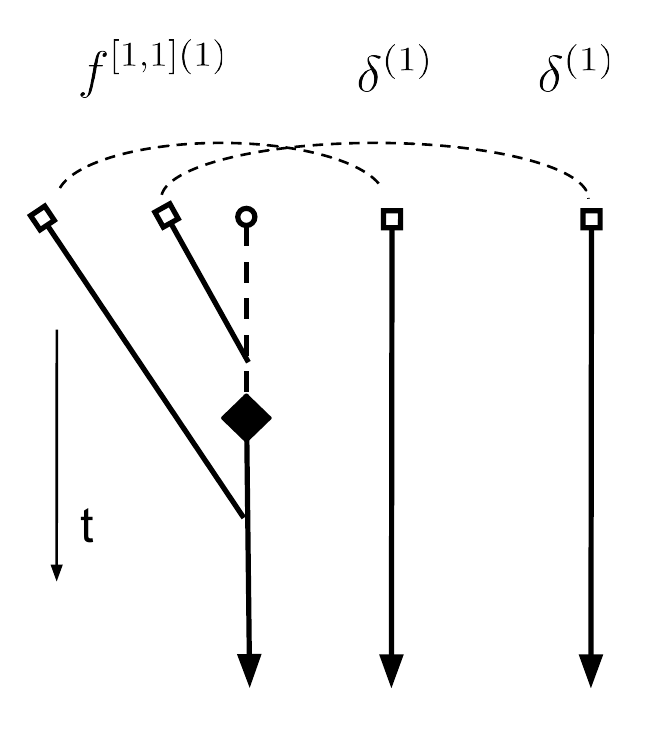}
\caption{Diagrammatic representation of the non-integral diagram. We contract $\td_{\rm dm}^{\lbrack 1,1\rbrack(1)}$ with two linear dark matter fields, $\td_{\rm dm}^{(1)}$, to obtain $\Delta B^{ \lbrack 1 , 1 \rbrack (1) \lvert (1) \lvert (1)}_{\mathrm{dm, dm, dm}}$. It is important to notice that the sum over all possible insertions of $\td_{\rm dm \leftarrow \nu }$ and $\theta_{\rm dm \leftarrow \nu }$ has to be considered. \label{fig:B_dm_dm_dm_nonintegral_diag}}
\end{figure}

The other contributing diagram is obtained by using the perturbations induced by $\delta_{\nu}$ in $\delta_{\mathrm{dm}}$ and $\theta_{\mathrm{dm}}$, which are given by equation \eqref{eq:nu_induced_pert}, denoted by $\delta_{\mathrm{dm \leftarrow \nu}}^{(n)}$ and $\Theta_{\mathrm{dm \leftarrow \nu}}^{(n)}$, and then evolving these fields non-linearly at the relevant order. This leads, at tree-level, to one non-integral diagram shown in figure \ref{fig:B_dm_dm_dm_nonintegral_diag}. In our notation, we denote these solutions by $\td_{\rm dm}^{\lbrack q, i_1 \dots i_q \rbrack (m)}$, with $i_1 + \dots + i_q = n$, and construct a dark matter solution of order $(n+m)$. In other words, we consider a $\delta_{\mathrm{dm \leftarrow \nu}}^{(n)}$ or $\Theta_{\mathrm{dm \leftarrow \nu}}^{(n)}$ induced by the neutrino distribution $f^{\lbrack q, i_1 \dots i_q \rbrack}$, with $i_1 + \dots + i_q = n$, and insert additional dark matter interactions to $m$-th order. The desired diagram is obtained by contracting this solution, $\td_{\rm dm}^{\lbrack q, i_1 \dots i_q \rbrack (m)}$, with two dark matter fields at order $p_1$ and $p_2$, given by $\td_{\rm dm}^{(p_1)}$ and $\td_{\rm dm}^{(p_2)}$, respectively, yielding $\Delta B^{\left[q, i_1 \dots i_q \right](m) \lvert (p_1)\lvert (p_2)}_{\mathrm{dm, dm, dm}}$.

In order to compute the diagram shown in figure \ref{fig:B_dm_dm_dm_nonintegral_diag}, we use the solution $\td^{\lbrack 1,1\rbrack (1)}_{\rm dm}$, for which we use $\td_{\rm dm \leftarrow \nu}^{\lbrack 1,1\rbrack}$ and $\Theta_{\rm dm \leftarrow \nu}^{\lbrack 1,1\rbrack}$, respectively, and contract it with two linear dark matter fields, $\td^{(1)}_{\rm dm}$. To compute this diagram, we apply the following the trick: first, we solve equation \eqref{eq:EoM_delta_B_dmdmdm} by using equation \eqref{eq:dtGreen}. Then, we contract the three density fields: $\td^{\lbrack 1,1\rbrack (1)}_{\rm dm}$ in the presence of massive neutrinos and the two linear order dark matter fields, $\td^{(1)}_{\rm dm}$. From the resulting diagram, we subtract the same diagram in the absence of neutrinos in the same cosmology. This procedure gives us the desired diagram $\Delta B^{ \lbrack 1 , 1 \rbrack (1) \lvert (1) \lvert (1)}_{\mathrm{dm, dm, dm}}$. It is important to note, that we have to account for the correct time- and $k$-dependence for $\td_{\rm dm}^{(1)}$ and $\Theta_{\rm dm}^{(1)}$, as in the presence of massive neutrinos the linear growth factor is $k$-dependent. 

Now, we address the aforementioned arising technical complication when using the expressions for linear dark matter fields induced at linear level by $\td_{\rm diff}$, denoted by $\td_{\rm dm \leftarrow \nu}^{\lbrack 1,1\rbrack}$ and $\Theta_{\rm dm \leftarrow \nu}^{\lbrack 1,1\rbrack}$ in equation \eqref{eq:dtGreen}. As mentioned earlier, these perturbations are not only used to compute the diagram in figure \ref{fig:B_dm_dm_dm_4}, but also, to create $\td_{\rm dm \leftarrow \nu}^{\lbrack 1,1\rbrack (1)}$ for the diagram in figure \ref{fig:B_dm_dm_dm_nonintegral_diag}. The diagrams $\Delta B^{ \lbrack 1 , 1 \rbrack  \lvert (1) \lvert (2)}_{\mathrm{dm, dm, dm}}$, shown in figure \ref{fig:B_dm_dm_dm_4}, and $\Delta B^{ \lbrack 1 , 1 \rbrack (1) \lvert (1) \lvert (1)}_{\mathrm{dm, dm, dm}}$, shown in figure \ref{fig:B_dm_dm_dm_nonintegral_diag}, are logarithmically sensitive to the initial time of evaluation because the Green's function behaves as $\sim 1/a^2$ at early times. This leads to a \textit{logarithmic enhancement}. 
It originates from a \textit{back-reaction} of neutrinos on dark matter. This leads to a correction of the dark matter growth rate\footnote{The correction to the dark matter growth factor is derived in App. \ref{sec:growth_factor_DM_appendix} in an Einstein-de Sitter universe.}. This phenomenon, typical when treating perturbatively a term in the linear equations of motion, is known as \textit{secular enhancement} and leads to a break down of the perturbative solution of linear equations of motion after long enough times. Since the effect is still relatively small in our case perturbation theory still holds. In our universe, this logarithmic sensitivity is regulated at the epoch of matter-radiation equality, where, however, our non-relativistic equations do not hold anymore. In order to compute these diagrams, we use the following trick: we construct $\td_{\rm dm \leftarrow diff}^{\lbrack 1,1\rbrack}$ and $\Theta_{\rm dm \leftarrow diff}^{\lbrack 1,1\rbrack}$, which are the linearly induced dark matter fields by $\td_{\rm diff}$, by using the difference of two exact solutions given by the relativistic Boltzmann solver \textsc{camb}\footnote{\url{http://camb.info}} \cite{Lewis:1999bs}. This guarantees that our equations are correct at early times. We then compute each diagram in the presence of massive neutrinos and subtract from the resulting diagram the one in absence of neutrinos but with the same cosmology. This workaround: using a relativistic Boltzmann solver instead of equation \eqref{eq:nu_induced_pert}, allows us to correctly compute the diagram. 

Having computed the integral and non-integral diagrams, we conclude the evaluation of the diagrams that contribute to $\Delta B_{\mathrm{dm, dm, dm}}$. As mentioned earlier, we are doing the calculations at tree-level and, hence, do not include counterterms. 

\section{Results of the Bispectrum Computations \label{sec:BispectrumResults}}
\begin{figure}[t!]
\centering
\subfigure[Linear total-matter power spectrum generated with \textsc{camb} for massive and massless neutrinos \label{fig:PowerspecMassiveNeutrinos}]{\includegraphics[width = 0.49\textwidth]{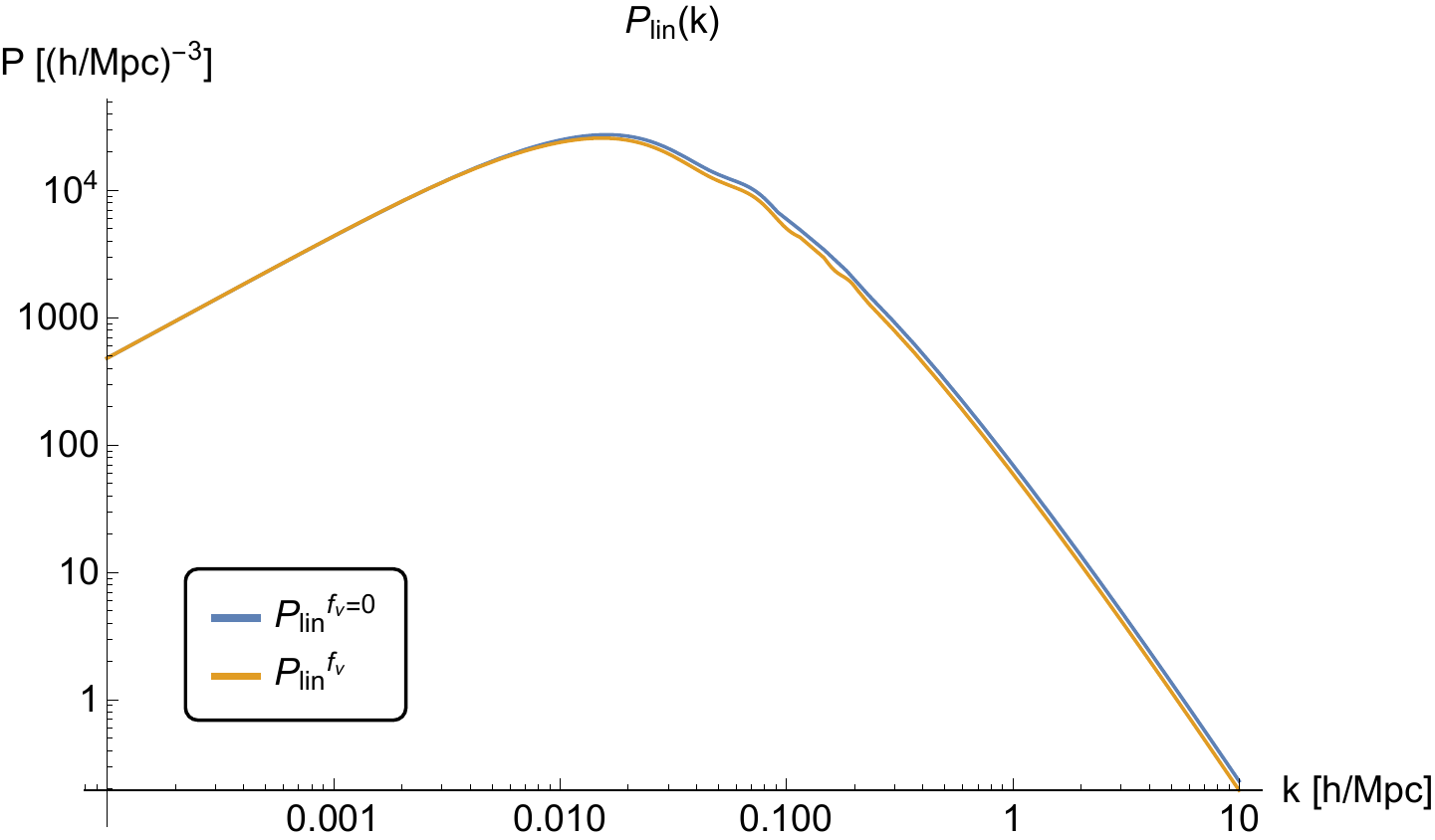}}
\subfigure[Damping of the linear dark matter power spectrum due to massive neutrinos with $\Sigma\ m_{\nu, i}=0.23$eV. \label{fig:PowerspecMassiveNeutrinosRatio}]{\includegraphics[width = 0.49\textwidth]{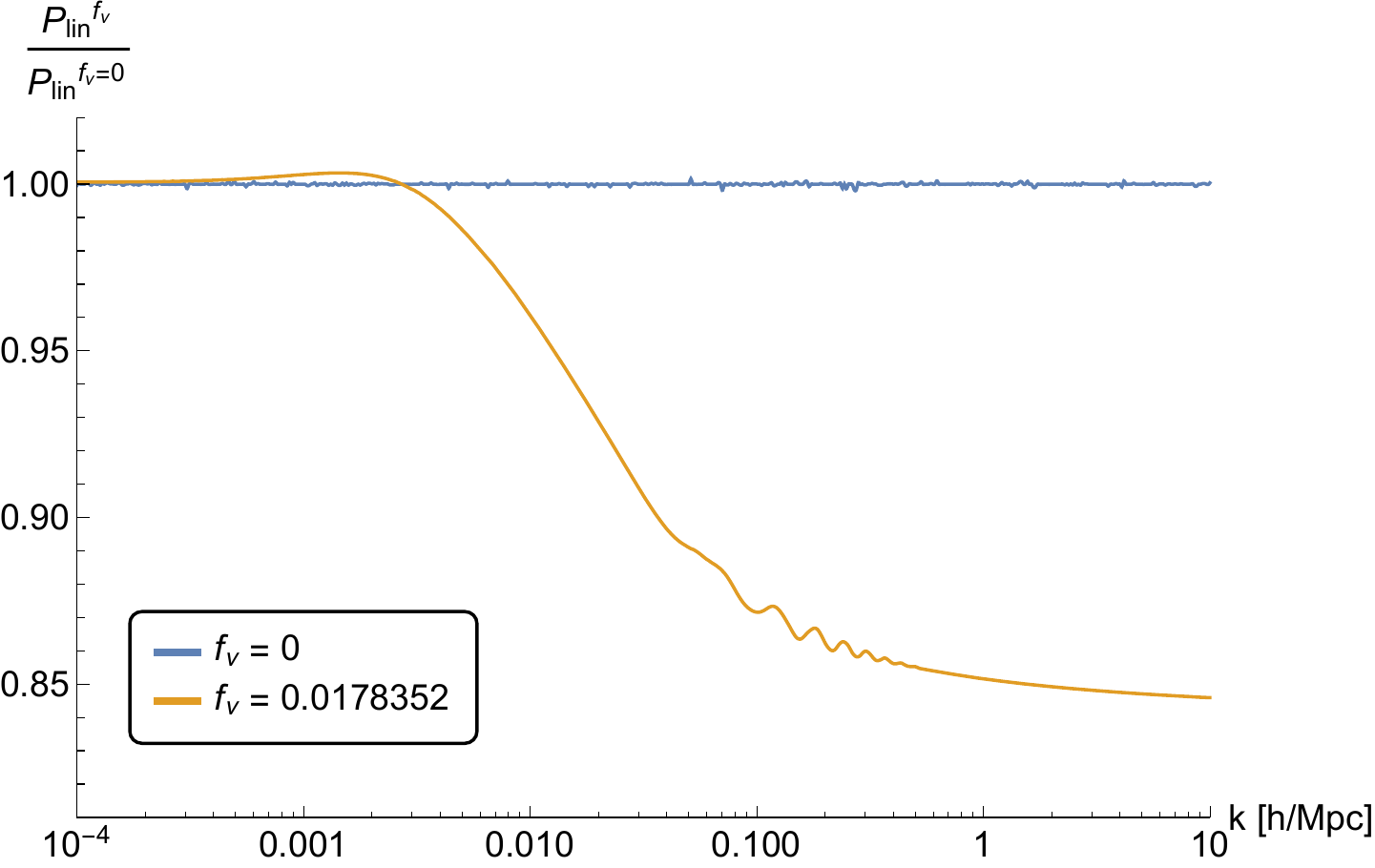}}
\caption{The linear power spectrum with massless and massive neutrinos generated by the relativistic Boltzmann solver \textsc{camb} is given in figure \ref{fig:PowerspecMassiveNeutrinos}. The suppression of the linear matter power spectrum due to massive neutrinos ($P_{\rm lin}^{f_{\nu}}/P_{\rm lin}^{f_{\nu}=0}$) is shown in figure \ref{fig:PowerspecMassiveNeutrinosRatio}. The suppression sets in at $k \sim 5 \cdot 10^{-3} \hinvMpc$. The damping factor depends on the neutrino energy density fraction, here taken to be $f_{\nu} \approx 0.018$. The effect is of order $\sim f_{\nu}\log(\frac{a_{\rm eq}}{a_0}) \sim 8f_{\nu}$ for $k>k_{\rm fs}$. \label{fig:PowerspecMassiveNeutrinosPlot}}
\end{figure}
In this section we discuss the obtained results when numerically integrating the tree-level diagrams. The correction to the leading order total-matter bispectrum is presented and discussed. The cosmological parameters used in order to generate the linear power spectrum and perform the computations have been kept the same as in \cite{Senatore:massive_neutrinos}, for ease of comparison:
\begin{eqnarray}
&&\Omega_{\mathrm{dm}}=0.2430308\,,\quad \Omega_b = 0.0468\,,\quad \Omega_{\Lambda}=0.704948\,,\quad \Omega_\nu=0.0051692\,,\quad H_0=68.8\,, \nonumber\\ 
&&P_\zeta = 2.187 \cdot 10^{-9}\,,\quad n_s=0.9676\,,\quad k_{\rm pivot}=0.05 h \rm Mpc^{-1} \ .
\end{eqnarray}
For the neutrinos the following masses have been chosen, which lay within the Planck $2\sigma$ bound \cite{Ade:2015xua}. The sum of the neutrino masses is $\sum m_{\nu_i}=0.23 {\rm eV}$ and the individual masses are given by the following hierarchy:
\begin{equation}
m_{\nu_1}=0.0712909 {\rm eV}, \quad m_{\nu_2}=0.0718178 {\rm eV}, \quad m_{\nu_3}=0.0868913 {\rm eV} ,\quad f_\nu=0.0178352\ .
\end{equation}

\begin{figure}[t!]
\centering
\includegraphics[width=0.75\textwidth]{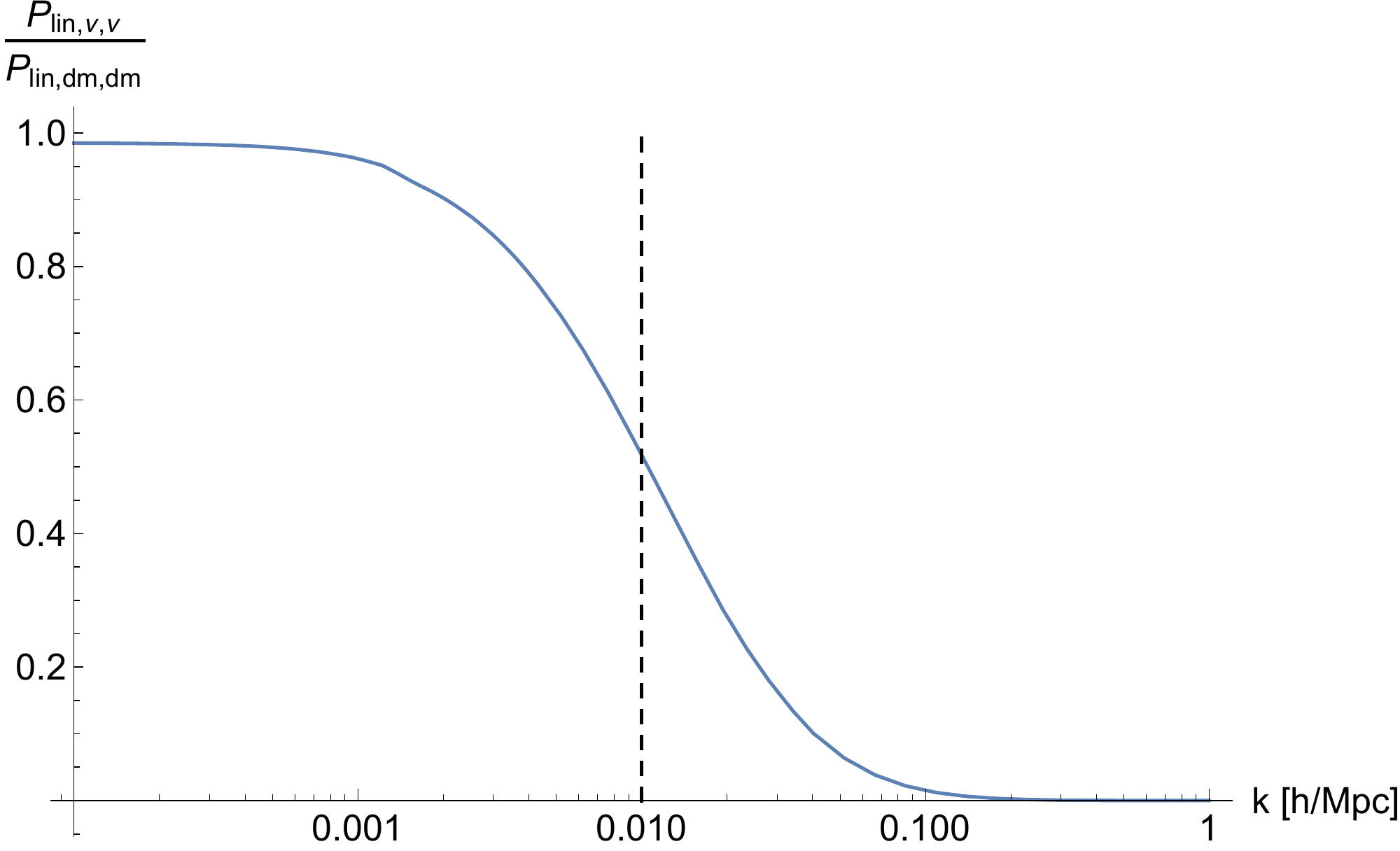}
\caption{Ratio of the linear power spectra for massive neutrinos and dark matter at low and high wavenumber $k$. The linear power spectrum of neutrinos is suppressed at higher wavenumbers due to neutrino free streaming. Therefore, for long wavelengths $\td_{\nu} \simeq \td_{\rm dm}$ and for short wavelengths $\td_{\nu} \simeq 0$. The dashed black line indicates the transition between the two limits and is the wavenumber ($k_{\rm fs}$) associated to the free streaming length. \label{fig:RatioPowerspec}}
\end{figure}

Figure \ref{fig:PowerspecMassiveNeutrinos} shows the linear power spectrum for massless and massive neutrinos. As the neutrinos are massive, they become non-relativistic at late times and cluster which, in turn, affects the total-matter clustering. This can be seen in the power spectrum since massive neutrinos damp it on scales below their free streaming length. This expected damping effect is illustrated in figure \ref{fig:PowerspecMassiveNeutrinosRatio} and augments with increasing $f_{\nu}$. On large scales (i.e. for small wavenumbers $k$) the power spectra are identical for both cases because on distances longer than the free streaming length, which is shorter than the Hubble length because the neutrinos are non-relativistic, neutrinos cluster. The effect of the neutrino masses becomes visible below their free streaming length at $k \sim 5 \cdot 10^{-3} \hinvMpc$. The amount of damping of the power spectrum is proportional to the total energy density of neutrinos, and, therefore, the total mass of all the neutrino species. However, the exact shape and the scale at which the damping sets in, depends on the individual mass of each eigenstate. With the chosen $f_{\nu}$ the power spectrum of purely dark matter is damped by approximately 15\%: this corresponds to a damping of $\sim \log(\frac{a_{\rm eq}}{a_0})f_{\nu} \sim 8f_{\nu}$ for $k \gtrsim 0.2 \hinvMpc$. This is what we referred to earlier as secular enhancement from which the main contribution stems. This gives an effect enhanced in the log of the two scale factors.

\begin{figure}[t!]
\centering
\includegraphics[width=\textwidth]{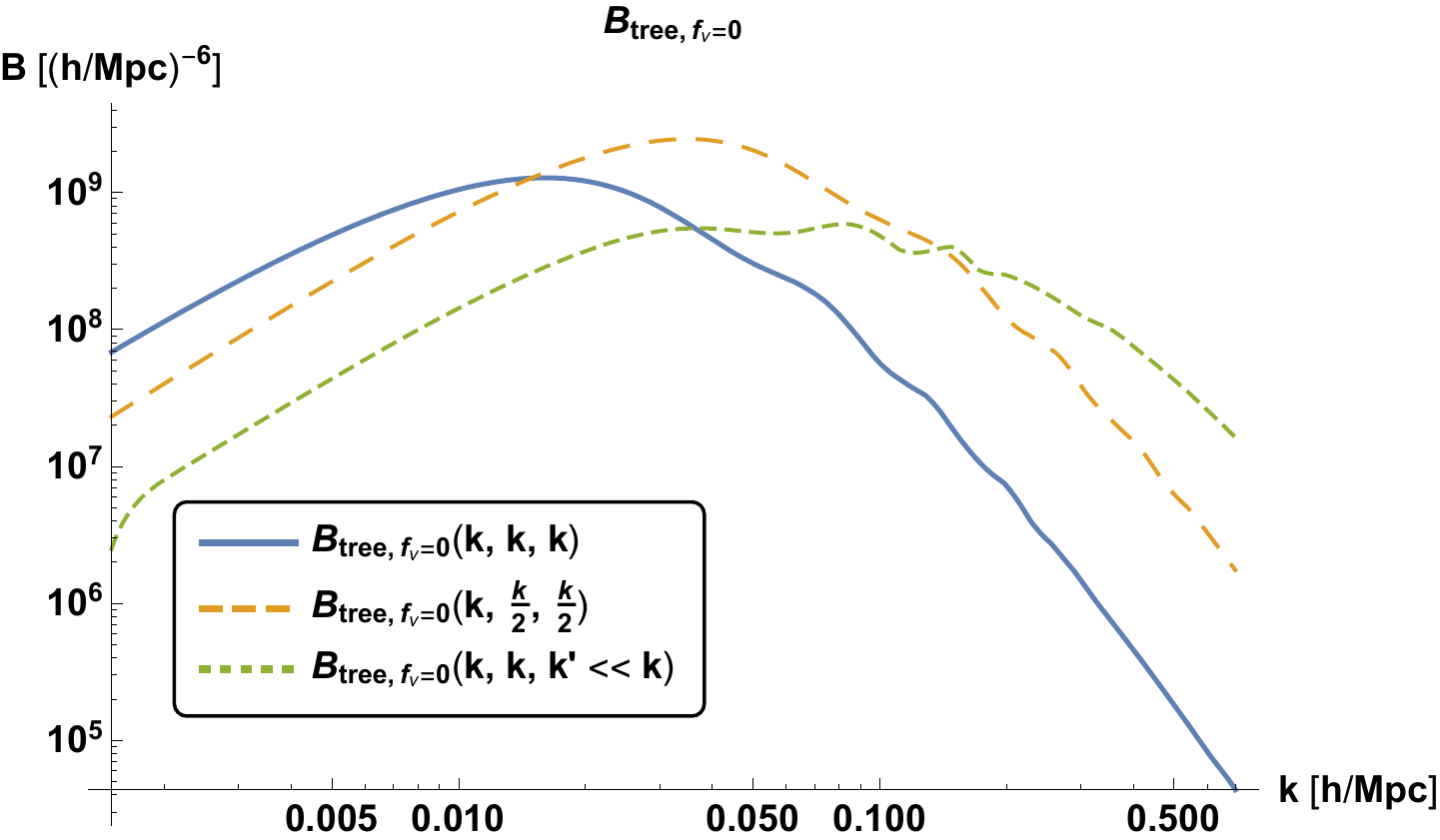}
\caption{Shape dependence of the tree-level dark matter bispectrum $B_{\rm tree, \ f_{\nu}  =  0}$ with varying triangular configurations. The solid blue curve represents the equilateral configuration, $B_{\rm tree, \ f_{\nu}  =  0}(k,k,k)$, the dashed orange one the flat, $B_{\rm tree, \ f_{\nu}  =  0}(k,k/2,k/2)$, and the green dashed line the squeezed limit configuration, $B_{\rm tree, \ f_{\nu}  =  0}(k,k,k' \ll k)$ with $k'/k = 0.1$. \label{fig:Bispectrum_shape_dependence_2D}}
\end{figure}

Figure \ref{fig:RatioPowerspec} shows the ratio between the linear power spectra for massive neutrinos and dark matter. For low wavenumbers, $k \lesssim 5\cdot 10^{-3}\hinvMpc$, the neutrino density fluctuations behave like CDM perturbations. In this regime the neutrino free streaming is negligible. With the chosen cosmology and neutrino masses, the suppression starts at $k \sim 5\cdot 10^{-3} \hinvMpc$. The break, although not sharp, is indicated by the dashed vertical black line in the figure. For $k \gtrsim 10^{-2} \hinvMpc$ the linear power spectrum of neutrinos is highly suppressed due to neutrino free streaming leading to $\td_{\nu}(k, a) \simeq 0$.  This plot allows us to simply derive the behavior of $\td_{\rm diff}$ on all scales at linear order: for long wavelengths, $\td_{\rm diff}$ goes to zero and for short wavelengths $\td_{\rm diff}$ approaches $-\td_{\rm dm}$. The behavior of the linear power spectrum at different scales is important for the interpretation of all perturbative results.

Figure \ref{fig:Bispectrum_shape_dependence_2D} shows the triangular shape dependence in one dimension of the tree-level dark matter bispectrum. Therefore, we plot the equilateral configuration, $B_{\rm tree, \ f_{\nu}  =  0}(k,k,k)$, the flat triangular configuration, $B_{\rm tree, \ f_{\nu}  =  0}(k,k/2,k/2)$, and the squeezed limit, $B_{\rm tree, \ f_{\nu}  =  0}(k,k,k' \ll k)$, with $\frac{k'}{k}\simeq 0.1$.

\begin{figure}[t!]
\centering
\subfigure[$k_1 = 0.05 \hinvMpc$\label{fig:Bispectrum_3D_k1_005}]{\includegraphics[width = 0.48\textwidth]{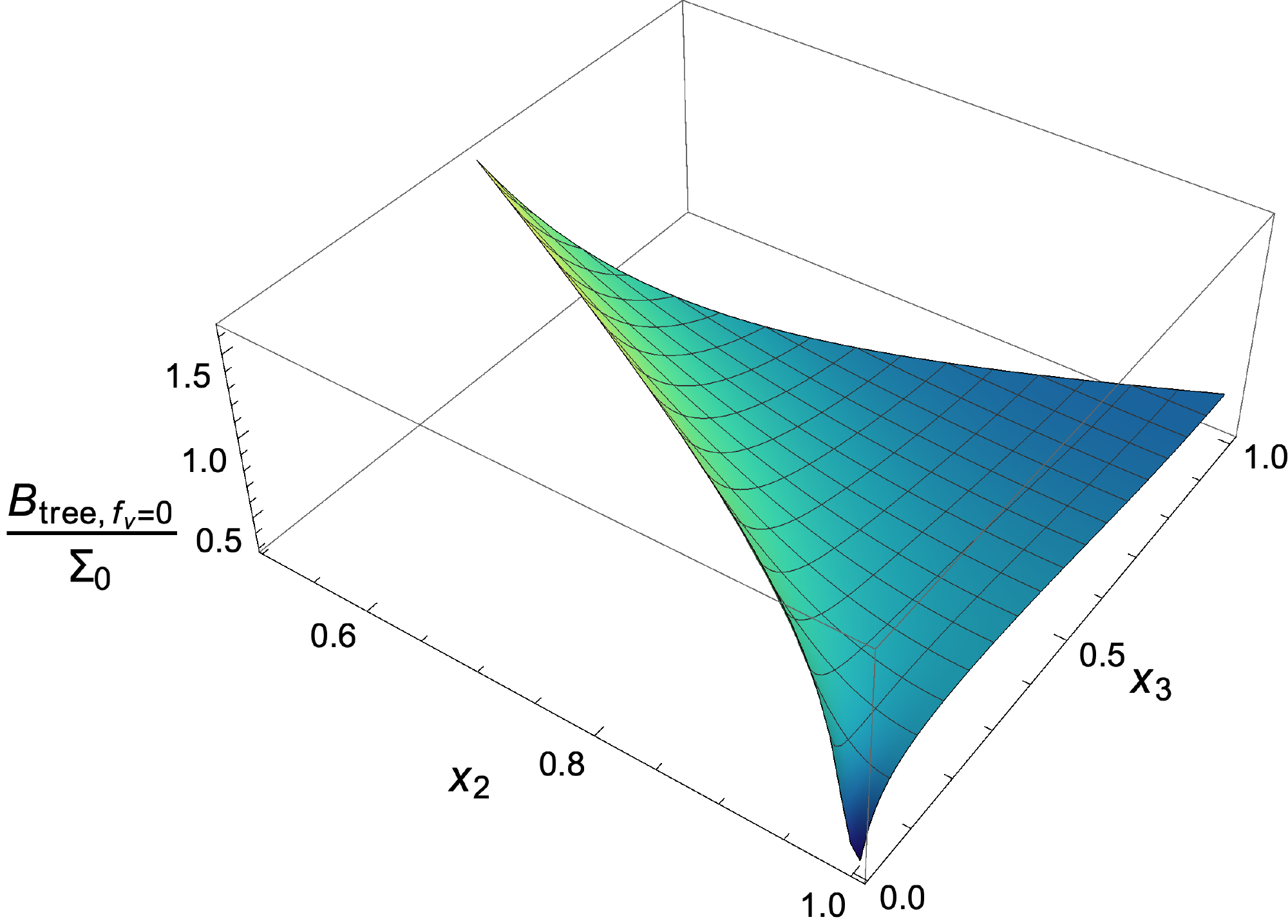}} \quad
\subfigure[$k_1 = 0.2 \hinvMpc$\label{fig:Bispectrum_3D_k1_02}]{\includegraphics[width = 0.48\textwidth]{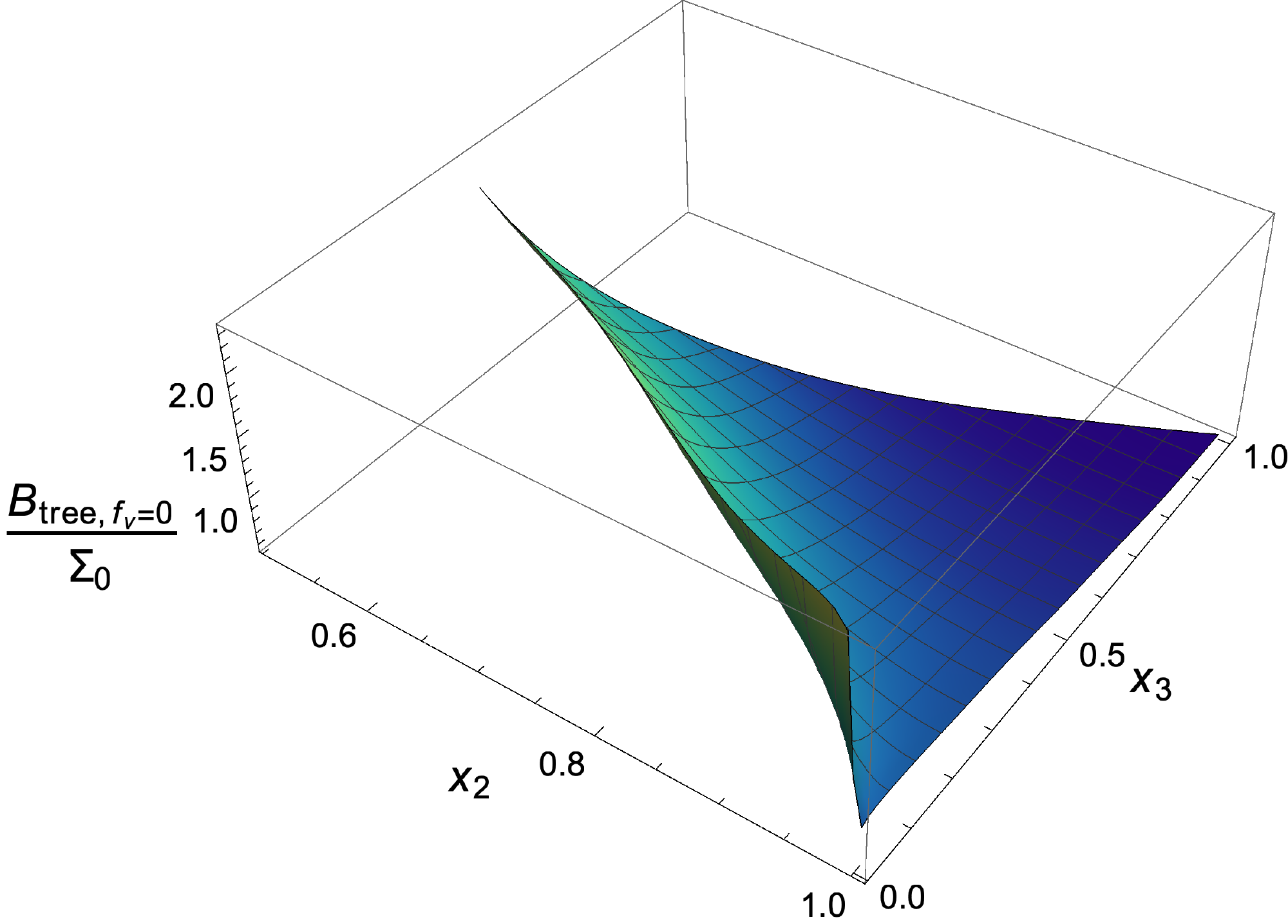}}
\caption{The shape dependence of the reduced bispectrum $Q_{\rm tree}$, which is the ratio of the tree-level dark matter bispectrum $B_{\rm tree, f_{\nu}=0}$ and the sum over the permutations of the linear power spectra $\Sigma_0$, is plotted. $Q_{\rm tree}$ is plotted as function of $x_2 \equiv \frac{k_2}{k_1}$ and $x_3 \equiv \frac{k_3}{k_1}$. The shape is constrained to the fundamental domain $x_2 \geq x_3 \geq 1-x_2$. The triangular shape dependence is shown for a constant $k_1 = 0.05 \hinvMpc$, showing long wavelengths, in figure \ref{fig:Bispectrum_3D_k1_005} and for a constant  $k_1 = 0.2 \hinvMpc$, showing short wavelengths, in figure \ref{fig:Bispectrum_3D_k1_02}. \label{fig:Bispectrum_shape_dependence}}
\end{figure}

In order to better visualize the shape dependence of $B_{\rm tree, f_{\nu}=0}$, we plot the reduced dark matter tree-level bispectrum, $Q_{\rm tree}$, in two dimensions in figure \ref{fig:Bispectrum_shape_dependence}, as suggested in \cite{Bernardeau:2001CPT}. Therefore, the tree-level bispectrum has been divided by the sum over the permutations of the linear power spectra
\begin{equation}
Q_{\rm tree} \equiv \dfrac{B_{\rm tree, f_{\nu}=0}(k_1, k_2, k_3,a)}{\Sigma_0} = \dfrac{2F_2(\vk_1, \vk_2)P_{\rm lin}(k_1, a)P_{\rm lin}(k_2, a) {\rm \ + \ 2 \ cyclic \ perm.}}{P_{\rm lin}(k_1, a)P_{\rm lin}(k_2, a) {\rm \ + \ 2 \ cyclic \ perm.}} \ .
\end{equation}
Translational invariance forces the three-point correlation function to conserve momentum. Rotational invariance further constrains the number of independent variables to just two. Hence, it is instructive to plot the bispectrum as a function of $x_2 \equiv \frac{k_2}{k_1}$ and $x_3 \equiv \frac{k_3}{k_1}$ with constant $k_1$ values. The variables $x_2$ and $x_3$ are constrained to the region $x_2 \geq x_3 \geq 1-x_2$ in order to avoid considering the same configuration twice, with $x_3 \geq 1-x_2$ following from the triangular inequality. 

\begin{figure}[t!]
\centering
\includegraphics[width=\textwidth]{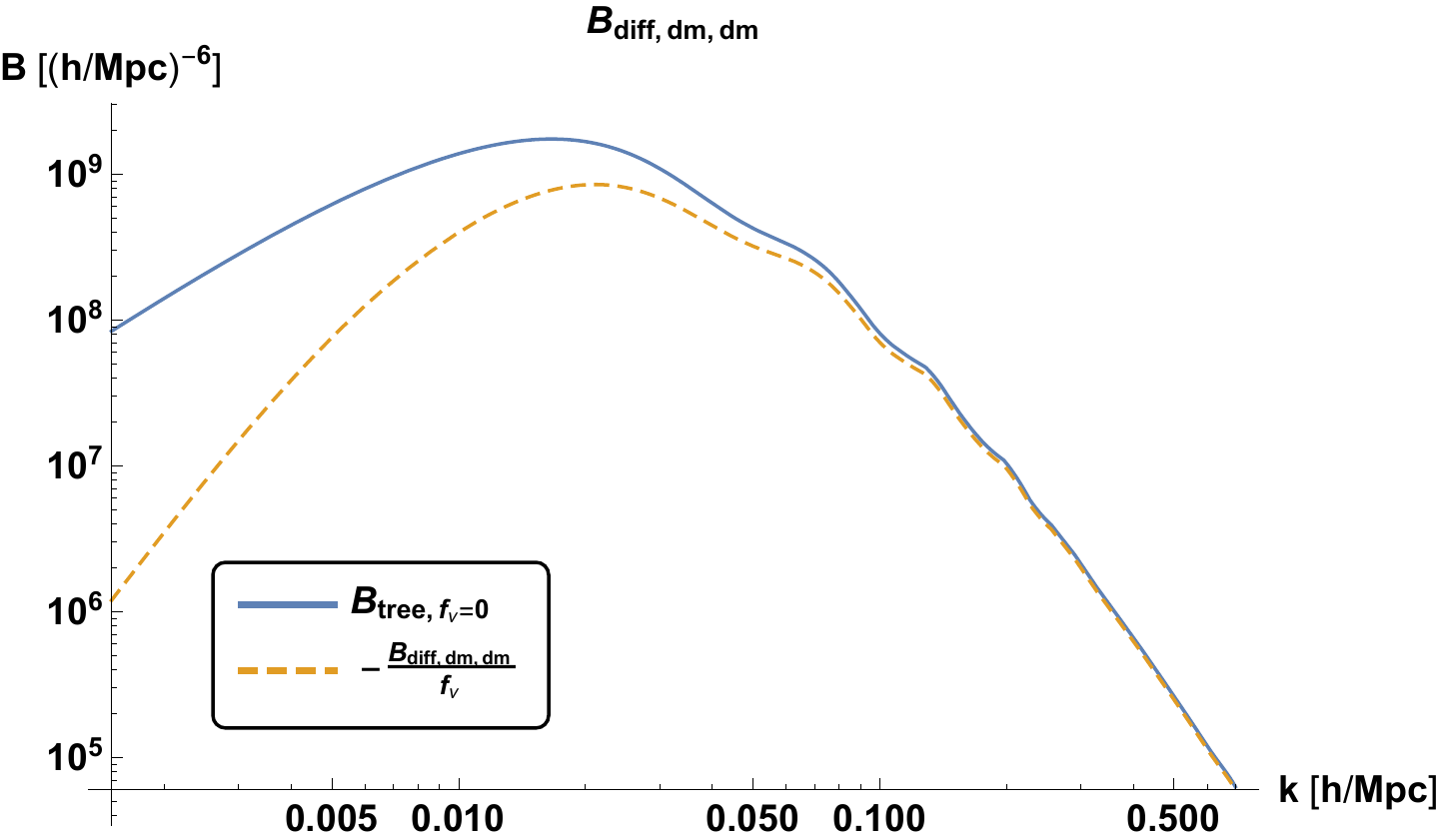}
\caption{Comparison of the absolute value of $B_{\rm diff, dm, dm}$ to the tree-level dark matter bispectrum $B_{\rm tree, f_{\nu}=0}$ for equilateral configurations. On small scales, for $k \gtrsim 0.1 \hinvMpc$, $B_{\rm diff, dm, dm}$ approaches the dark matter tree-level bispectrum by a factor of $-f_{\nu}$. At large scales, for $k< k_{\rm fs}$, $B_{\rm diff, dm, dm}$ is suppressed by up to two orders of magnitude because $\td_{\rm diff}$ is highly suppressed at these scales. \label{fig:B_diff_dm_dm}}
\end{figure}

In figure \ref{fig:Bispectrum_3D_k1_005} the shape dependence of the tree-level dark matter bispectrum, normalized by $\Sigma_0$, with a constant $k_1 = 0.05 \hinvMpc$ is shown in two dimensions. The flat configuration is enhanced by a factor of 1.5.

In figure \ref{fig:Bispectrum_3D_k1_02} the shape dependence of the tree-level dark matter bispectrum, normalized by $\Sigma_0$, with a constant $k_1 = 0.2 \hinvMpc$ is shown in two dimensions. It is predicted that the reduced bispectrum is correlated to the triangular configuration; the flat configuration ($x_2 + x_3 \approx 1$) and the squeezed triangle configuration ($x_2 \rightarrow 1$ and $x_3 \rightarrow 0$) show an enhancement of up to a factor of 2.5 with respect to the equilateral configuration ($x_2 = x_3 = 1$). This is due to the fact that gravity generates anisotropies in LSS, as explained in \cite{Bernardeau:2001CPT}. The equilateral configuration is suppressed by approximately a third. 

Figure \ref{fig:B_diff_dm_dm} compares the dark matter tree-level bispectrum, $B_{\rm tree, f_{\nu}=0}$, to $f_{\nu}B_{\rm diff, dm, dm}$ that we computed in Sec. \ref{sec:OrdinaryB_diff_dm_dm}. For $k \gtrsim 0.1\hinvMpc$, $B_{\rm diff, dm, dm}$ approaches the dark matter bispectrum up to a factor of $-f_{\nu}$ on equilateral configurations. For low wavenumbers, $k \lesssim 10^{-2}\hinvMpc$, $B_{\rm diff, dm, dm}$ is smaller by up to two orders of magnitude. This effect has been anticipated because $\td_{\rm diff}$ is highly suppressed at linear order for small wavenumbers, as one can easily infer from figure \ref{fig:RatioPowerspec}. Similarly, it approaches $-\td_{\rm dm}$ at high $k$.

\begin{figure}[t!]
\centering
\includegraphics[width=\textwidth]{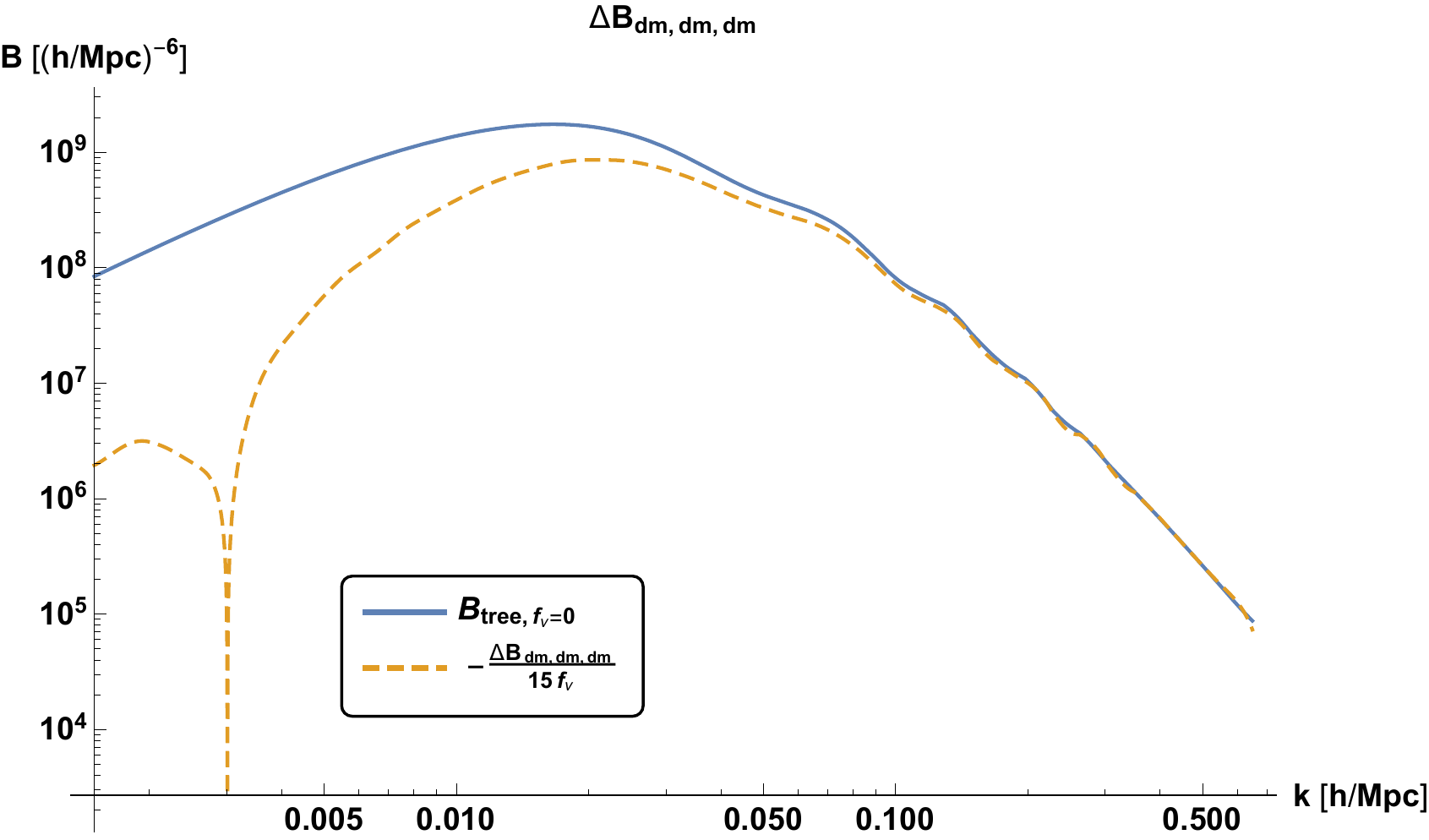}
\caption{Comparison of the absolute value of $\Delta B_{\rm dm, dm, dm}$ to the tree-level dark matter bispectrum $B_{\rm tree, f_{\nu}=0}$ for equilateral configurations. On small scales, for $k \gtrsim 0.1h \rm \ Mpc^{-1}$, $\Delta B_{\rm dm, dm, dm}$ approaches the dark matter tree-level bispectrum by a factor of $-15f_{\nu}$. At large scales, for $k< k_{\rm fs}$, $\Delta B_{\rm dm, dm, dm}$ is suppressed by up to two orders of magnitude because $\td_{\rm diff}$ is higly supressed at these scales. \label{fig:B_dm_dm_dm}}
\end{figure}

Figure \ref{fig:B_dm_dm_dm} shows the influence massive neutrinos have on $B_{\rm dm, dm ,dm }$, denoted by $\Delta B_{\rm dm, dm, dm}$. The overall structure is very similar to the previously discussed case, $B_{\rm diff, dm, dm}$. This comes from the fact that the matter perturbations are induced by $\td_{\rm diff}$ which affects $\td_{\rm dm}$ through the Poisson equation. However, there is a crucial difference: for $k \gtrsim 0.1h \rm \ Mpc^{-1}$, $\Delta B_{\rm dm, dm, dm}$ approaches the dark matter bispectrum up to a factor of $-15f_{\nu}$ for equilateral configurations. The main contribution stems from the influence of the linear $\td_{\rm diff}$ on the linear $\td_{\rm dm}$ which is logarithmically enhanced in the length of time from matter radiation equality to the present time. As previously discussed, this effect is called secular enhancement\footnote{The dark matter growth factor in the presence of massive neutrinos is explicitly derived in App. \ref{sec:growth_factor_DM_appendix}.}.
The sharp drop at $k \sim 3 \cdot 10^{-3}\hinvMpc$ comes from a sign change of the bispectrum. For long wavelengths, for $k \lesssim 10^{-2}\hinvMpc$, the correction is smaller by up to two orders of magnitude due to the suppression of $\td_{\rm diff}$ on large scales.

\begin{figure}[t!]
\centering
\includegraphics[width=\textwidth]{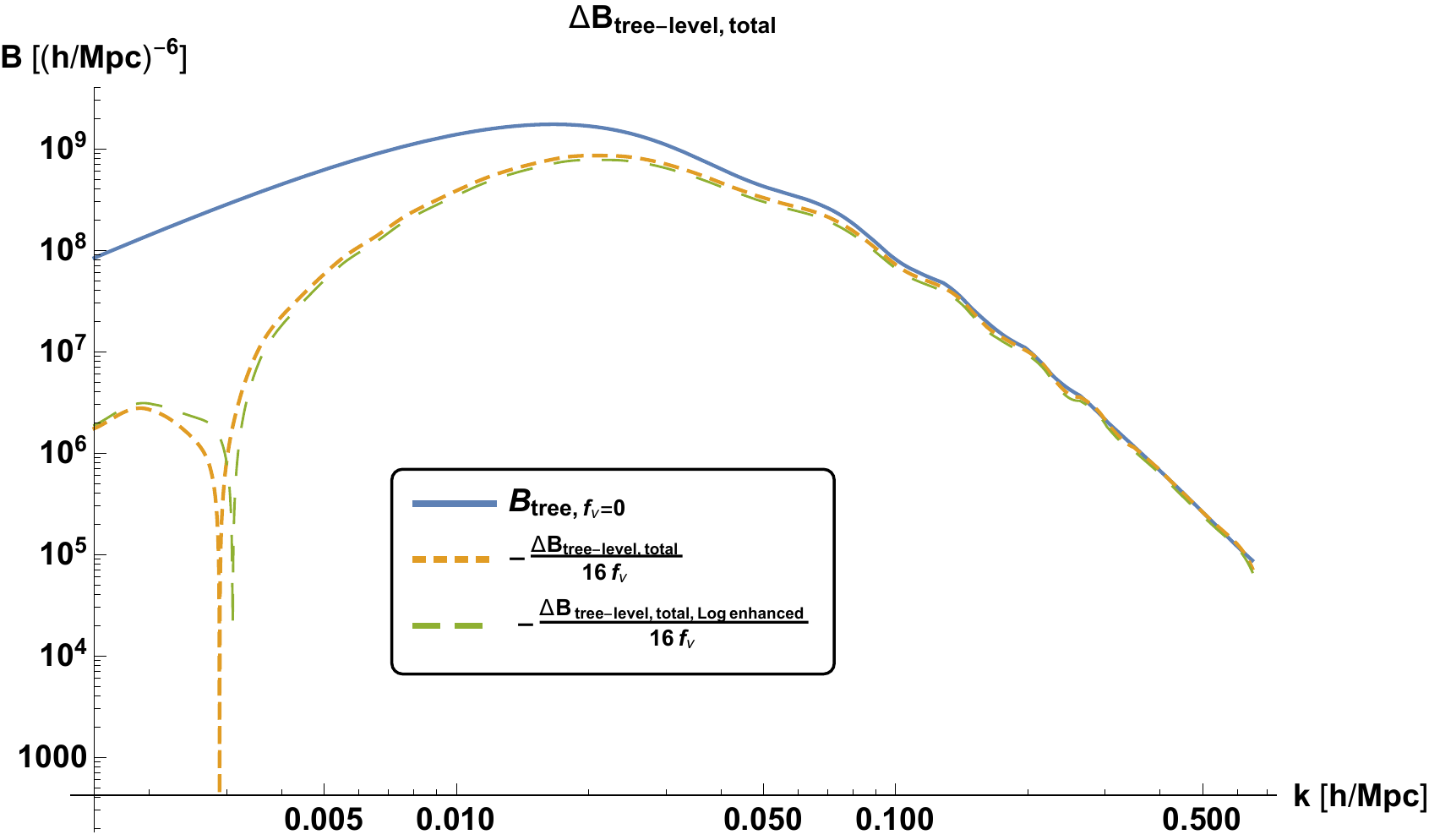}
\caption{Comparison of the absolute value of the tree-level correction to the total-matter bispectrum in the presence of massive neutrinos, $\Delta B_{\rm tree-level, total}$, which is the sum of $B_{\rm diff, dm, dm}$ and $\Delta B_{\rm dm, dm, dm}$ to the dark matter bispectrum $B_{\rm tree, f_{\nu}=0}$ for equilateral configurations. On small scales, for $k \gtrsim 0.1h \rm \ Mpc^{-1}$, $\Delta B_{\rm tree-level, total}$ approaches the dark matter tree-level bispectrum by a factor of $-16f_{\nu}$. At large scales, for $k < k_{\rm fs}$, $\Delta B_{\rm tree-level, total}$ is suppressed by up to two orders of magnitude because $\td_{\rm diff}$ is higly supressed at these scales.\label{fig:B_total}}
\end{figure}

Figure \ref{fig:B_total} shows the sum of the contributions $B_{\rm diff, dm, dm}$ and $\Delta B_{\rm dm, dm, dm}$ to the total-matter tree-level bispectrum, called $\Delta B_{\rm tree-level, total}$, explicitly given by: 
\begin{align}
\Delta B_{\rm tree-level, total} (k_1, k_2, k_3,a) &= \Delta B_{\rm dm, dm, dm}(k_1, k_2, k_3,a) + B_{\rm diff, dm, dm}(k_1, k_2, k_3,a) \\
& \quad + B_{\rm dm, diff, dm}(k_1, k_2, k_3,a)+B_{\rm dm, dm, diff}(k_1, k_2, k_3,a) \nn \ .
\end{align}
The sum of the contributions approaches $-16 f_{\nu}$ times the dark matter bispectrum for large wavenumbers, $k \gtrsim 0.1\hinvMpc$, on equilateral configurations. As explained previously, the logarithmic enhancement comes from the effect of $\td_{\rm diff}$ on $\td_{\rm dm}$ at linear level. The overall shape of $\Delta B_{\rm tree-level, total}$ is very similar to the previous figures. The green dashed line indicates the contribution purely from the logarithmically enhanced diagrams which account for approximately 90\% of the final result\footnote{We point out that, although the contribution of the back-reaction of neutrinos on dark matter accounts for 90\% of the effect, it is a less accurate part of the answer for the one loop power spectrum \cite{Senatore:massive_neutrinos}. More importantly, it is not easy to imagine that a perturbative expansion exists where one can recover the effect of the non log-enhanced diagrams by perturbing around the solution with just the log-enhanced diagrams. Therefore, the full Boltzmann equation should be solved.}.

\begin{figure}[t!]
\centering
\subfigure[$k_1 = 0.05 \hinvMpc$\label{fig:Bispectrum_3D_relative_k1_005}]{\includegraphics[width = 0.47\textwidth]{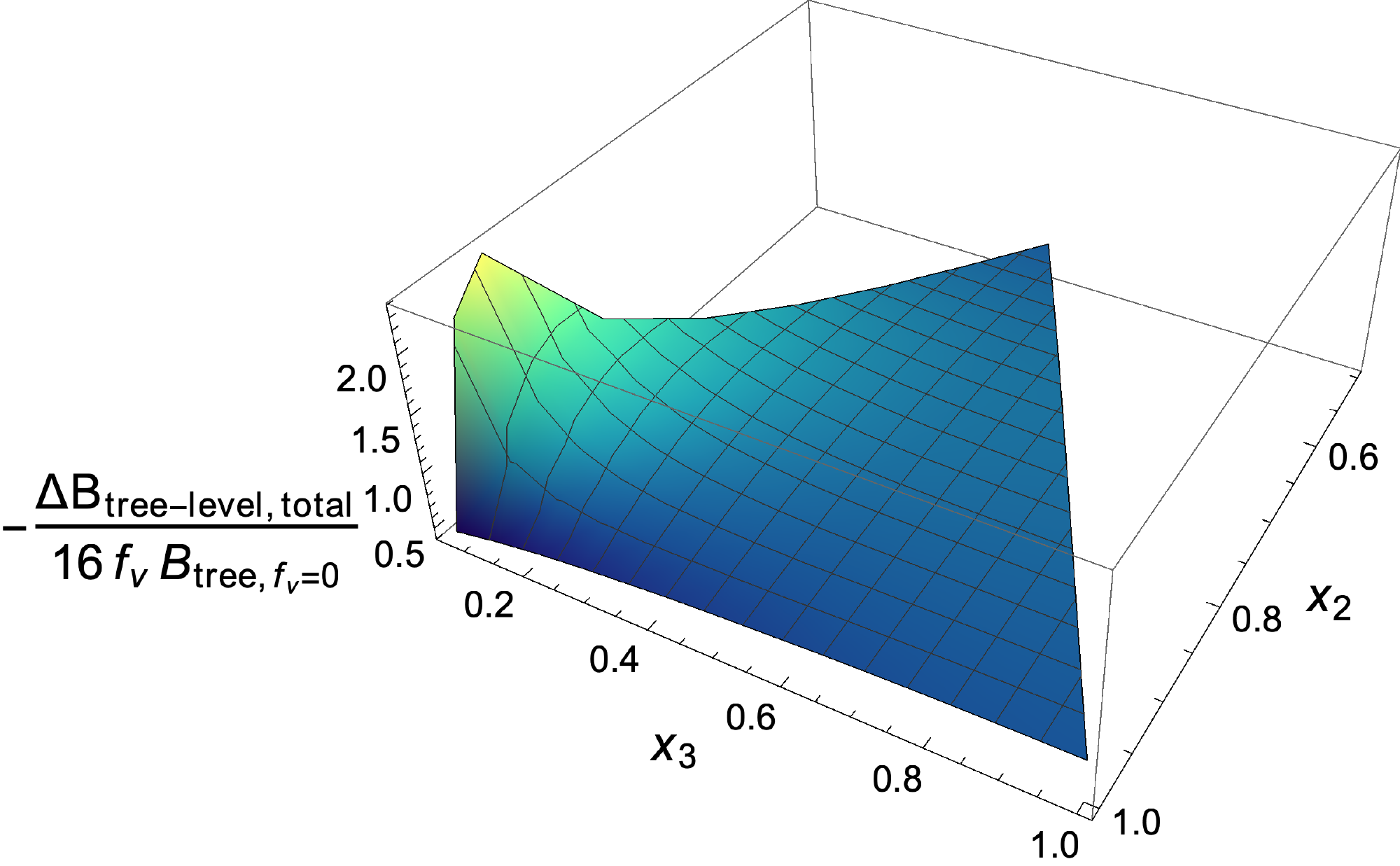}}
\subfigure[$k_1 = 0.2 \hinvMpc$\label{fig:Bispectrum_3D_relative_k1_02}]{\includegraphics[width = 0.47\textwidth]{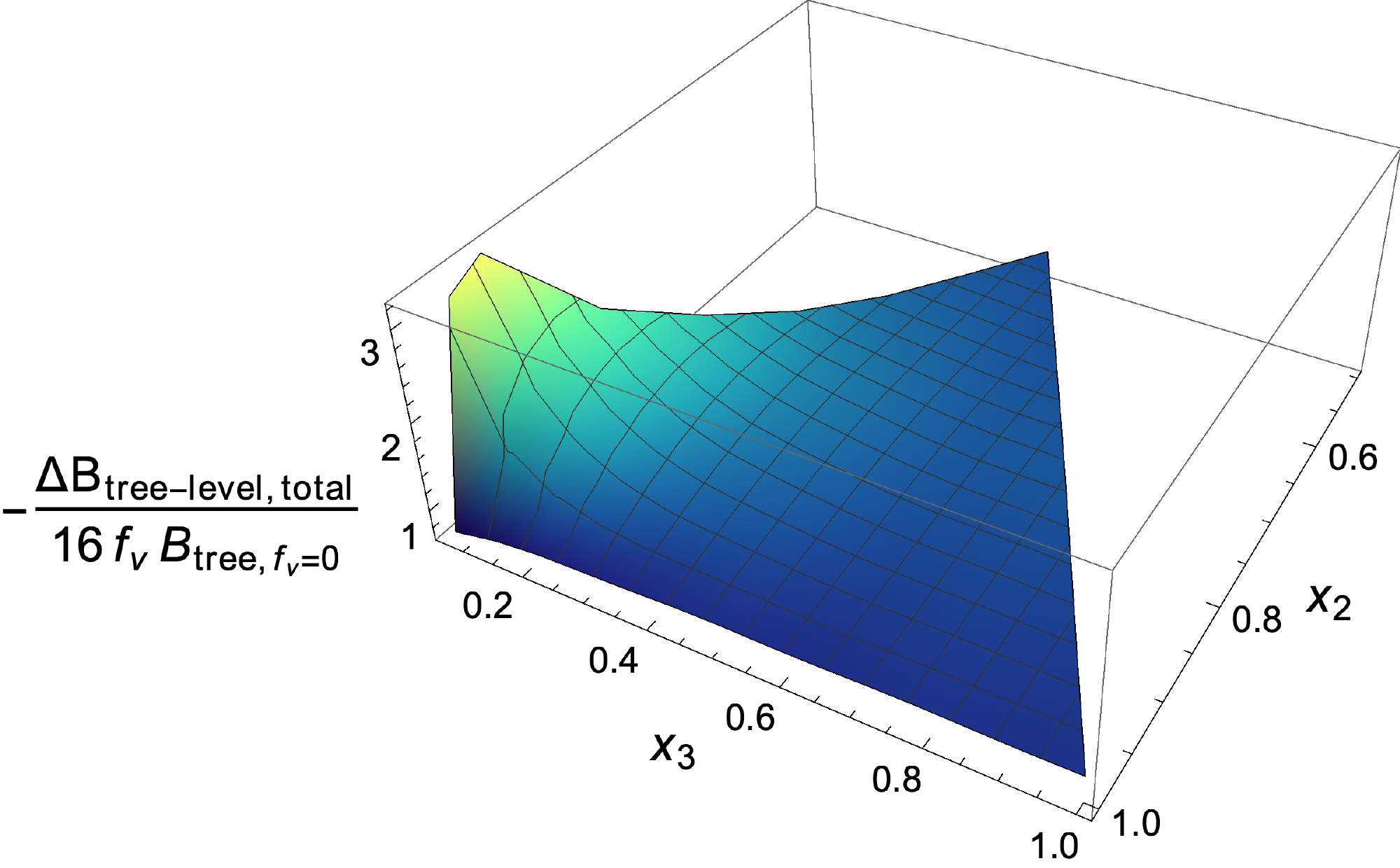}}
\caption{Ratio of the tree-level correction to the total-matter bispectrum in the presence of massive neutrinos, $\Delta B_{\rm tree-level, total}$, and the tree-level dark matter bispectrum: $ - \frac{ \Delta B_{\rm tree-level, total}}{16f_{\nu}B_{\rm tree, f_{\nu}=0}}$. Figure \ref{fig:Bispectrum_3D_relative_k1_005} is plotted with a constant wavenumber $k_1 = 0.05 \hinvMpc$ and figure \ref{fig:Bispectrum_3D_relative_k1_02} with constant $k_1 = 0.2 \hinvMpc$. \label{fig:Bispectrum_relative}}
\end{figure}

In figure \ref{fig:Bispectrum_relative} the ratio of the correction induced by the massive neutrinos to the total-matter tree-level bispectrum, $\Delta B_{\rm tree-level, total}$, and $B_{\rm tree, f_{\nu}=0}$ is shown and extends figure \ref{fig:B_total} to two dimensions. Therefore, we plot, for two different constant $k_1$ values, the ratio $-\frac{1}{16f_{\nu}}\frac{ \Delta B_{\rm tree-level, total}}{B_{\rm tree, f_{\nu}=0}}$.

In figure \ref{fig:Bispectrum_3D_relative_k1_005} the ratio $-\frac{1}{16f_{\nu}}\frac{ \Delta B_{\rm tree-level, total}}{B_{\rm tree, f_{\nu}=0}}$ for a constant wavenumber $k_1 = 0.05 \hinvMpc$ is shown. As we are looking at longer wavelengths, the agreement between the total-matter tree-level bispectrum and $16f_{\nu}$ times the dark matter tree-level bispectrum is about 70\% for the equilateral and the flat triangular configurations, as anticipated from the discussion of figure \ref{fig:B_total}. Additionally, we obtain an enhancement by a factor of two times $-16f_{\nu}$ for flat triangles in the squeezed limit ($x_2 =0.9$ and $x_3 = 0.1$) which we will discuss in more depth later.  

In figure \ref{fig:Bispectrum_3D_relative_k1_02} the same ratio for a constant wavenumber $k_1 = 0.2 \hinvMpc$ is shown. The agreement for shorter wavelengths, in particular, for the equilateral and flat configurations are, as expected from the discussion of figure \ref{fig:B_total}, of order of a few percent. For the flat triangular configuration in the squeezed limit ($x_2 = 0.9$ and $x_3 = 0.1$) we obtain, analogously to the previous case, an enhancement by a factor of 2.5 times $-16f_{\nu}$.

\begin{figure}[t!]
\centering
\subfigure[Comparison of $\Delta B_{\rm tree-level, total}$ and $B_{\rm tree, f_{\nu}=0}$. \label{fig:B_squeezed_limit}]{\includegraphics[width = 0.49\textwidth]{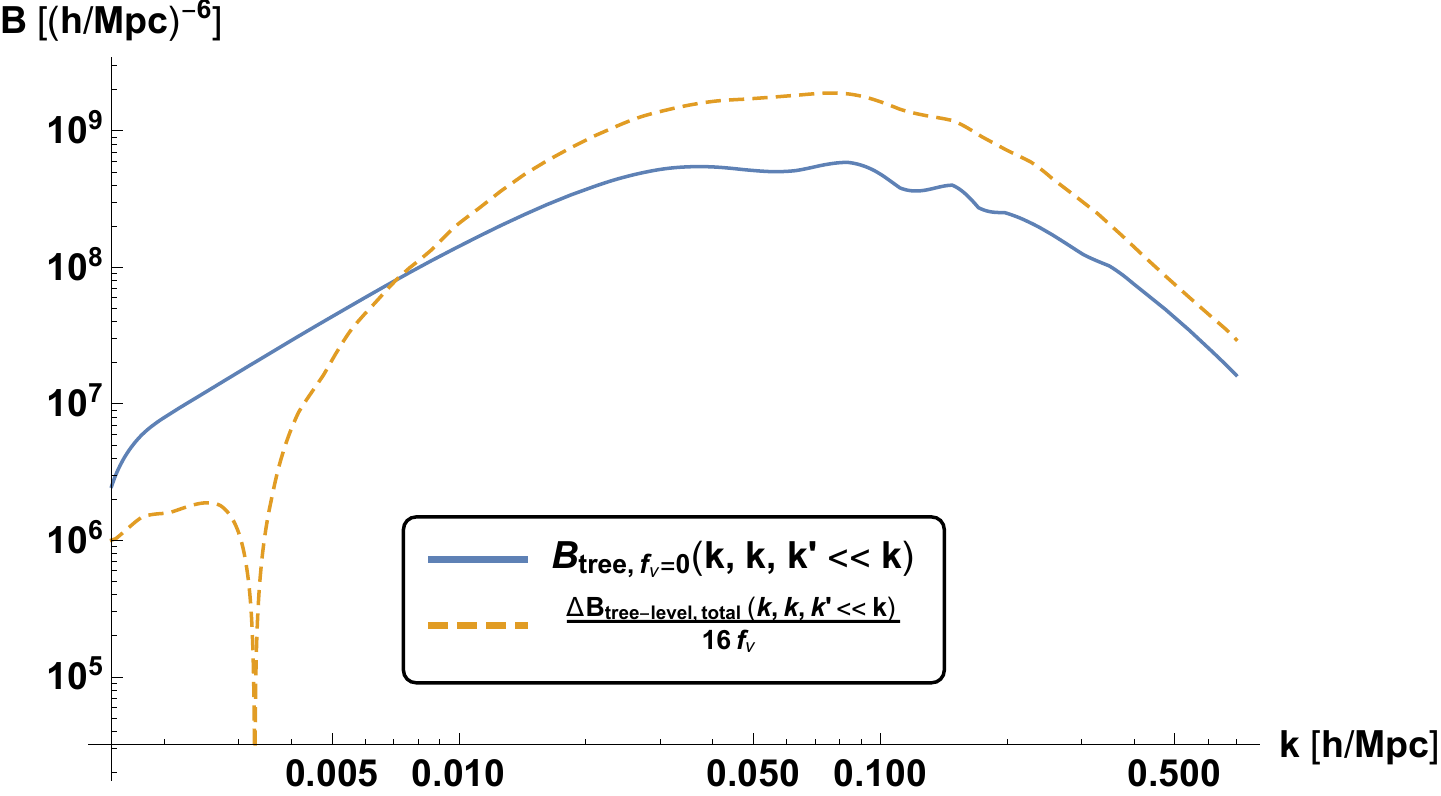}}
\subfigure[Ratio of $\Delta B_{\rm tree-level, total}$ and $B_{\rm tree, f_{\nu}=0}$. \label{fig:B_squeezed_limit_ratio}]{\includegraphics[width = 0.49\textwidth]{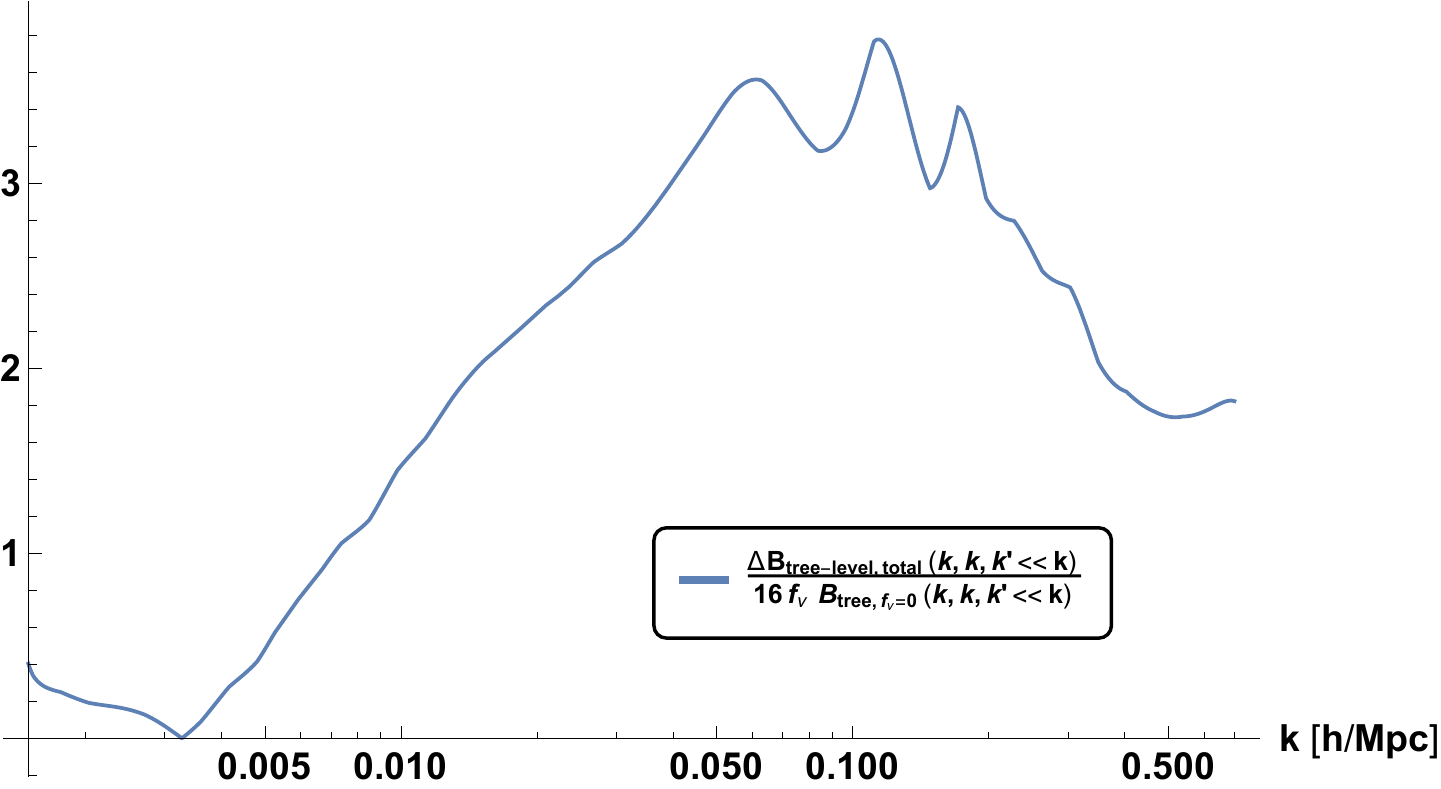}}
\caption{Comparison of the absolute value of $\Delta B_{\rm tree-level, total}$ and $B_{\rm tree, f_{\nu}=0}$ in the flat triangular configuration in the squeezed limit ($x_2 =0.9$ and $x_3=0.1$). We observe an enhancement of $\Delta B_{\rm tree-level, total}$ with respect to $B_{\rm tree, f_{\nu}=0}$ for $k>k_{\rm fs}$ in figure \ref{fig:B_squeezed_limit}. In particular, we plot the absolute value of the ratio $\frac{1}{16f_{\nu}}\frac{\Delta B_{\rm tree-level, total}}{B_{\rm tree, f_{\nu}=0}}$ in figure \ref{fig:B_squeezed_limit_ratio} to quantify this enhancement more precisely. We observe an enhancement by a factor of approximately two, for $k_1 = 0.05 \hinvMpc$, and 2.5, for $k_1 = 0.2 \hinvMpc$, times $16f_{\nu}\cdot B_{\rm tree, f_{\nu}=0}$ in the chosen triangular configuration. \label{fig:Bispectrum_relative_squeezed}}
\end{figure}

In the following we want to develop some intuition on why the flat triangular configuration in the squeezed limit for high wavenumbers is non-zero, and also why for low wavenumbers $\Delta B_{\rm tree-level,total}$ is non-zero. Therefore, let us consider a triangle with two short modes at linear order, $\td_s^{(1)}$, and one long mode at second order, $\td_l^{(2)}$. In particular, we want to compute $\langle \td_s^{(1)} \td_s^{(1)} \td_l^{(2)} \rangle$: naively, one would expect this configuration, for wavenumbers longer than the free streaming length, to be zero, but, as visible in figures \ref{fig:Bispectrum_relative} and \ref{fig:Bispectrum_relative_squeezed}, not only we get non-zero, but actually we obtain an enhancement of $\Delta B_{\rm tree-level, total}$ with respect to the dark matter bispectrum in the flat triangular configuration in the squeezed limit ($x_2 = 0.9$ and $x_3 = 0.1$). This comes from the fact that $\td_l^{(2)}$ is sourced by two short modes: at linear level, $\td_{l, \rm diff}^{(1)}$ is vanishingly small for long wavelengths, as discussed in figure \ref{fig:RatioPowerspec}, while, at non-linear level, $\td_{l, \rm diff}^{(2)}$ has a non-vanishing contribution thanks to the contribution of the product of two short modes. Therefore, we plot the absolute value of $\Delta B_{\rm tree-level, total}$ in figure \ref{fig:B_squeezed_limit} and the dark matter bispectrum in the flat triangular configuration in the squeezed limit with $x_2=0.9$ and $x_3=0.1$ and observe an enhancement for $k>k_{\rm fs}$ which comes from the log-enhanced diagrams. In figure \ref{fig:B_squeezed_limit_ratio} we plot the absolute value of the ratio $\frac{1}{16f_{\nu}}\frac{\Delta B_{\rm tree-level, total}}{B_{\rm tree, f_{\nu}=0}}$ in the same limit and observe an enhancement by a factor of approximately two, for $k_1 = 0.05 \hinvMpc$, and approximately 2.5, for $k_1 = 0.2 \hinvMpc$, times $16f_{\nu}$ the dark matter tree-level bispectrum. The above explanation enlightens the fact that the contributions in the low-$k$ regime and in the flat-squeezed configuration in the high-$k$ regime are non-zero, but does not yet explain why it is actually boosted by a factor of approximately two in the squeezed-flat limit, and not so in the completely squeezed limit. Due to different contributions from each individual permutation in the flat squeezed limit the suppression in this calculation is boosted by a factor of two as explained in the following:

First, for the tree-level dark matter bispectrum in the flat-squeezed configuration, we notice that there is a cancellation by a factor of approximately three when summing over all possible permutations. The main contribution comes from the configuration $B_{\rm 121, dm}$ and the main cancellation from the configuration $B_{\rm 211, dm}$. This cancellation is not present in the completely squeezed limit.

Instead, for the neutrino case, some permutations or contractions give a negligible contribution because of the fact that some diagrams would require $\td_{\rm diff}$ at long wavelengths. This lack of contribution from some diagrams spoils the cancellation that we find in the dark matter case, explaining the enhancement by a factor of approximately two in the flat squeezed limit and the reduced supresion in the completely squeezed limit (where the dark matter bispectrum does not have a cancellation). When only looking at the very squeezed limit ($x_2=1$ and $x_3=0$), we observe a reduced supression, corresponding to $\sim 12 f_{\nu}$. These results are in approximate quantitative agreement with \cite{Ruggeri:2017dda}, the physical interpretation is quite similar for the completely squeezed limit, but it is somewhat more elaborate for the flat squeezed limit.

\subsection*{Cross Checks \label{sec:cross_checks}}
In order to check if the diagrams are correct, a series of powerful checks have been successfully performed. Although the integrals are prone to errors, the cross checks give a high level of confidence on the correctness of the results. 

First, for $B_{\mathrm{diff, dm, dm}}$ we can set the neutrino masses, $m_{\nu}$, arbitrarily high which equals to setting $\tilde{f}_{\nu}^{[0]} = 1$ and $\td_{\rm diff} = 0$. Substituting in our computations $\tilde{f}^{[0]}$ with $\tilde{f}_{\nu}^{[0]}$ reproduces the tree-level bispectrum $B_{\rm tree, f_{\nu}=0}$ for dark matter to percent level. 

Second, we check the two log-enhanced diagrams: $\Delta B^{ \lbrack 1 , 1 \rbrack  \lvert (1) \lvert (2)}_{\mathrm{dm, dm, dm}}$ and $\Delta B^{ \lbrack 1 , 1 \rbrack (1) \lvert (1) \lvert (1)}_{\mathrm{dm, dm, dm}}$. We replace $\delta_{\rm diff}$ with $\delta_{\rm dm}$ and obtain with percent level accuracy the tree-level bispectrum for dark matter for each log-enhanced diagram\footnote{This is a non-trivial check because we have to account for the $k$-dependence of the growth factor and the combinatorics of the different diagrams.}. Moreover, we check that when we evaluate the diagrams in a cosmology with massless neutrinos, that the log-enhanced diagrams, which are given by the difference of two diagrams, yield, as expected, zero.

Furthermore, we have two different possibilities to compute the log-enhanced diagrams: on one hand, we can calculate, as described earlier, the log-enhanced diagrams as a sum of $\Delta B^{ \lbrack 1 , 1 \rbrack  \lvert (1) \lvert (2)}_{\mathrm{dm, dm, dm}}$ and $\Delta B^{ \lbrack 1 , 1 \rbrack (1) \lvert (1) \lvert (1)}_{\mathrm{dm, dm, dm}}$. On the other hand, we obtain the same result, when taking the difference of the total-matter tree-level bispectrum with massive neutrinos\footnote{We computed $B_{\rm tree}^{f_{\nu}}$ with the $k$-dependent Green's functions and with $P_{\rm lin}^{f_{\nu}}$ in a given cosmology taken from the relativistic Boltzmann solver \textsc{camb}.}, $B_{\rm tree}^{f_{\nu}}$, and the one in the absence of neutrinos, $B_{\rm tree}^{f_{\nu}=0}$. This further increases our level of confidence about our result.

We did not check the non log-enhanced integral diagrams that are contributing to $\Delta B_{\mathrm{dm, dm, dm}}$, but, given that the structure only differs from $B_{\mathrm{diff, dm, dm}}$ by an additional integration over time and that $B_{\mathrm{diff, dm, dm}}$ has been checked, it is unlikely that these diagrams are erroneous.

The computations passed all the above mentioned checks which makes us confident that the result is correct. The computations are based on the publicly available \textsc{Mathematica} notebook for the one-loop power spectrum computed in \cite{Senatore:massive_neutrinos} and are published in the EFTofLSS code repository\footnote{\url{http://web.stanford.edu/~senatore/}}.

\section{Conclusion and Summary \label{sec:Conclusion}}
The EFTofLSS continuously improves our theoretical understanding of the large-scale structure of the universe. In the present paper we further developed an analytic formalism to describe dark matter clustering in the presence of massive neutrinos by computing the leading order three-point correlation function, i.e. the tree-level bispectrum in Fourier space, in the mildly non-linear regime. We can describe dark matter as en effective fluid-like system for $k\lesssim k_{\rm NL}$ since its free streaming length is about the non-linear scale. For neutrinos, instead, we split the neutrino population into two families when perturbatively solving the neutrino Boltzmann equation based on their different expansion parameters: fast and slow neutrinos. Fast neutrinos have a free streaming wavenumber smaller than the non-linear scale, $k_{\rm fs} \lesssim k_{\rm NL}$, for which each interaction corresponds to a small correction. On the contrary, slow neutrinos have a free streaming wavenumber larger than the non-linear scale, $k_{\rm fs} \gtrsim k_{\rm NL}$. When considering interactions with gravitational fields with wavenumbers shorter than the non-linear scale, the expansion parameter for slow neutrinos is larger than order one and the perturbative solution does not hold anymore. However, for slow-neutrino perturbations with wavenumbers longer than the non-linear scale, one can consider slow neutrinos as an effective fluid-like system similar to dark matter, and, therefore, the effect of short distance physics can be corrected by the inclusion of a speed-of-sound like counterterm ($c_s$). In our tree-level calculations no counterterms are required as no UV sensitive terms arise when performing the integrals, contrary to the case of the one-loop power spectrum computed in \cite{Senatore:massive_neutrinos}. We solve this coupled set of equations perturbatively by expanding in the smallness of their relative energy density, $f_{\nu}$, and in the ratio of the wavenumber of interest over the wavenumber associated to the non-linear scale, $k/k_{\rm NL}$. 

For our calculations we assumed massive neutrinos with a total mass of $\sum m_{\nu_i} = 0.23 \ {\rm eV}$. As the neutrinos are massive, they become non-relativistic at late times and cluster which, in turn, affects the total-matter clustering. Through the gravitational force the neutrino clustering affects the dynamics of dark matter. This is visible in a suppression of the linear power spectrum for wavenumbers larger than the free streaming wavenumber, $k \gtrsim k_{\rm fs}$, by a factor of $\sim 8f_{\nu}$. We computed, at linear order in $f_{\nu}$, the correction induced by massive neutrinos to the total-matter tree-level bispectrum, called $\Delta B_{\rm tree-level, total}$. Therefore we computed the following corrections: first, the cross-correlation function between the density fields $\td_{\rm diff}=\td_{\nu} - \td_{\rm dm}$ and $\td_{\rm dm}$ which yields $B_{\rm diff, dm, dm}$. The latter approaches the tree-level dark matter bispectrum up to a factor of $-f_{\nu}$ for $k>k_{\rm fs}$. For low wavenumbers, $k< k_{\rm fs}$, the correction was suppressed by up to two orders of magnitude. Second, the correction of neutrinos to the matter bispectrum due to the different evolution of dark matter in the presence of massive neutrinos, called $\Delta B_{\rm dm,dm,dm}$, was computed. It is much more significant and approaches $-15f_{\nu}$ times the tree-level dark matter bispectrum for short wavelengths. Similarly to the previous case, for long wavelengths the correction is suppressed by up to two orders of magnitude due to $\td_{\rm diff}\simeq 0$. The total correction to the tree-level bispectrum amounts to summing all the computed diagrams yielding approximately $-16f_{\nu}$ times the tree-level dark matter bispectrum and is denoted $\Delta B_{\rm tree-level, total}$ for $k>k_{\rm fs}$. For lower wavenumbers the correction is smaller by up to two orders of magnitude. These results hold for equilateral and flat triangular configurations. For the flat triangular configuration in the squeezed limit, instead, we observe an enhancement by a factor of approximately two times $-16f_{\nu}B_{\rm tree, f_{\nu}=0}$. This further enhancement by a factor of two is due to the fact that the tree-level dark matter bispectrum experiences a cancellation in this configuration, which is absent for the contribution due to neutrinos.

The main contribution to the enhancement, as discussed earlier, with respect to the tree-level dark matter bispectrum stems from the modification of the Poisson equation at linear level: the effect of the linear $\td_{\rm diff}$ on the linear $\td_{\rm dm}$ is logarithmically enhanced in length of time from matter-radiation equality until present time.
The logarithmically enhanced diagrams account for approximately 90\% of the final result.

The further development of the analytic description of dark matter clustering opens up many novel research possibilities such as the computation of higher-order correlation functions (i.e. the one-loop bispectrum including its counterterms). Since the upcoming cosmological LSS surveys will potentially become accurate enough to measure the effect of massive neutrinos, it would be interesting to analyze the impact of massive neutrinos on observable quantities, such as the lensing potential or redshift-space galaxy clustering.

\section*{Acknowledgments}
We thank Ethan Nadler for his help with \textsc{camb}. R.B. thanks Alexandre Refregier and Adam Amara for guidance and support throughout the project. R.B. is grateful for the kind hospitality at the Stanford Institute for Theoretical Physics (SITP) at Stanford University and at the Kavli Institute for Particle Astrophysics and Cosmology (KIPAC) at SLAC National Accelerator Laboratory. L.S. is partially supported by NSF award 1720397.

\appendix
\section{Renormalization of Neutrino Correlation Functions \label{sec:Renormalization_Correlation}}
In a perturbative expansion with $\r_{\rm dm}\ll 1$ and  $k \ll k_{\rm NL}$, we can write for a 
two-point correlation function \cite{Senatore:massive_neutrinos}:
\begin{equation}\label{eq:two_point_renorm}
\lbrack\delta(\vec x_1,\tau_1)\,\delta(\vec x_2,\tau_2)\rbrack_{\rm R}=\lbrack\delta(\vec x_1,\tau_1)\rbrack_{\rm R}\,\lbrack\delta(\vec x_2,\tau_2)\rbrack_{\rm R}
+\lbrack\delta(\vec x_1,\tau_1)\,\delta(\vec x_2,\tau_2)\rbrack_{\rm C}\ ,
\end{equation}
and extend this to the case of a 
three-point correlation function, which is given by \cite{Senatore:massive_neutrinos}:
\begin{align}\label{eq:three_point_renorm}
&\lbrack\delta(\vec x_1,\tau_1)\,\delta(\vec x_2,\tau_2)\delta(\vec x_3,\tau_3)\rbrack_{\rm R}\\ \nn
&\qquad =\left\lbrace \lbrack\delta(\vec x_1,\tau_1)\,\delta(\vec x_2,\tau_2)\rbrack_{\rm R}\,\lbrack\delta(\vec x_3,\tau_3)\rbrack_{\rm R}+{\rm permutations}\right\rbrace
+\lbrack\delta(\vec x_1,\tau_1)\,\delta(\vec x_2,\tau_2)\,\delta(\vec x_3,\tau_3)\rbrack_{\rm C}\ ,
\end{align}
where we omitted for clarity the subscript for dark matter. For $k \ll k_{\rm NL}$ the product of fields can be redefined as a renormalized product $[\dots]_{\rm R}$. The addition of suitable counterterms $[\dots]_{\rm C}$ cancels the contributions from short-distance physics. The first term on the right hand side of equation \eqref{eq:three_point_renorm} represents the renormalized $\td(\vx, \tau)$ coming from perturbation theory which includes the counterterms from the stress tensor of the effective equations that describe the long-wavelength dark matter dynamics. The term $\lbrack\dots\rbrack_{\rm C}$ stands for the counterterms that renormalize the products of two density fields at the same location, which is given by:
\begin{align}\label{eq:counter2}
&\lbrack\delta(\vec{x}_1,\tau_1)\delta(\vec{x}_2,\tau_2)\rbrack_{\rm C}=\\
& \qquad \int^{{\rm Max}(\tau_1,\tau_2)} d\tau' \left\lbrack C^{(2)}_1(\tau_1,\tau_2,\tau')+\epsilon^{(2)}_{\rm stoch, 1}(\tau_1,\tau_2,x_{\rm fl}(\vec{x}_1,\tau_1,\tau'),\tau') \right. \nn \\ 
&\left.\qquad +\ \d ^i\d ^j\Phi (x_{\rm fl}(\vec{x}_1, \tau,\tau'),\tau')\left(C^{(2)}_2(\tau,\tau') \delta^{ij}+C^{(2)}_3(\tau,\tau')\frac{\d_i\d_j}{\d^2}\right)+\right. \nn\\ 
&\left. \qquad \quad +\ \epsilon_{{\rm stoch, 2}}^{(2), ij}(\tau_1,\tau_2,x_{\rm fl}(\vec{x}_1,\tau_1,\tau'),\tau') \frac{\d_i\d_j}{\d^2}+\ldots\right\rbrack\ \    \frac{1}{k_{\rm NL}^3}\delta^{(3)}_D\left(x_{\rm fl}(\vec{x}_1,\tau_1,\tau')-x_{\rm fl}(\vec{x}_2,\tau_2,\tau')\right)\ , \nn
\end{align}
The counterterms that renormalize the product of three density fields at the same location are given by:
\begin{align}\label{eq:counter3}
&\lbrack\delta(\vec{x}_1,\tau_1)\,\delta(\vec{x}_2,\tau_2)\,\delta(\vec{x}_3,\tau_3)\rbrack_{\rm C}=\\ \nn
&\qquad \int^{{\rm Max}(\tau_1,\tau_2,\tau_3)} d\tau'\;  \left[C^{(3)}_1(\tau_1,\tau_2,\tau_3,\tau')+\epsilon^{(3)}_{\rm stoch, 1}(\tau_1,\tau_2,\tau_3,x_{\rm fl}(\vec{x}_1,\tau_1,\tau'),\tau')\right. \\ \nn
&\left.\qquad +\left(\d^i\d^j\Phi(x_{\rm fl}(\vec{x}_1, \tau,\tau'),\tau')\left(C^{(3)}_2(\tau,\tau') \delta^{ij}+C^{(3)}_3(\tau,\tau')\frac{\d_{x_1^i}\d_{x_1^j}}{\d_{\vec{x}_1}^2}\right)+{\rm perm.} \right)+\ldots\right\rbrack\\ \nn
&\qquad\qquad   \frac{1}{k_{\rm NL}^6}\cdot\delta^{(3)}_D\left(x_{\rm fl}(\vec{x}_1,\tau_1,\tau')-x_{\rm fl}(\vec{x}_2,\tau_2,\tau'_{2})\right)\delta^{(3)}_D\left(x_{\rm fl}(\vec{x}_1,\tau_1,\tau')-x_{\rm fl}(\vec{x}_3,\tau_3,\tau')\right)\ , \nn
\end{align}
with the terms hierarchically organized in an expansion of the smallness of the $\td$ fields and the ratio of the wavenumber of interest over the non-linear scale. The counterterms for the products of fields in eq. \eqref{eq:counter3} have support only when, first, the three fields are at the same location, and, second, as the theory is non-local in time, when they have support on the past-light cone of the resulting point \cite{Carrasco:2013mua, Senatore:massive_neutrinos, Senatore:2014eva}. The terms $\epsilon^{(3)}_{\rm stoch, 1}(\tau_1,\tau_2,\tau_3,x_{\rm fl}(\vec{x}_1,\tau_1,\tau'),\tau')$ and $\epsilon^{(3)}_{\rm stoch, 1}(\tau_1,\tau_2,\tau_3,x_{\rm fl}(\vec x_1,\tau_1,\tau'),\tau')$ represent a stochastic term with zero expectation value and Poisson statistics. It accounts for the difference in a given realization between the expectation value and the actual value. The tidal tensor field of gravity $\d_i\d_j\Phi$ incorporates the effect of the deformation of the fields. The term $C^{(3)}_1$ corrects for the expectation value of the correlation function and the terms $C^{(3)}_i(\tau,\tau')$ with $i = 1,2,3$ are the coefficients of the counterterms. In particular, when computing higher order corrections it is important to understand the size of the mistake. At tree-level no counterterms are required but when computing higher order corrections such as the one-loop bispectrum, they will become crucial \cite{Senatore:massive_neutrinos, Angulo:2014tfa, Baldauf:Bispectrum}.  

\section{Dark Matter Growth Factor in the Presence of Massive Neutrinos \label{sec:growth_factor_DM_appendix}}

We derive the logarithmic correction to the growth factor of dark matter in the presence of massive neutrinos. Therefore, we take the linearized equations for a cosmic fluid:
\begin{align}
\d_t\td + \frac{1}{a}\vec{\nabla} \cdot \vec{v} &= 0 \ , \label{eq:A_continuity}\\
\d_t \vec{v} + \frac{\dot{a}}{a} \vec{v}  &= - \frac{1}{a}\vec{\nabla}\phi \ , \label{eq:A_Euler}\\
\nabla^2 \phi &= 4\pi G a^2 \td \r \ . \label{eq:A_Poisson}
\end{align}
Solving them allows us to derive the time evolution of the density perturbations. Taking the divergence of the Euler equation \eqref{eq:A_Euler} and combining the result with the continuity equation \eqref{eq:A_continuity} and the Poisson equation \eqref{eq:A_Poisson}, we obtain a second order partial differential equation for the linearized equation for the growth of density perturbations $\td(\vec{x},t)$ in a pure dark matter universe, given by: 
\begin{equation} \label{eq:lin_growth_pertub}
\ddot{\td}_{\rm dm} + 2\frac{\dot{a}}{a}\dot{\td}_{\rm dm} = 4\pi G a^2\td\r \ ,
\end{equation}
where $\td \r = \bar{\r} \td_{\rm dm}$ is the total density perturbation. Since we are trying to solve analytically a second order partial differential equation, we already know two things about $\td$: first, the density perturbation has two solutions $\td(\vx, t) =\td_1(\vx, t) +\td_2(\vx, t)$. Second, as the equation only includes differentials in time, its time evolution must be independent of location $\vx$. Hence, we can separate the solution into a spatial, $\td_i(\vx)$, and a temporal, $D_i(t)$, part, given by $\td_i(\vx, t) = D_i(t)\td_i(\vx)$ with $i=1,2$. The general solution is given by a superposition of the temporal and spatial solution:
\begin{equation}
\td(\vx, t) = D_1(t)\td_1(\vx) + D_2(t)\td_2(\vx) \ ,
\end{equation}
where $D_i(t)$, with $i=1,2$, represents the time dependence of our solution. $\td_i(\vx)$, with $i=1,2$, represents the corresponding spatial configuration of the cosmic primordial matter distribution. 

We are going to derive the logarithmically enhanced correction to the growth factor by using a simple toy model. Assuming that the universe is filled with two non-relativistic populations: one clustering, one non-clustering. In particular, consider a flat universe only filled with non-relativistic matter; an Einstein-de Sitter (EdS) universe, with $\Omega_{\rm m} = 1$, Hubble parameter $H(t) = H_0$ and a scale factor that behaves as $a(t) \propto t^{\frac{2}{3}}$. For the expansion rate we get $\frac{3}{2}\left( \frac{\dot{a}}{a}\right)^2 = 4 \pi G a^2 \bar{\r}$. Hence, for $\frac{\dot{a}}{a} = \frac{2}{3t}$. Plugging these parameters for our toy model into equation \eqref{eq:lin_growth_pertub}, yields:
\begin{equation} \label{eq:lin_grow_perturb_EdS}
\ddot{\td}_{\rm dm} + \frac{4}{3t}\dot{\td}_{\rm dm} - \frac{2}{3t^2} (1-f_{\nu}) \td_{\rm dm} = 0\ ,
\end{equation}
where we assume that a fraction $f_{\nu}$ of the dark matter does not cluster in the presence of massive neutrinos. From figure \ref{fig:RatioPowerspec}, we know that for $k < k_{\rm fs}$ the neutrino perturbations behave as CDM perturbations at linear order and that they vanish for $k > k_{\rm fs}$. Therefore, equation \eqref{eq:lin_grow_perturb_EdS} represents the correct evolution of dark matter for wavenumbers higher than the free streaming length, which is the regime we want to explore in this appendix. For the linearized equation for the growth of density perturbations in an EdS universe we can assume a simple power-law for the density perturbations $\td_i(t) \propto t^{\alpha}$ with $i=1,2$. Plugging this Ansatz into equation \eqref{eq:lin_grow_perturb_EdS}, yields for the growing mode solution:
\begin{equation}
\alpha = \frac{1}{6} \left( 5\sqrt{1-\frac{24}{25}f_{\nu}} -1\right) \approx \frac{2}{3} \left( 1-\frac{3}{5}f_{\nu}\right) \ ,
\end{equation}
with $f_{\nu} \simeq 10^{-2} \ll 1$. Since density perturbations only start to grow after matter-radiation equality, the correction to the evolution equations, as described by equation \eqref{eq:lin_grow_perturb_EdS}, starts at $a_{\rm eq}$ and endures until present time, giving rise to the following expression:
\begin{equation}
\frac{\td(a_0)}{\td(a_{\rm eq})} = \frac{a_0}{a_{\rm eq}}e^{-\frac{3}{5}f_{\nu}\ln\left(\frac{a_0}{a_{\rm eq}}\right)}\ .
\end{equation}
Matter-radiation equality occurred at redshift $a_{\rm eq} \approx 3500$ and gives a correction to the growth factor of $-\frac{3}{5} \ln(a_0/a_{\rm eq}) \sim 4-5$. In the present paper we Taylor expand in $f_{\nu}$ and, hence, obtain $D(t) \sim (a_0/a_{\rm eq})(1- \frac{3}{5}f_{\nu} \ln(a_0/a_{\rm eq}))$. For our estimates we use $\sim -\frac{1}{2} \ln(a_0/a_{\rm eq})\sim 4$.

\bibliography{references}

\providecommand{\href}[2]{#2}\begingroup\raggedright\begin{thebibliography}{10}

\bibitem{Baumann:2010tm}
D.~Baumann, A.~Nicolis, L.~Senatore and M.~Zaldarriaga, \emph{{Cosmological
  Non-Linearities as an Effective Fluid}},
  \href{https://doi.org/10.1088/1475-7516/2012/07/051}{\emph{JCAP} {\bfseries
  1207} (2012) 051} [\href{https://arxiv.org/abs/1004.2488}{{\ttfamily
  1004.2488}}].

\bibitem{Carrasco:2012cv}
J.~J.~M. Carrasco, M.~P. Hertzberg and L.~Senatore, \emph{{The Effective Field
  Theory of Cosmological Large Scale Structures}},
  \href{https://doi.org/10.1007/JHEP09(2012)082}{\emph{JHEP} {\bfseries 09}
  (2012) 082} [\href{https://arxiv.org/abs/1206.2926}{{\ttfamily 1206.2926}}].

\bibitem{Porto:2013qua}
R.~A. Porto, L.~Senatore and M.~Zaldarriaga, \emph{{The Lagrangian-space
  Effective Field Theory of Large Scale Structures}},
  \href{https://doi.org/10.1088/1475-7516/2014/05/022}{\emph{JCAP} {\bfseries
  1405} (2014) 022} [\href{https://arxiv.org/abs/1311.2168}{{\ttfamily
  1311.2168}}].

\bibitem{Carrasco:2013sva}
J.~J.~M. Carrasco, S.~Foreman, D.~Green and L.~Senatore, \emph{{The 2-loop
  matter power spectrum and the IR-safe integrand}},
  \href{https://doi.org/10.1088/1475-7516/2014/07/056}{\emph{JCAP} {\bfseries
  1407} (2014) 056} [\href{https://arxiv.org/abs/1304.4946}{{\ttfamily
  1304.4946}}].

\bibitem{Carrasco:2013mua}
J.~J.~M. Carrasco, S.~Foreman, D.~Green and L.~Senatore, \emph{{The Effective
  Field Theory of Large Scale Structures at Two Loops}},
  \href{https://doi.org/10.1088/1475-7516/2014/07/057}{\emph{JCAP} {\bfseries
  1407} (2014) 057} [\href{https://arxiv.org/abs/1310.0464}{{\ttfamily
  1310.0464}}].

\bibitem{Carroll:2013oxa}
S.~M. Carroll, S.~Leichenauer and J.~Pollack, \emph{{Consistent effective
  theory of long-wavelength cosmological perturbations}},
  \href{https://doi.org/10.1103/PhysRevD.90.023518}{\emph{Phys. Rev.}
  {\bfseries D90} (2014) 023518}
  [\href{https://arxiv.org/abs/1310.2920}{{\ttfamily 1310.2920}}].

\bibitem{Senatore:2014via}
L.~Senatore and M.~Zaldarriaga, \emph{{The IR-resummed Effective Field Theory
  of Large Scale Structures}},
  \href{https://doi.org/10.1088/1475-7516/2015/02/013}{\emph{JCAP} {\bfseries
  1502} (2015) 013} [\href{https://arxiv.org/abs/1404.5954}{{\ttfamily
  1404.5954}}].

\bibitem{Baldauf:2015zga}
T.~Baldauf, E.~Schaan and M.~Zaldarriaga, \emph{{On the reach of perturbative
  methods for dark matter density fields}},
  \href{https://doi.org/10.1088/1475-7516/2016/03/007}{\emph{JCAP} {\bfseries
  1603} (2016) 007} [\href{https://arxiv.org/abs/1507.02255}{{\ttfamily
  1507.02255}}].

\bibitem{Foreman:2015lca}
S.~Foreman, H.~Perrier and L.~Senatore, \emph{{Precision Comparison of the
  Power Spectrum in the EFTofLSS with Simulations}},
  \href{https://doi.org/10.1088/1475-7516/2016/05/027}{\emph{JCAP} {\bfseries
  1605} (2016) 027} [\href{https://arxiv.org/abs/1507.05326}{{\ttfamily
  1507.05326}}].

\bibitem{Baldauf:2015aha}
T.~Baldauf, L.~Mercolli and M.~Zaldarriaga, \emph{{Effective field theory of
  large scale structure at two loops: The apparent scale dependence of the
  speed of sound}},
  \href{https://doi.org/10.1103/PhysRevD.92.123007}{\emph{Phys. Rev.}
  {\bfseries D92} (2015) 123007}
  [\href{https://arxiv.org/abs/1507.02256}{{\ttfamily 1507.02256}}].

\bibitem{Cataneo:2016suz}
M.~Cataneo, S.~Foreman and L.~Senatore, \emph{{Efficient exploration of
  cosmology dependence in the EFT of LSS}},
  \href{https://doi.org/10.1088/1475-7516/2017/04/026}{\emph{JCAP} {\bfseries
  1704} (2017) 026} [\href{https://arxiv.org/abs/1606.03633}{{\ttfamily
  1606.03633}}].

\bibitem{Lewandowski:2017kes}
M.~Lewandowski and L.~Senatore, \emph{{IR-safe and UV-safe integrands in the
  EFTofLSS with exact time dependence}},
  \href{https://doi.org/10.1088/1475-7516/2017/08/037}{\emph{JCAP} {\bfseries
  1708} (2017) 037} [\href{https://arxiv.org/abs/1701.07012}{{\ttfamily
  1701.07012}}].

\bibitem{Pajer:2013jj}
E.~Pajer and M.~Zaldarriaga, \emph{{On the Renormalization of the Effective
  Field Theory of Large Scale Structures}},
  \href{https://doi.org/10.1088/1475-7516/2013/08/037}{\emph{JCAP} {\bfseries
  1308} (2013) 037} [\href{https://arxiv.org/abs/1301.7182}{{\ttfamily
  1301.7182}}].

\bibitem{Abolhasani:2015mra}
A.~A. Abolhasani, M.~Mirbabayi and E.~Pajer, \emph{{Systematic Renormalization
  of the Effective Theory of Large Scale Structure}},
  \href{https://doi.org/10.1088/1475-7516/2016/05/063}{\emph{JCAP} {\bfseries
  1605} (2016) 063} [\href{https://arxiv.org/abs/1509.07886}{{\ttfamily
  1509.07886}}].

\bibitem{McQuinn:2015tva}
M.~McQuinn and M.~White, \emph{{Cosmological perturbation theory in 1+1
  dimensions}},
  \href{https://doi.org/10.1088/1475-7516/2016/01/043}{\emph{JCAP} {\bfseries
  1601} (2016) 043} [\href{https://arxiv.org/abs/1502.07389}{{\ttfamily
  1502.07389}}].

\bibitem{Senatore:2014eva}
L.~Senatore, \emph{{Bias in the Effective Field Theory of Large Scale
  Structures}},
  \href{https://doi.org/10.1088/1475-7516/2015/11/007}{\emph{JCAP} {\bfseries
  1511} (2015) 007} [\href{https://arxiv.org/abs/1406.7843}{{\ttfamily
  1406.7843}}].

\bibitem{Mercolli:2013bsa}
L.~Mercolli and E.~Pajer, \emph{{On the velocity in the Effective Field Theory
  of Large Scale Structures}},
  \href{https://doi.org/10.1088/1475-7516/2014/03/006}{\emph{JCAP} {\bfseries
  1403} (2014) 006} [\href{https://arxiv.org/abs/1307.3220}{{\ttfamily
  1307.3220}}].

\bibitem{Senatore:2014vja}
L.~Senatore and M.~Zaldarriaga, \emph{{Redshift Space Distortions in the
  Effective Field Theory of Large Scale Structures}},
  \href{https://arxiv.org/abs/1409.1225}{{\ttfamily 1409.1225}}.

\bibitem{Baldauf:2015xfa}
T.~Baldauf, M.~Mirbabayi, M.~Simonović and M.~Zaldarriaga, \emph{{Equivalence
  Principle and the Baryon Acoustic Peak}},
  \href{https://doi.org/10.1103/PhysRevD.92.043514}{\emph{Phys. Rev.}
  {\bfseries D92} (2015) 043514}
  [\href{https://arxiv.org/abs/1504.04366}{{\ttfamily 1504.04366}}].

\bibitem{Senatore:2017pbn}
L.~Senatore and G.~Trevisan, \emph{{On the IR-Resummation in the EFTofLSS}},
  \href{https://arxiv.org/abs/1710.02178}{{\ttfamily 1710.02178}}.

\bibitem{Lewandowski:2018ywf}
M.~Lewandowski and L.~Senatore, \emph{{An analytic implementation of the
  IR-resummation for the BAO peak}},
  \href{https://arxiv.org/abs/1810.11855}{{\ttfamily 1810.11855}}.

\bibitem{Lewandowski:2014rca}
M.~Lewandowski, A.~Perko and L.~Senatore, \emph{{Analytic Prediction of
  Baryonic Effects from the EFT of Large Scale Structures}},
  \href{https://doi.org/10.1088/1475-7516/2015/05/019}{\emph{JCAP} {\bfseries
  1505} (2015) 019} [\href{https://arxiv.org/abs/1412.5049}{{\ttfamily
  1412.5049}}].

\bibitem{Angulo:2014tfa}
R.~E. Angulo, S.~Foreman, M.~Schmittfull and L.~Senatore, \emph{{The One-Loop
  Matter Bispectrum in the Effective Field Theory of Large Scale Structures}},
  \href{https://doi.org/10.1088/1475-7516/2015/10/039}{\emph{JCAP} {\bfseries
  1510} (2015) 039} [\href{https://arxiv.org/abs/1406.4143}{{\ttfamily
  1406.4143}}].

\bibitem{Baldauf:2014qfa}
T.~Baldauf, L.~Mercolli, M.~Mirbabayi and E.~Pajer, \emph{{The Bispectrum in
  the Effective Field Theory of Large Scale Structure}},
  \href{https://doi.org/10.1088/1475-7516/2015/05/007}{\emph{JCAP} {\bfseries
  1505} (2015) 007} [\href{https://arxiv.org/abs/1406.4135}{{\ttfamily
  1406.4135}}].

\bibitem{Bertolini:2016bmt}
D.~Bertolini, K.~Schutz, M.~P. Solon and K.~M. Zurek, \emph{{The Trispectrum in
  the Effective Field Theory of Large Scale Structure}},
  \href{https://doi.org/10.1088/1475-7516/2016/06/052}{\emph{JCAP} {\bfseries
  1606} (2016) 052} [\href{https://arxiv.org/abs/1604.01770}{{\ttfamily
  1604.01770}}].

\bibitem{Baldauf:2015tla}
T.~Baldauf, E.~Schaan and M.~Zaldarriaga, \emph{{On the reach of perturbative
  descriptions for dark matter displacement fields}},
  \href{https://doi.org/10.1088/1475-7516/2016/03/017}{\emph{JCAP} {\bfseries
  1603} (2016) 017} [\href{https://arxiv.org/abs/1505.07098}{{\ttfamily
  1505.07098}}].

\bibitem{Foreman:2015uva}
S.~Foreman and L.~Senatore, \emph{{The EFT of Large Scale Structures at All
  Redshifts: Analytical Predictions for Lensing}},
  \href{https://doi.org/10.1088/1475-7516/2016/04/033}{\emph{JCAP} {\bfseries
  1604} (2016) 033} [\href{https://arxiv.org/abs/1503.01775}{{\ttfamily
  1503.01775}}].

\bibitem{Mirbabayi:2014zca}
M.~Mirbabayi, F.~Schmidt and M.~Zaldarriaga, \emph{{Biased Tracers and Time
  Evolution}}, \href{https://doi.org/10.1088/1475-7516/2015/07/030}{\emph{JCAP}
  {\bfseries 1507} (2015) 030}
  [\href{https://arxiv.org/abs/1412.5169}{{\ttfamily 1412.5169}}].

\bibitem{Angulo:2015eqa}
R.~Angulo, M.~Fasiello, L.~Senatore and Z.~Vlah, \emph{{On the Statistics of
  Biased Tracers in the Effective Field Theory of Large Scale Structures}},
  \href{https://doi.org/10.1088/1475-7516/2015/09/029,
  10.1088/1475-7516/2015/9/029}{\emph{JCAP} {\bfseries 1509} (2015) 029}
  [\href{https://arxiv.org/abs/1503.08826}{{\ttfamily 1503.08826}}].

\bibitem{Fujita:2016dne}
T.~Fujita, V.~Mauerhofer, L.~Senatore, Z.~Vlah and R.~Angulo, \emph{{Very
  Massive Tracers and Higher Derivative Biases}},
  \href{https://arxiv.org/abs/1609.00717}{{\ttfamily 1609.00717}}.

\bibitem{Perko:2016puo}
A.~Perko, L.~Senatore, E.~Jennings and R.~H. Wechsler, \emph{{Biased Tracers in
  Redshift Space in the EFT of Large-Scale Structure}},
  \href{https://arxiv.org/abs/1610.09321}{{\ttfamily 1610.09321}}.

\bibitem{Nadler:2017qto}
E.~O. Nadler, A.~Perko and L.~Senatore, \emph{{On the Bispectra of Very Massive
  Tracers in the Effective Field Theory of Large-Scale Structure}},
  \href{https://doi.org/10.1088/1475-7516/2018/02/058}{\emph{JCAP} {\bfseries
  1802} (2018) 058} [\href{https://arxiv.org/abs/1710.10308}{{\ttfamily
  1710.10308}}].

\bibitem{McDonald:2009dh}
P.~McDonald and A.~Roy, \emph{{Clustering of dark matter tracers: generalizing
  bias for the coming era of precision LSS}},
  \href{https://doi.org/10.1088/1475-7516/2009/08/020}{\emph{JCAP} {\bfseries
  0908} (2009) 020} [\href{https://arxiv.org/abs/0902.0991}{{\ttfamily
  0902.0991}}].

\bibitem{Lewandowski:2015ziq}
M.~Lewandowski, L.~Senatore, F.~Prada, C.~Zhao and C.-H. Chuang, \emph{{On the
  EFT of Large Scale Structures in Redshift Space}},
  \href{https://arxiv.org/abs/1512.06831}{{\ttfamily 1512.06831}}.

\bibitem{Lewandowski:2016yce}
M.~Lewandowski, A.~Maleknejad and L.~Senatore, \emph{{An effective description
  of dark matter and dark energy in the mildly non-linear regime}},
  \href{https://doi.org/10.1088/1475-7516/2017/05/038}{\emph{JCAP} {\bfseries
  1705} (2017) 038} [\href{https://arxiv.org/abs/1611.07966}{{\ttfamily
  1611.07966}}].

\bibitem{Cusin:2017wjg}
G.~Cusin, M.~Lewandowski and F.~Vernizzi, \emph{{Dark Energy and Modified
  Gravity in the Effective Field Theory of Large-Scale Structure}},
  \href{https://doi.org/10.1088/1475-7516/2018/04/005}{\emph{JCAP} {\bfseries
  1804} (2018) 005} [\href{https://arxiv.org/abs/1712.02783}{{\ttfamily
  1712.02783}}].

\bibitem{Bose:2018orj}
B.~Bose, K.~Koyama, M.~Lewandowski, F.~Vernizzi and H.~A. Winther,
  \emph{{Towards Precision Constraints on Gravity with the Effective Field
  Theory of Large-Scale Structure}},
  \href{https://doi.org/10.1088/1475-7516/2018/04/063}{\emph{JCAP} {\bfseries
  1804} (2018) 063} [\href{https://arxiv.org/abs/1802.01566}{{\ttfamily
  1802.01566}}].

\bibitem{Assassi:2015jqa}
V.~Assassi, D.~Baumann, E.~Pajer, Y.~Welling and D.~van~der Woude,
  \emph{{Effective theory of large-scale structure with primordial
  non-Gaussianity}},
  \href{https://doi.org/10.1088/1475-7516/2015/11/024}{\emph{JCAP} {\bfseries
  1511} (2015) 024} [\href{https://arxiv.org/abs/1505.06668}{{\ttfamily
  1505.06668}}].

\bibitem{Assassi:2015fma}
V.~Assassi, D.~Baumann and F.~Schmidt, \emph{{Galaxy Bias and Primordial
  Non-Gaussianity}},
  \href{https://doi.org/10.1088/1475-7516/2015/12/043}{\emph{JCAP} {\bfseries
  1512} (2015) 043} [\href{https://arxiv.org/abs/1510.03723}{{\ttfamily
  1510.03723}}].

\bibitem{Bertolini:2015fya}
D.~Bertolini, K.~Schutz, M.~P. Solon, J.~R. Walsh and K.~M. Zurek,
  \emph{{Non-Gaussian Covariance of the Matter Power Spectrum in the Effective
  Field Theory of Large Scale Structure}},
  \href{https://doi.org/10.1103/PhysRevD.93.123505}{\emph{Phys. Rev.}
  {\bfseries D93} (2016) 123505}
  [\href{https://arxiv.org/abs/1512.07630}{{\ttfamily 1512.07630}}].

\bibitem{Bertolini:2016hxg}
D.~Bertolini and M.~P. Solon, \emph{{Principal Shapes and Squeezed Limits in
  the Effective Field Theory of Large Scale Structure}},
  \href{https://doi.org/10.1088/1475-7516/2016/11/030}{\emph{JCAP} {\bfseries
  1611} (2016) 030} [\href{https://arxiv.org/abs/1608.01310}{{\ttfamily
  1608.01310}}].

\bibitem{Lesgourgues:2006nd}
J.~Lesgourgues and S.~Pastor, \emph{{Massive neutrinos and cosmology}},
  \href{https://doi.org/10.1016/j.physrep.2006.04.001}{\emph{Phys. Rept.}
  {\bfseries 429} (2006) 307}
  [\href{https://arxiv.org/abs/astro-ph/0603494}{{\ttfamily
  astro-ph/0603494}}].

\bibitem{Audren:2012vy}
B.~Audren, J.~Lesgourgues, S.~Bird, M.~G. Haehnelt and M.~Viel, \emph{{Neutrino
  masses and cosmological parameters from a Euclid-like survey: Markov Chain
  Monte Carlo forecasts including theoretical errors}},
  \href{https://doi.org/10.1088/1475-7516/2013/01/026}{\emph{JCAP} {\bfseries
  1301} (2013) 026} [\href{https://arxiv.org/abs/1210.2194}{{\ttfamily
  1210.2194}}].

\bibitem{Cerbolini:2013uya}
M.~C.~A. Cerbolini, B.~Sartoris, J.-Q. Xia, A.~Biviano, S.~Borgani and M.~Viel,
  \emph{{Constraining neutrino properties with a Euclid-like galaxy cluster
  survey}}, \href{https://doi.org/10.1088/1475-7516/2013/06/020}{\emph{JCAP}
  {\bfseries 1306} (2013) 020}
  [\href{https://arxiv.org/abs/1303.4550}{{\ttfamily 1303.4550}}].

\bibitem{Senatore:massive_neutrinos}
L.~Senatore and M.~Zaldarriaga, \emph{{The Effective Field Theory of
  Large-Scale Structure in the presence of Massive Neutrinos}},
  \href{https://arxiv.org/abs/1707.04698}{{\ttfamily 1707.04698}}.

\bibitem{Baldauf:Bispectrum}
T.~Baldauf, L.~Mercolli, M.~Mirbabayi and E.~Pajer, \emph{{The Bispectrum in
  the Effective Field Theory of Large Scale Structure}},
  \href{https://doi.org/10.1088/1475-7516/2015/05/007}{\emph{JCAP} {\bfseries
  1505} (2015) 007} [\href{https://arxiv.org/abs/1406.4135}{{\ttfamily
  1406.4135}}].

\bibitem{Villaescusa-Navarro:2013pva}
F.~Villaescusa-Navarro, F.~Marulli, M.~Viel, E.~Branchini, E.~Castorina,
  E.~Sefusatti et~al., \emph{{Cosmology with massive neutrinos I: towards a
  realistic modeling of the relation between matter, haloes and galaxies}},
  \href{https://doi.org/10.1088/1475-7516/2014/03/011}{\emph{JCAP} {\bfseries
  1403} (2014) 011} [\href{https://arxiv.org/abs/1311.0866}{{\ttfamily
  1311.0866}}].

\bibitem{Baldi:2013iza}
M.~Baldi, F.~Villaescusa-Navarro, M.~Viel, E.~Puchwein, V.~Springel and
  L.~Moscardini, \emph{{Cosmic degeneracies – I. Joint N-body simulations of
  modified gravity and massive neutrinos}},
  \href{https://doi.org/10.1093/mnras/stu259}{\emph{Mon. Not. Roy. Astron.
  Soc.} {\bfseries 440} (2014) 75}
  [\href{https://arxiv.org/abs/1311.2588}{{\ttfamily 1311.2588}}].

\bibitem{Castorina:2013wga}
E.~Castorina, E.~Sefusatti, R.~K. Sheth, F.~Villaescusa-Navarro and M.~Viel,
  \emph{{Cosmology with massive neutrinos II: on the universality of the halo
  mass function and bias}},
  \href{https://doi.org/10.1088/1475-7516/2014/02/049}{\emph{JCAP} {\bfseries
  1402} (2014) 049} [\href{https://arxiv.org/abs/1311.1212}{{\ttfamily
  1311.1212}}].

\bibitem{Castorina:2015bma}
E.~Castorina, C.~Carbone, J.~Bel, E.~Sefusatti and K.~Dolag, \emph{{DEMNUni:
  The clustering of large-scale structures in the presence of massive
  neutrinos}}, \href{https://doi.org/10.1088/1475-7516/2015/07/043}{\emph{JCAP}
  {\bfseries 1507} (2015) 043}
  [\href{https://arxiv.org/abs/1505.07148}{{\ttfamily 1505.07148}}].

\bibitem{Zennaro:2016nqo}
M.~Zennaro, J.~Bel, F.~Villaescusa-Navarro, C.~Carbone, E.~Sefusatti and
  L.~Guzzo, \emph{{Initial Conditions for Accurate N-Body Simulations of
  Massive Neutrino Cosmologies}},
  \href{https://doi.org/10.1093/mnras/stw3340}{\emph{Mon. Not. Roy. Astron.
  Soc.} {\bfseries 466} (2017) 3244}
  [\href{https://arxiv.org/abs/1605.05283}{{\ttfamily 1605.05283}}].

\bibitem{Banerjee:2016zaa}
A.~Banerjee and N.~Dalal, \emph{{Simulating nonlinear cosmological structure
  formation with massive neutrinos}},
  \href{https://doi.org/10.1088/1475-7516/2016/11/015}{\emph{JCAP} {\bfseries
  1611} (2016) 015} [\href{https://arxiv.org/abs/1606.06167}{{\ttfamily
  1606.06167}}].

\bibitem{Banerjee:2018bxy}
A.~Banerjee, D.~Powell, T.~Abel and F.~Villaescusa-Navarro, \emph{{Reducing
  Noise in Cosmological N-body Simulations with Neutrinos}},
  \href{https://arxiv.org/abs/1801.03906}{{\ttfamily 1801.03906}}.

\bibitem{Hu:1998kj}
W.~Hu, \emph{{Structure formation with generalized dark matter}},
  \href{https://doi.org/10.1086/306274}{\emph{Astrophys. J.} {\bfseries 506}
  (1998) 485} [\href{https://arxiv.org/abs/astro-ph/9801234}{{\ttfamily
  astro-ph/9801234}}].

\bibitem{Saito:2008bp}
S.~Saito, M.~Takada and A.~Taruya, \emph{{Impact of massive neutrinos on
  nonlinear matter power spectrum}},
  \href{https://doi.org/10.1103/PhysRevLett.100.191301}{\emph{Phys. Rev. Lett.}
  {\bfseries 100} (2008) 191301}
  [\href{https://arxiv.org/abs/0801.0607}{{\ttfamily 0801.0607}}].

\bibitem{Wong:2008ws}
Y.~Y.~Y. Wong, \emph{{Higher order corrections to the large scale matter power
  spectrum in the presence of massive neutrinos}},
  \href{https://doi.org/10.1088/1475-7516/2008/10/035}{\emph{JCAP} {\bfseries
  0810} (2008) 035} [\href{https://arxiv.org/abs/0809.0693}{{\ttfamily
  0809.0693}}].

\bibitem{Lesgourgues:2009am}
J.~Lesgourgues, S.~Matarrese, M.~Pietroni and A.~Riotto, \emph{{Non-linear
  Power Spectrum including Massive Neutrinos: the Time-RG Flow Approach}},
  \href{https://doi.org/10.1088/1475-7516/2009/06/017}{\emph{JCAP} {\bfseries
  0906} (2009) 017} [\href{https://arxiv.org/abs/0901.4550}{{\ttfamily
  0901.4550}}].

\bibitem{Shoji:2009gg}
M.~Shoji and E.~Komatsu, \emph{{Third-order Perturbation Theory With Non-linear
  Pressure}},
  \href{https://doi.org/10.1088/0004-637X/700/1/705}{\emph{Astrophys. J.}
  {\bfseries 700} (2009) 705}
  [\href{https://arxiv.org/abs/0903.2669}{{\ttfamily 0903.2669}}].

\bibitem{Shoji:2010hm}
M.~Shoji and E.~Komatsu, \emph{{Massive Neutrinos in Cosmology: Analytic
  Solutions and Fluid Approximation}},
  \href{https://doi.org/10.1103/PhysRevD.81.123516,
  10.1103/PhysRevD.82.089901}{\emph{Phys. Rev.} {\bfseries D81} (2010) 123516}
  [\href{https://arxiv.org/abs/1003.0942}{{\ttfamily 1003.0942}}].

\bibitem{Lesgourgues:2011rh}
J.~Lesgourgues and T.~Tram, \emph{{The Cosmic Linear Anisotropy Solving System
  (CLASS) IV: efficient implementation of non-cold relics}},
  \href{https://doi.org/10.1088/1475-7516/2011/09/032}{\emph{JCAP} {\bfseries
  1109} (2011) 032} [\href{https://arxiv.org/abs/1104.2935}{{\ttfamily
  1104.2935}}].

\bibitem{Upadhye:2013ndm}
A.~Upadhye, R.~Biswas, A.~Pope, K.~Heitmann, S.~Habib, H.~Finkel et~al.,
  \emph{{Large-Scale Structure Formation with Massive Neutrinos and Dynamical
  Dark Energy}}, \href{https://doi.org/10.1103/PhysRevD.89.103515}{\emph{Phys.
  Rev.} {\bfseries D89} (2014) 103515}
  [\href{https://arxiv.org/abs/1309.5872}{{\ttfamily 1309.5872}}].

\bibitem{Blas:2014hya}
D.~Blas, M.~Garny, T.~Konstandin and J.~Lesgourgues, \emph{{Structure formation
  with massive neutrinos: going beyond linear theory}},
  \href{https://doi.org/10.1088/1475-7516/2014/11/039}{\emph{JCAP} {\bfseries
  1411} (2014) 039} [\href{https://arxiv.org/abs/1408.2995}{{\ttfamily
  1408.2995}}].

\bibitem{Fuhrer:2014zka}
F.~F{\"{u}}hrer and Y.~Y.~Y. Wong, \emph{{Higher-order massive neutrino
  perturbations in large-scale structure}},
  \href{https://doi.org/10.1088/1475-7516/2015/03/046}{\emph{JCAP} {\bfseries
  1503} (2015) 046} [\href{https://arxiv.org/abs/1412.2764}{{\ttfamily
  1412.2764}}].

\bibitem{Levi:2016tlf}
M.~Levi and Z.~Vlah, \emph{{Massive neutrinos in nonlinear large scale
  structure: A consistent perturbation theory}},
  \href{https://arxiv.org/abs/1605.09417}{{\ttfamily 1605.09417}}.

\bibitem{AliHaimoud:2012vj}
Y.~Ali-Haimoud and S.~Bird, \emph{{An efficient implementation of massive
  neutrinos in non-linear structure formation simulations}},
  \href{https://doi.org/10.1093/mnras/sts286}{\emph{Mon. Not. Roy. Astron.
  Soc.} {\bfseries 428} (2012) 3375}
  [\href{https://arxiv.org/abs/1209.0461}{{\ttfamily 1209.0461}}].

\bibitem{Ruggeri:2017dda}
R.~Ruggeri, E.~Castorina, C.~Carbone and E.~Sefusatti, \emph{{DEMNUni: Massive
  neutrinos and the bispectrum of large scale structures}},
  \href{https://doi.org/10.1088/1475-7516/2018/03/003}{\emph{JCAP} {\bfseries
  1803} (2018) 003} [\href{https://arxiv.org/abs/1712.02334}{{\ttfamily
  1712.02334}}].

\bibitem{Bertschinger:1988mu}
E.~Bertschinger, \emph{{Cosmic Strings and Galaxy Formation}},
  \href{https://doi.org/10.1111/j.1749-6632.1989.tb50503.x}{\emph{Annals N. Y.
  Acad. Sci.} {\bfseries 571} (1989) 151}.

\bibitem{Lewandowski2016:Greensfunction}
M.~Lewandowski, A.~Maleknejad and L.~Senatore, \emph{{An effective description
  of dark matter and dark energy in the mildly non-linear regime}},
  \href{https://doi.org/10.1088/1475-7516/2017/05/038}{\emph{JCAP} {\bfseries
  1705} (2017) 038} [\href{https://arxiv.org/abs/1611.07966}{{\ttfamily
  1611.07966}}].

\bibitem{Bernardeau:2001CPT}
F.~Bernardeau, S.~Colombi, E.~Gaztanaga and R.~Scoccimarro, \emph{{Large scale
  structure of the universe and cosmological perturbation theory}},
  \href{https://doi.org/10.1016/S0370-1573(02)00135-7}{\emph{Phys. Rept.}
  {\bfseries 367} (2002) 1}
  [\href{https://arxiv.org/abs/astro-ph/0112551}{{\ttfamily
  astro-ph/0112551}}].

\bibitem{Lewis:1999bs}
A.~Lewis, A.~Challinor and A.~Lasenby, \emph{{Efficient computation of CMB
  anisotropies in closed FRW models}},
  \href{https://doi.org/10.1086/309179}{\emph{Astrophys. J.} {\bfseries 538}
  (2000) 473} [\href{https://arxiv.org/abs/astro-ph/9911177}{{\ttfamily
  astro-ph/9911177}}].

\bibitem{Ade:2015xua}
{\scshape Planck} collaboration, P.~A.~R. Ade et~al., \emph{{Planck 2015
  results. XIII. Cosmological parameters}},
  \href{https://doi.org/10.1051/0004-6361/201525830}{\emph{Astron. Astrophys.}
  {\bfseries 594} (2016) A13}
  [\href{https://arxiv.org/abs/1502.01589}{{\ttfamily 1502.01589}}].

\end{thebibliography}\endgroup
\bibliographystyle{JHEP}

\end{document}